\documentclass[twocolumn,superscriptaddress,floatfix,preprintnumbers,prd,nofootinbib]{revtex4-2}
\usepackage[colorlinks=true,breaklinks=true]{hyperref}
\usepackage{multirow}
\usepackage[utf8]{inputenc}
\hypersetup{urlcolor=Blue,
	    citecolor=Blue,
	    linkcolor=Blue}
\usepackage{url}
\usepackage{mathtools}
\usepackage{amssymb}
\usepackage[usenames,dvipsnames]{xcolor}

\usepackage{orcidlink}
\usepackage{enumitem}

\long\def\exclude#1{}
\def\hp{\hphantom{1}}
\def\hm{\hphantom{$-$}}

\newcommand{\Nhit}{N_{\rm hit}}

\bibpunct{[}{]}{,}{n}{}{,}

\makeatletter
\def\l@subsection#1#2{}
\def\l@subsubsection#1#2{}
\makeatother

\begin{document}

\title{Supernova Simulations Confront SN 1987A Neutrinos}

\author{Damiano F.\ G.\ Fiorillo \orcidlink{0000-0003-4927-9850}} 
\affiliation{Niels Bohr International Academy, Niels Bohr Institute,
University of Copenhagen, 2100 Copenhagen, Denmark}

\author{Malte Heinlein \orcidlink{0009-0003-5147-6105}}
\affiliation{Max-Planck-Institut f\"ur Astrophysik, 
Karl-Schwarzschild-Str.~1, 85748 Garching, Germany}
\affiliation{Technische Universität München, TUM School of Natural Sciences, Physics Department, James-Franck-Str.~1, 85748 Garching, Germany}

\author{\hbox{Hans-Thomas Janka \orcidlink{0000-0002-0831-3330}}}
\affiliation{Max-Planck-Institut f\"ur Astrophysik, Karl-Schwarzschild-Str.~1, 85748 Garching, Germany}

\author{Georg Raffelt \orcidlink{0000-0002-0199-9560}}
\affiliation{Max-Planck-Institut f\"ur Physik (Werner-Heisenberg-Institut), 
Boltzmannstr.~8,  85748 Garching, Germany}

\author{Edoardo Vitagliano
\orcidlink{0000-0001-7847-1281}}
\affiliation{Racah Institute of Physics, Hebrew University of Jerusalem, Jerusalem 91904, Israel}

\author{Robert Bollig \orcidlink{0000-0003-2354-2454}}
\affiliation{Max-Planck-Institut f\"ur Astrophysik, Karl-Schwarzschild-Str.~1, 85748 Garching, Germany}

\date{August 4, 2023}

\begin{abstract}
We return to interpreting the historical SN~1987A neutrino data from a modern perspective. To this end, we construct a suite of spherically symmetric supernova models with the \textsc{Prometheus-Vertex} code, using four different equations of state and five choices of final baryonic neutron-star (NS) mass in the 1.36--1.93\,M$_\odot$ range. Our models include muons and proto-neutron star (PNS) convection by a mixing-length approximation.  The time-integrated signals of our 1.44\,M$_\odot$ models agree reasonably well with the combined data of the four relevant experiments, IMB, Kam-II, BUST, and LSD, but the high-threshold IMB detector alone favors a NS mass of 1.7--1.8\,M$_\odot$, whereas \hbox{Kam-II} alone prefers a mass around 1.4\,M$_\odot$. The cumulative energy distributions in these two detectors are well matched by models for such NS masses, and the previous tension between predicted mean neutrino energies and the combined measurements is gone, with and without flavor swap. Generally, our predicted signals do not strongly depend on assumptions about flavor mixing, because the PNS flux spectra depend only weakly on antineutrino flavor. 
While our models show compatibility with the events detected during the first seconds, PNS convection and nucleon correlations in the neutrino opacities lead to short PNS cooling times of 5--9\,s, in conflict with the late event bunches in Kam-II and BUST after 8--9\,s, which are also difficult to explain by background. Speculative interpretations include the onset of fallback of transiently ejected material onto the NS, a late phase transition in the nuclear medium, e.g.\ from hadronic to quark matter, or other effects that add to the standard PNS cooling emission and either stretch the signal or provide a late source of energy. More research, including systematic 3D simulations, is needed to assess these open issues.
\end{abstract}

\maketitle

\renewcommand{\baselinestretch}{1}\normalsize
\tableofcontents
\renewcommand{\baselinestretch}{1.0}\normalsize

\section{Introduction}
\label{sec:intro}

After almost four decades, the historical SN~1987A of $23^{\rm rd}$~February 1987 remains the only case of a measured neutrino signal from stellar core collapse. Today, many large-scale detectors are running or in preparation so that the neutrino signal from the next nearby supernova (SN) will provide a bonanza of high-statistics information on the dynamics of core collapse (CC) and SN explosion, on neutrinos and their flavor-dependent interaction and propagation, the nuclear equation of state, and hypothetical feebly interacting particles. Also multi-messenger information including gravitational waves will yield new insights. Standard and nonstandard astrophysical and particle-physics ideas will be put to the test \cite{Raffelt:1996wa, Dighe:1999bi, Dighe:2003jg,Raffelt:2010zza, Scholberg:2012id, 
IceCube:2011cwc, Serpico:2011ir, Kuroda:2013rga,Gossan:2015xda, Mirizzi:2015eza, Fischer:2016cyd, Meyer:2016wrm,Bustamante:2016ciw, Jaeckel:2017tud, Horowitz:2018ndv, Horiuchi:2018ofe, DeRocco:2019jti, DeRocco:2019njg, Ge:2020zww,Li:2020ujl, Caputo:2021kcv, Baryakhtar:2022hbu, Akita:2022etk, Lin:2022dbl, Jana:2022tsa,Jana:2023ufy, Antel:2023hkf, Muller:2023vjm}.

Until that time, however, the SN~1987A legacy data remain the only direct test of such questions. Broadly, the data agree with expectations, as confirmed in a number of studies following the event~\cite{Bahcall:1987ua, Burrows:1988ba, Janka+1989, Jegerlehner:1996kx, Loredo:2001rx, Mirizzi:2005tg, Costantini:2004ry, Costantini:2006xd, Pagliaroli:2008ur, Ianni:2009bd, Vissani:2014doa}, but at that time, theoretical understanding and numerical SN modeling were in their infancy, and after almost four decades of progress, this question deserves a fresh look, a sentiment also shared by other recent authors \cite{Olsen:2021uvt, Li:2023ulf, DedinNeto:2023hhp}. How well do modern SN models agree with the old data and are there open issues? One motivation to return to this subject is the role of SN~1987A as a particle-physics laboratory, a topic that has gained a fresh boost of activity over the past few years \cite{Payez:2014xsa, Chang:2016ntp, Jaeckel:2017tud, Chang:2018rso, DeRocco:2019njg, Carenza:2019pxu, Croon:2020lrf, Carenza:2020cis, Camalich:2020wac,Caputo:2021rux, Fiorillo:2022cdq, Caputo:2022rca, Ferreira:2022xlw, Caputo:2022mah, Hoof:2022xbe, Diamond:2023scc, Lella:2023bfb,Manzari:2023gkt}. However, here we do not delve into question of new physics, but simply ask about the match between new models with old data.

One frontier of CCSN modeling has been the progress from spherically symmetric (1D) to fully three-dimen\-sional (3D) simulations. Today, it is widely agreed that the delayed neutrino-driven explosion mechanism proposed by Bethe and Wilson \cite{Bethe+1985} only a few years before SN~1987A is the correct paradigm, supported by the successful explosions obtained in multi-D simulations \cite{Janka:2012wk, Janka:2016fox, Mueller2016, Janka2017Handbooka, Mueller2020, Burrows+2020, Mezzacappa+2020, Burrows+2021}. Nevertheless, to compare with the measured neutrino signal from SN~1987A, self-consistently exploding SN models, carried through all of the proto-neutron star (PNS) cooling evolution in 3D with different nuclear equations of state (EoSs), are not yet available. However, such detailed modeling is probably not required for comparison with the sparse SN~1987A neutrino data, which do not capture details of the time structure of the neutrino emission, but mainly test the phases of post-bounce accretion and PNS Kelvin-Helmholtz cooling. 

Therefore, our study will be parametric in the sense that we perform 1D simulations with artificially triggered explosions, choosing the progenitor profile and instant of explosion such that the final neutron star (NS) reaches a chosen baryonic mass, where we specifically investigate the range 1.36--1.93\,M$_\odot$. Moreover, we use four different nuclear EoSs that are widely applied in CCSN simulations and that are compatible with current experimental and astrophysical constraints. These two ``parameters'' define our model space.

The PNS mass is a more appropriate ordering parameter than the usually considered progenitor birth mass (the one on the zero-age main sequence or ZAMS) for two reasons. On the one hand, mass loss or gain (by stellar winds and binary interaction) imply a severe uncertainty in the connection between ZAMS mass and the stellar mass before collapse and thus of the compact remnant. On the other hand, the outcome of stellar evolution and core collapse depends very non-monotonically on the ZAMS and pre-collapse mass \cite{OConnor+2011,Ugliano+2012,Sukhbold+2014,Sukhbold+2018}. The PNS mass, however, correlates closely with the progenitor's core compactness\footnote{The compactness value for a certain enclosed mass $M$ is defined by $\xi_M = (M/\mathrm{M}_\odot)/(R(M)/1000\,\mathrm{km})$, where $R(M)$ is the radius enclosing mass $M$~\cite{OConnor+2011}.} \cite{Nakamura+2015}, which governs the neutrino emission of collapsing stars in the pre-explosion phase~\cite{OConnor+2013}, as well as the post-explosion neutrino emission of PNSs in 2D simulations~\cite{Vartanyan+2023}. Therefore the PNS mass, which lacks the ambiguity of the chosen $M$-value for the core compactness, serves well as an ordering parameter for the neutrino signal in the pre-explosion as well as post-explosion phases. 

Our numerical simulations employ the \textsc{Pro\-metheus-Vertex} neutrino-hydrodynamics code, which solves the fully energy and velocity dependent neutrino transport for all six species of neutrinos and antineutrinos with a state-of-the-art implementation of the neutrino interactions. In particular, the models take into account the presence of muons in the hot PNS, including the corresponding muonic neutrino interactions~\cite{Bollig:2017lki,Bollig:2020xdr}, although  the $\mu$ and $\tau$ flavored spectral differences turn out to be small in practice. Therefore, we will usually only distinguish between $\bar\nu_e$ and $\bar\nu_x$ flux spectra, the latter being an average of $\bar\nu_\mu$ and $\bar\nu_\tau$. 

Numerical models at the time of SN~1987A suggested a large flavor-dependent hierarchy of the average neutrino energies and thus potentially large flavor-conversion effects. Present-day models, on the other hand, yield only relatively small differences so that the issues of flavor dependence are not a dominant concern for our work.

The effects of flavor conversion are not included in our SN and PNS simulations: neutrino masses and mixing angles are effectively set to zero, following the standard treatment of most SN simulations (with few and constrained exceptions~\cite{Stapleford+2020}). The phenomenon of fast-flavor conversion, which does not directly depend on neutrino vacuum properties, is also left out as in all other SN simulations except for recent parametric studies~\cite{Ehring+2023,Ehring+2023a}.  Perhaps the most fundamental development since SN~1987A has been the experimental and theoretical progress in neutrino flavor conversion physics, for which then only the Home\-stake solar neutrino experiment provided first and only preliminary evidence. Ironically, ab-initio implementations in CCSN simulations remain elusive because of our uncertain understanding of collective effects caused by neutrino-neutrino refraction.

One intriguing 3D phenomenon that modifies the neutrino emission during the post-bounce phase is asymmetric neutrino emission by anisotropic accretion of postshock matter onto the PNS and by the LESA (Lepton-number Emission Self-sustained Asymmetry) effect, which is a hemispheric asymmetry of lepton-number emission and thus especially of the $\bar\nu_e$ flux and spectra due to low-order spherical harmonics asymmetries of convection inside the PNS \cite{Tamborra+2014, Tamborra+2014a, Janka:2016fox, OConnor+2018a, Glas+2019, Walk+2019, Powell+2019, Vartanyan+2019}. In other words, the observed $\bar\nu_e$ signal depends on the direction of observation relative to the SN. This is a random and time-dependent variable~\cite{Tamborra+2014a} and as such an unavoidable uncertainty of the expected species-dependent neutrino fluxes. Once more, because the differences between the $\bar\nu_e$ and $\bar\nu_x$ flux spectra are not large, this effect should be seen as an opportunity for future high-statistics observations, but not as a major issue of the SN~1987A data interpretation.

There is one important 3D effect that we cannot ignore: Ledoux convection in the interior of the PNS and its long-time impact on the PNS evolution by the lepton-number and energy loss in neutrinos \cite{Epstein1979, Burrows+1988, Keil+1996, Janka+2001proc, Buras+2006a, Dessart+2006, Nagakura+2020}. Ledoux convection is driven by entropy and lepton-number gradients and as such is generic. In our 1D simulations, we use a mixing-length treatment to describe the accelerated energy and lepton-number transport by this fluid-dynamic effect~\cite{Mirizzi+2016}. Actually, we will see that the cooling timescales of 5--9~s of our models and the associated short signal durations are not easily compatible with the SN~1987A data except in the sense of a rare signal or background fluctuation of the sparse data. This question has not been systematically explored in previous analyses of SN~1987A data and many numerical models do not include PNS convection (e.g.~Refs.~\cite{Nakazato+2020, Fischer+2020, Sumiyoshi+2023}, but contrariwise Refs.~\cite{Roberts+2012a, Mirizzi:2015eza, Nagakura+2021b,Pascal+2022}). Many traditional SN~1987A particle-physics constraints rely on the long signal duration that would be reduced by novel channels of energy loss \cite{Raffelt:1987yt, Turner:1987by, Mayle:1987as, Burrows:1988ah, Keil:1996ju, Raffelt:2006cw}.

One of our main findings is that the questions of signal duration, PNS convection, and novel energy losses require further systematic studies. Such explorations should also
include 3D CCSN modeling for long evolution periods and with different theories of the nuclear EoS, which affects PNS convection \cite{Roberts+2012a}, because only 3D simulations can provide confirmation for the viability of the mixing-length approximation for PNS convection over seconds. Moreover, only 3D CCSN models can yield reliable answers for the duration of the post-bounce accretion phase, for accretion and LESA asymmetries in the neutrino emission, and for later fallback of initial SN ejecta. Clearly, such aspects reach far beyond the scope of the present work.

To perform a direct comparison between models and data, we have systematically collected the experimental information that is somewhat scattered in the older literature. We provide many details in a long Appendix that may also be useful for other researchers. We include data from all four relevant experiments, IMB (Irvine-Michigan-Brookhaven), Kamiokande-II
(Kam-II), BUST (Baksan), and LSD (Mont Blanc), in the latter case the non-observation at the time when the other detectors registered the neutrino signal. For the largest detector IMB, for the first time we explicitly include the large uncertainty of the trigger efficiency, leading roughly to a $\pm50\%$ uncertainty of the expected event number, although the usual tension between the average neutrino energies seen in IMB and Kam-II persists in somewhat reduced form.

We will perform a variety of different analyses. First, we consider a generic time-integrated neutrino signal, assuming it is represented by a quasi-thermal spectrum, described by the total energy in electron antineutrinos $\bar\nu_e$ arriving at the detectors, their average energy, and the pinching parameter $\alpha$ of the spectral shape. In this way we construct confidence regions in parameter space, similar to those in the previous literature, and compare them with the parameter values implied by our numerical models. We provide similar results under the assumption that the late events (after 6\,s) in Kam-II and BUST are not associated with PNS cooling. Moreover, we compare cumulative energy distributions of the time-integrated signals with those measured by three detectors.

The next type of analysis uses the energy-integrated but time-dependent signal, which is compared with the time structure of the detected events as well as with respect to the overall predicted cooling periods and signal durations. We defined the latter as the periods over which 95\% of the expected number of events would have arrived in a given detector. It is this analysis that reveals a huge tension between expected and observed signal durations and leads us to speculate that the late events might have an origin other than PNS cooling.

Finally, we perform a global maximum-likelihood comparison of our models, treating their PNS mass and EoS as fit parameters. We believe that such a model comparison makes only sense under the assumption that this class of models can actually explain the data, which does not seem to be the case for the late events. Therefore, we compare the models under the assumption that within our model space, the late events in Kam-II and BUST have a different explanation than PNS cooling. In this case, the smaller-mass models are clearly favored, driven by the small number of events in Kam-II. Looking at individual detectors, IMB alone would favor models with larger masses.

The remainder of our paper begins in Sec.~\ref{sec:models} with a description of our numerical models, their global properties, and the overall characteristics of their flavor-dependent neutrino outputs. In Sec.~\ref{sec:data} we provide a brief overview of the detectors and data, leaving most of the details to Appendix~\ref{sec:SN-Observations}. Next, in Sec.~\ref{sec:Time-Integrated-Analysis}, we turn to the time-integrated analysis with or without the late events as well as assuming Maxwell-Boltzmann or quasi-thermal pinched flux spectra. In Sec.~\ref{sec:Time-Dependent-Analysis} a time-dependent analysis follows and in Sec.~\ref{sec:Overall-Comparison} an overall model comparison. Section~\ref{sec:speculations} is devoted to possible interpretations of the tension in the observed time structure with our convective models before summarizing our findings in Sec.~\ref{sec:Discussion}. Many technical details and tables are relegated to a series of appendices.

\onecolumngrid

\clearpage

\section{Numerical Supernova Models}
\label{sec:models}

\twocolumngrid

\subsection{Modeling Tools and Inputs}

The neutrino transport module \textsc{Vertex} of the \textsc{Pro\-me\-theus-Ver\-tex} neutrino-hydrodynamics code integrates the velocity-dependent (order $v/c$ of the fluid velocity $v$) neutrino energy and momentum equations, discretized in space, time, and neutrino energy, for neutrinos and antineutrinos of all flavors. This set of angular-moment equations of the Boltzmann equation is closed by a variable Eddington factor obtained from the solution of a model Boltzmann equation~\cite{Rampp+2002}. The time integration of the coupled moment equations and Boltzmann equations is performed implicitly in an iterative procedure to achieve convergence up to a predefined precision. The solution of the transport problem provides the source terms for lepton number, energy, and momentum (pressure when neutrinos are trapped) needed in the (1D and multi-D) \textsc{Prometheus} hydrodynamics module, which is an explicit, finite-volume Eulerian multifluid code~\cite{Fryxell+1991,Mueller+1991}, based on the piecewise
parabolic method~\cite{Colella+1984} and employing an exact, iterative Riemann solver for real gases~\cite{Colella+1985}. 

Although \textsc{Prometheus} solves the conservation equations of Newtonian hydrodynamics (for mass, energy, momentum, and composition variables, i.e., nuclear species and electrons and muons as charged leptons), the source term for gravity includes general relativistic corrections~\cite{Rampp+2002,Marek+2006}, and the \textsc{Vertex} transport solver accounts for effects of general relativistic redshifting and time dilation~\cite{Rampp+2002}. The finite-volume discretization of nearly all terms in the transport and hydrodynamics equations permits an almost ideal global conservation of lepton number and energy with numerical errors in the percent range for the full CCSN problem~\cite{Rampp+2002,Mueller+2010}. The 1D and multi-D versions of \textsc{Prometheus-Vertex} were favorably tested against other codes used by the community of CCSN modelers \cite{Liebendoerfer+2005,Mueller+2010,OConnor+2018,Just+2018}. 

The CCSN and PNS models employed in this paper, though computed in 1D, account for the effects of PNS convection by a mixing-length treatment as described, applied, and also tested against long-time 2D PNS cooling simulations \cite{Mirizzi+2016}. Neutrinos, which are trapped at the high-density conditions in the PNS convection layer, are taken into account in the criterion for Ledoux convection and the convective fluxes by their beta-equilibrium conditions. PNS convection has been shown to be a dominant factor accelerating the electron lepton number and energy loss and thus deleptonization and cooling of the hot PNS~\cite{Roberts+2012a,Mirizzi+2016}, a fact that has recently been confirmed~\cite{Pascal+2022}.

The EoS of the stellar medium is treated as a combination of different regimes. At densities above a certain threshold $\rho_\mathrm{th}$, a tabulated EoS for the interacting nuclear components at high densities is applied, coupled to the EoS contributions from photons, electrons and positrons as well as (anti-)muons, which are described as ideal boson or fermion gases, respectively. The plasma modifications of the photon and electron dispersion relations are ignored, but all charged leptons take on their appropriate degree of degeneracy, and the use of relativistic energy-momentum relations with vacuum masses ensures a consistent transition to the limiting case of nonrelativistic particles at low densities or temperatures.\break 
The threshold density $\rho_\mathrm{th}$ is chosen to be $10^{11}\,$g\,cm$^{-3}$ after core bounce, where matter in the postshock region is safely in nuclear statistical equilibrium (NSE), and it is set to a lower value (down to $10^8\,$g\,cm$^{-3}$) prior to bounce in order to connect the high-density composition smoothly into the low-density regime for high-density EoS models that assume a representative, density-dependent heavy nucleus in NSE.

At densities below $\rho_\mathrm{th}$, the EoS is considered to be a mix of ideal gases of photons, electrons, positrons, and a chosen set of selected nuclear species, including heavy nuclei, alpha particles, neutrons, protons, and light isotopes of He and H (all treated as Boltzmann gases), with Coulomb corrections taken into account approximately. Two regimes are discriminated here: for temperatures above a value of $T_\mathrm{NSE} = 0.5\,\mathrm{MeV} \approx 5.8\,$GK, the nuclear composition is assumed to be in NSE, whereas for lower temperatures, the composition is determined by nuclear burning with a small alpha network~\cite{Bollig+2021} or by a flashing approximation~\cite{Rampp+2002}. The latter is used in all of the 1D calculations performed for this study. 

In our study, we use four different versions of the high-density nuclear EoS, which are all widely used in present-day CCSN simulations, namely the classical LS220 EoS of Lattimer and Swesty~\cite{Lattimer+1991} with a nuclear incompressibility at saturation density of $K = 220$\,MeV, the SFHo and SFHx versions of Steiner, Fischer, and Hempel~\cite{Steiner+2013,Hempel+2010}, and DD2 of Typel et al.~\cite{Typel+2010,Hempel+2010,Hempel+2012}. Saturation density, binding energy, incompressibility, symmetry energy, and slope of the symmetry energy are compatible with current experimental and theoretical constraints or very close to them, and the radii and maximum masses of cold NSs computed with these EoS models comply with astronomical bounds~\hbox{\cite{Fischer+2014,Oertel+2017,Tews+2017}},\footnote{LS220 and SFHo are only marginally compatible with the lower limit for the maximum NS mass of 2.19 (2.09) M$_\odot$ at 1$\sigma$ (3$\sigma$) confidence recently determined by a joined analysis of black widow and redback pulsars~\cite{Romani+2022}. This is a special motivation to include both SFHo and SFHx in our study, although in many respects they yield very similar results.} including the gravitational-wave measurement of the NS merger of GW170817~\cite{Abbott+2018} and X-ray measurements with the {\em Neutron Star Interior Composition ExploreR} (NICER)~\cite{Riley+2019,Riley+2021,Miller+2019,Raaijmakers+2021}.

\textsc{Vertex} includes all neutrino reactions that have been identified as relevant for CCSNe and PNS cooling, obeying detailed-balance through the Kirchhoff-Planck relation, in a state-of-the-art implementation as previously documented~\cite{Buras+2006,Janka+2012}, upgraded more recently in various ways and supplemented by neutrino interactions involving muons~\cite{Lohs2015,Bollig:2017lki,Bollig2018}. Electron captures on a large set of heavy nuclei in NSE are included according to refined rate calculations including screening corrections~\cite{Juodagalvis+2010,Langanke+2003}. Inelastic neutrino-nucleus scattering~\cite{Langanke+2008} as well as coherent neutrino scattering with heavy nuclei and alpha-particles are included~\cite{Bruenn+1997}, too, accounting for ion-screening effects~\cite{Horowitz1997} and using a tabulated effective rate for the entire, large ensemble of nuclei with the integrated mass fraction renormalized to the nuclear composition provided by the high-density EoS. In the low-density EoS regime the coherent scattering rates are summed up over all considered species of heavy nuclei. Moreover, neutrino interactions with a possible (minor) admixture of lighter nuclei are taken into account approximately. The implementation of coherent neutrino scattering off light clusters follows that for alpha particles. Inelastic scattering and absorption are treated by applying the free-nucleon rates of Ref.~\cite{Bruenn1985} for the nucleons in these light clusters, assuming that these processes break up the light clusters; therefore the energies of interacting neutrinos are downshifted by the corresponding threshold values (of $\sim$\,2\,MeV for deuterium and $\sim$\,8\,MeV for tritium and $^3$He). Correlation effects of the nucleons inside the light nuclei, which further reduce the interaction rates at low neutrino energies, are ignored.   

Charged-current $\nu_e$ and $\bar\nu_e$ (direct URCA) interactions with nonrelativistic free nucleons are treated with their full reaction kinematics (see Ref.~\cite{Buras+2006} for the numerical handling), taking into account nucleon recoil and weak-magnetism corrections~\cite{Horowitz2002}, nucleon-nucleon correlations in the random-phase approximation (RPA)~\cite{Reddy+1998,Reddy+1999,Burrows+1999}, effective nucleon masses and axial-vector quenching~\cite{Buras+2006}, as well as nucleon mean-field potentials~\cite{Roberts+2012,Martinez-Pinedo+2012,Horowitz+2012}, where the medium-dependent parameters are adopted from the considered high-density EoS.\footnote{Our PNS simulations do not include neutron decays and their inverse for the production and emission properties of $\bar\nu_e$. These three-body direct URCA reactions might somewhat lift the luminosity and reduce the mean energy of the radiated $\bar\nu_e$ during the late PNS cooling phase, at least in models ignoring the effects of nucleon correlations and PNS convection~\cite{Fischer+2020}. These opposing trends suggest little relevance for the late-time neutrino detection. Test calculations with the neutron decays in our modeling setup are on the way.} Neutral-current neutrino-nucleon scattering is also implemented with its full reaction kinematics~\cite{Buras+2006}, including energy transfer by recoil and many-body effects (due to both density and spin correlations via RPA)~\cite{Reddy+1998,Reddy+1999,Burrows+1998}, weak-magnetism corrections~\cite{Horowitz2002}, and virial effects~\cite{Horowitz+2017}.

Neutrino-electron and positron scattering are implemented according to the rates of Refs.~\cite{Mezzacappa+1993,Chernohorsky1994}, $\nu\bar\nu$ pair production of all flavors by $e^-e^+$ annihilation follows the treatments of Refs.~\cite{Bruenn1985,Pons+1998}, and nucleon-nucleon bremsstrahlung adopts the one-pion exchange approximation~\cite{Hannestad+1998}, which was tested against improvements beyond this description~\cite{Bartl+2016}. \textsc{Vertex} also includes neutrino-pair annihilation between different flavors and elastic scattering of heavy-lepton neutrinos and antineutrinos with electron neutrinos and antineutrinos~\cite{Buras+2003,Bollig2018}. 

The implementation of neutrino reactions involving muons and anti-muons is guided by Ref.~\cite{Lohs2015} and detailed in Ref.~\cite{Bollig2018}. These reactions include neutrino scattering off muons, muon decay to $e^\pm$, charged-current muon neutrino and antineutrino absorption by neutrons and protons (accounting for nucleon recoil and in-medium effects~\cite{Burrows+1999}), respectively, pair annihilation of muons and anti-muons to neutrinos and anti-neutrinos of all flavors, and the conversion between $e^\pm$ and $\mu^\pm$ by neutrino absorption~\cite{Bollig:2017lki}.

For the 1D models considered in this paper, neutrino transport (hydrodynamics) was followed up to 10,000\,km ($6\times 10^7$\,km) with the number of radial zones dynamically increasing from 233 (800) after core bounce to more than 670 (1200) at the end of the simulations,\break 
ensuring sufficiently high resolution inside the PNS and in particular in the density gradient at its surface, which steepens with time. For the transport of neutrinos of all species, an energy grid with 21 coupled bins (the first six equidistant, then geometrically increasing) up to 380\,MeV was used.     

In summary, the main aspects discriminating our 1D simulations of CCSN explosions and PNS cooling from previous models used for comparison with the SN~1987A data are the inclusion of nucleon-nucleon correlations and medium effects in the neutrino-nucleon interactions, a mixing-length approximation for convection inside the PNS, and of muons in the EoS of the high-density medium and the neutrino transport. This implies a distinction of the transport of $\nu_\mu$ and $\nu_\tau$ in addition to the slight differences between heavy-lepton neutrinos and antineutrinos due to the weak-magnetism corrections in their neutral-current scattering reactions with nucleons~\cite{Horowitz2002,Buras+2006}.\break
The differences in the transport and emission properties of the four heavy-lepton neutrino species are, however, rather small and therefore we will consider an average of $\bar\nu_\mu$ and $\bar\nu_\tau$ when discussing flavor oscillations with $\bar\nu_e$. We did not take into account neutrino-flavor conversions in our neutrino-hydrodynamic simulations, thus adopting the standard approach in CCSN modeling (with few recent exceptions~\cite{Stapleford+2020,Ehring+2023,Ehring+2023a}).

\begin{table*}
 \caption{Global properties of our PNS formation models, usually based on six-species neutrino transport with muons and convection. The models denoted by ``-m'' use four-species transport and no muons, those with the name extension ``-c'' use six-species transport and muons but no convection.
 The detailed numerical neutrino outputs are available at the Garching Core-Collapse Supernova Archive \cite{JankaWeb}.
   \label{tab:GlobalProperties}}
 \vskip4pt
    \begin{tabular*}{\textwidth}{@{\extracolsep{\fill}}lrrrrrrrll}
    \hline\hline
Model      & $E_{\rm tot}^{\rm end}$ & $t_{\rm end}$& $t_{\rm acc}$ & $E_{\rm tot}^{\rm acc}$ & $\tau_E$ & $N_e^{\rm end}$ & $\tau_e$ & $N_{\bar\mu}^{\rm end}$ & $\tau_{\bar\mu}$\\
           &[B]          & [s]& [s]&[B]&[s]&[$10^{56}$]&[s]&[$10^{55}$]&[s]\\
\hline
1.36-DD2&180.34&8.69&0.32&44.50&5.19&5.68&4.33&2.77&1.37\\
1.36-LS220&187.22&12.36&0.33&48.27&6.44&5.60&4.94&3.34&2.33\\
1.36-SFHo&196.25&10.50&0.32&46.12&6.18&5.76&5.15&2.88&1.45\\
1.36-SFHx&197.09&10.06&0.32&46.03&6.28&5.75&5.02&2.89&2.00\\
\hline
1.44-DD2&205.33&13.72&0.22&39.66&5.48&6.16&4.71&3.08&1.42\\
1.44-LS220&215.36&14.84&0.23&43.09&6.85&6.02&5.23&3.96&2.60\\
1.44-SFHo&224.35&15.00&0.23&41.77&6.55&6.25&5.75&3.17&1.45\\
1.44-SFHx&225.07&11.72&0.22&40.68&6.66&6.23&5.48&3.17&2.05\\
\hline
1.62-DD2&262.53&10.75&0.51&80.87&5.99&7.06&5.10&3.84&1.56\\
1.62-LS220&272.63&13.58&0.51&88.32&7.22&6.77&4.76&5.56&3.80\\
1.62-SFHo&289.29&14.26&0.51&84.74&7.21&7.20&6.40&3.90&1.75\\
1.62-SFHx&291.33&13.45&0.51&84.73&7.36&7.19&6.23&3.91&2.27\\
\hline
1.77-DD2&314.41&11.26&0.62&105.19&6.44&7.78&5.53&4.50&1.69\\
1.77-LS220&328.32&16.33&0.62&116.04&7.91&7.40&4.71&7.05&4.47\\
1.77-SFHo&348.34&13.28&0.62&110.72&7.77&7.93&6.87&4.55&2.06\\
1.77-SFHx&351.38&13.91&0.62&110.82&7.97&7.91&6.80&4.56&2.29\\
\hline
1.93-DD2&375.79&12.81&0.60&121.90&6.93&8.57&6.02&5.24&1.99\\
1.93-LS220&396.17&19.95&0.60&135.34&8.69&8.09&4.57&8.96&5.53\\
1.93-SFHo&419.55&15.52&0.60&128.85&8.45&8.75&7.63&5.25&2.51\\
1.93-SFHx&425.18&16.38&0.60&129.17&8.74&8.74&7.63&5.33&2.57\\
\hline
1.62-DD2-c&256.90&13.95&0.47&71.39&8.78&6.93&9.44&3.77&0.78\\
1.61-LS220-c&264.67&20.92&0.47&78.17&12.67&6.70&12.72&5.55&0.94\\
1.62-SFHo-c&280.82&19.74&0.47&73.52&12.35&7.09&12.80&3.92&0.54\\
1.62-SFHx-c&279.80&18.75&0.47&73.26&12.87&7.04&12.48&3.97&0.65\\
\hline
1.62-DD2-m&260.43&9.58&0.51&80.73&5.67&6.94&5.03&0.03&0.01\\
1.62-SFHo-m&286.78&13.55&0.51&84.35&6.78&7.09&6.19&0.05&0.02\\
\hline
\end{tabular*}
\vskip4pt
\hbox{\hbox to3em{\hfil$E_{\rm tot}^{\rm end}$~~}\hbox to7cm{Total energy emitted in neutrinos up to $t_{\rm end}$\hfil}
\hbox to3em{\hfil$N_e^{\rm end}$~~}\hbox to8cm{Electron lepton number emitted up to $t_{\rm end}$\hfil}}
\vskip1pt
\hbox{\hbox to3em{\hfil$E_{\rm tot}^{\rm acc}$~~}\hbox to7cm{Same up to $t_{\rm acc}$\hfil}
\hbox to3em{\hfil$N_{\bar\mu}^{\rm end}$~~}\hbox to8cm{Same for anti-muon number\hfil}}
\hbox{\hbox to3em{\hfil$t_{\rm end}$~~}\hbox to7cm{Time at end of simulation\hfil}
\hbox to3em{\hfil$\tau_e$~~}\hbox to8cm{Period over which 95\% of $N_e^{\rm end}$ are emitted\hfil}}
\vskip1pt
\hbox{\hbox to3em{\hfil$t_{\rm acc}$~~}\hbox to7cm{Time at end of accretion phase\hfil}
\hbox to3em{\hfil$\tau_{\bar\mu}$~~}\hbox to8cm{Time when 95\% of $N_{\bar\mu}^{\rm end}$ are reached first\hfil}}
\vskip1pt
\hbox{\hbox to3em{\hfil$\tau_E$~~}\hbox to7cm{Period over which 95\% of \smash{$E_{\rm tot}^{\rm end}$} are emitted\hfil}
\hbox to3em{\hfil}\hbox to8cm{(non-monotonic evolution)\hfil}
}
\vskip8pt
\end{table*}

\begin{figure*}
    \centering
    \includegraphics[width=1.0\textwidth]{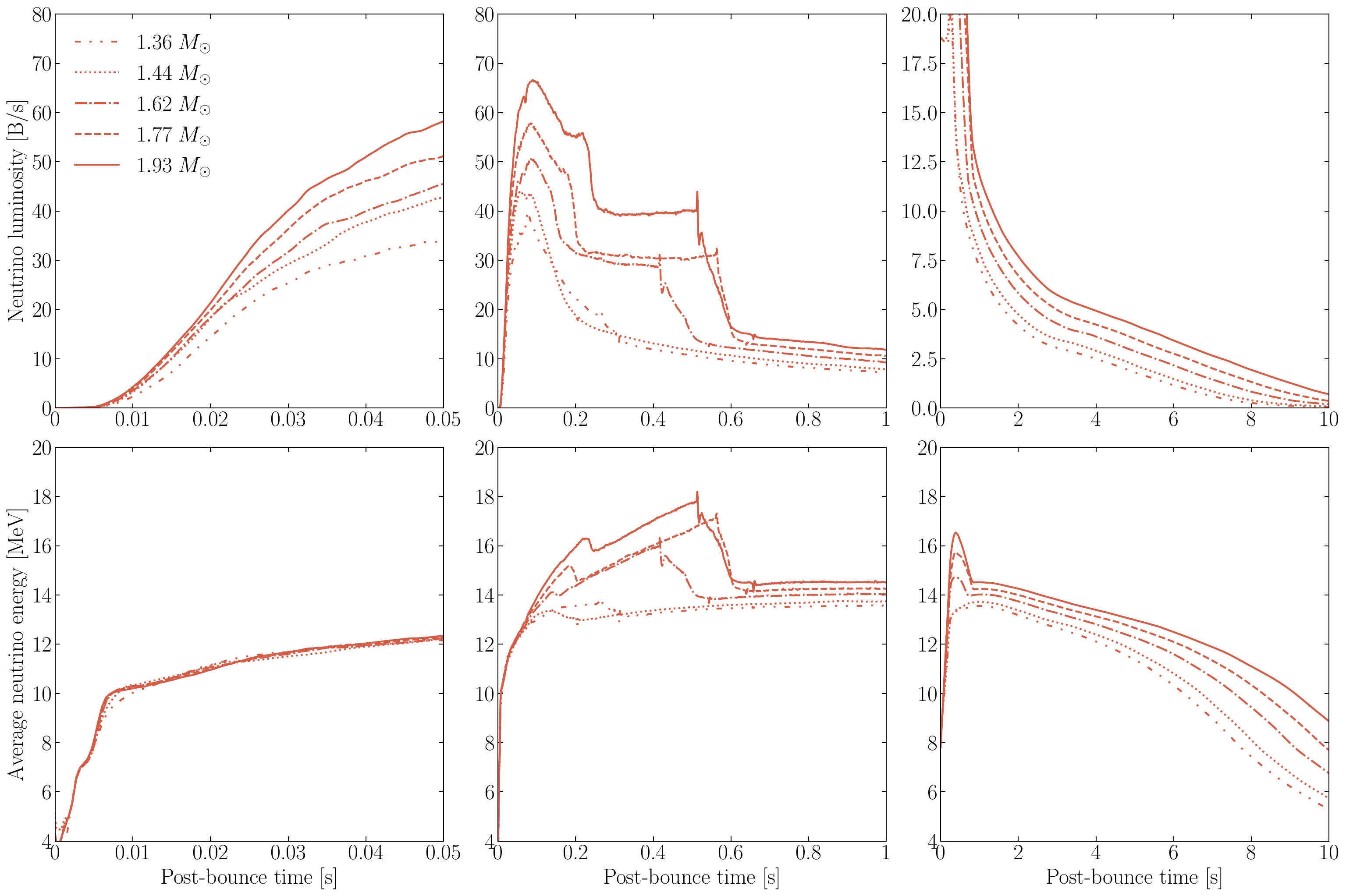}
    \caption{Evolution of the $\bar{\nu}_e$ luminosities (top) and average energies (bottom) for our models with SFHo EoS and different NS masses for the three main emission phases: shock breakout and prompt $\nu_e$ burst (left), post-bounce accretion and shock revival (middle), and PNS cooling (right). Note the different time and luminosity scales. Here and in the following similar plots, the neutrino emission properties are shown for a distant observer at rest in the reference frame of the source.} 
    \label{fig:SN-Model-PNSmass}
    \vskip4pt
\end{figure*}

\begin{figure*}
    \centering
    \includegraphics[width=1.0\textwidth]{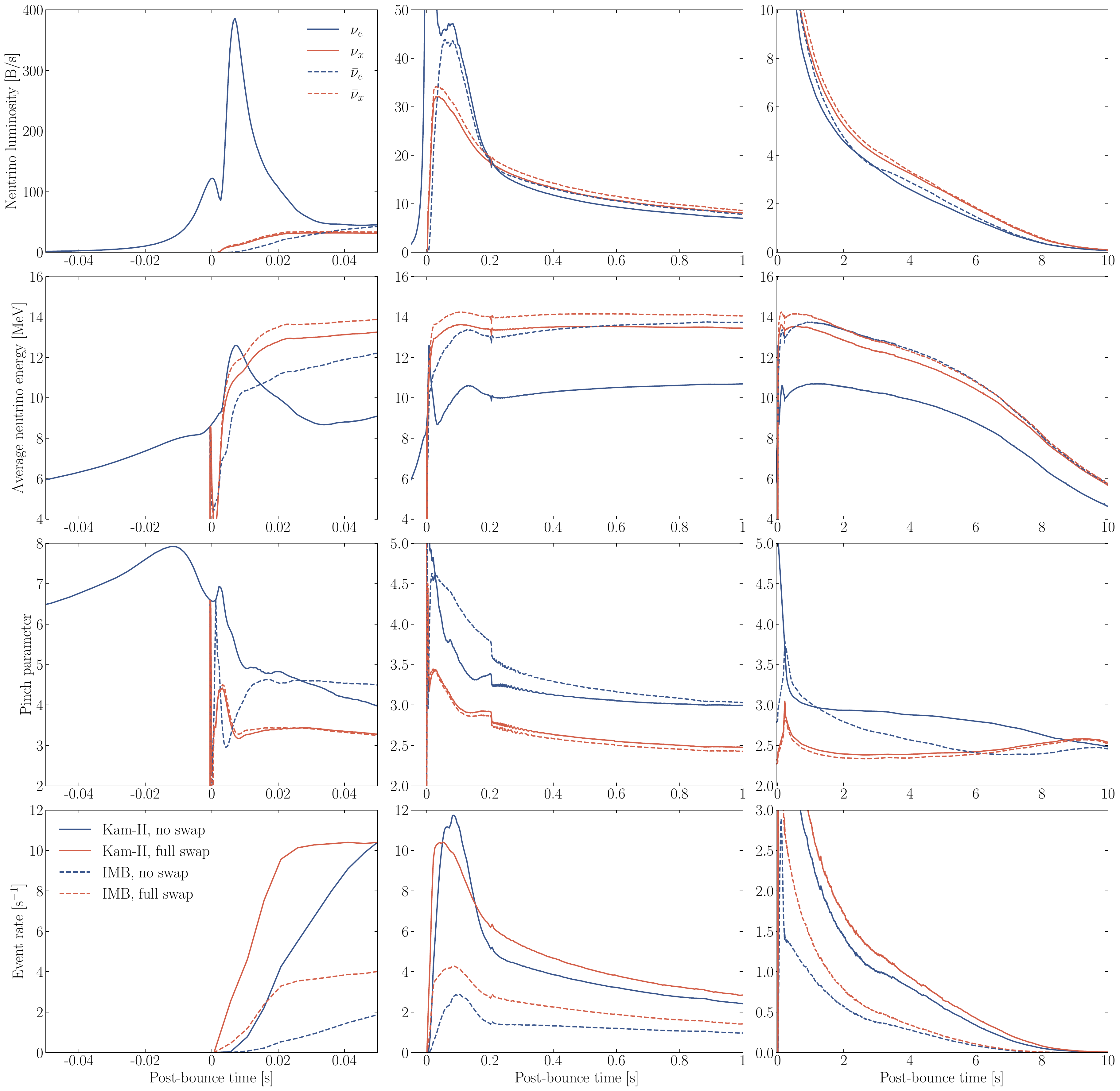}
    \caption{Evolution of the signal properties of different neutrino species for our model 1.44-SFHo during the three main emission phases of Fig.~\ref{fig:SN-Model-PNSmass}. The upper three rows display the luminosities, average energies, and spectral pinching parameters for the $\alpha$-fit  (Appendix~\ref{sec:Gamma}). While the model simulations were performed with six-species transport, the results are plotted here for $\nu_e$ and $\bar\nu_e$ as well as $\nu_x$ and $\bar\nu_x$, whose quantities are defined as arithmetic means of those of the $\mu$ and $\tau$ flavors. The bottom row shows the event rates by IBD predicted for Kam-II and IMB, once assuming no flavor conversion (i.e., detection of the $\bar{\nu}_e$ flux) and another time for a complete flavor swap (i.e., detection of the $\bar{\nu}_x$ flux). For IMB, the average trigger rate was used as discussed later in the text.}
    \label{fig:SN-Model-1}
\end{figure*}

\subsection{Brief Discussion of Model Results}

\subsubsection{Model Overview}

Our standard set of CCSN and PNS formation models employs the full physics described above, including muons, six-species neutrino transport, and convection. We performed simulations that yielded baryonic PNS masses of 1.36, 1.44, 1.62, 1.77, and 1.93\,M$_\odot$, in each case with the four different nuclear EoS implementations (DD2, LS220, SFHo, and SFHx) mentioned before. Correspondingly, we defined a naming convention for our models that specifies the mass and the EoS, e.g., 1.62-SFHo (see Table~\ref{tab:GlobalProperties}; the data files with the neutrino outputs are available at the Garching Core-Collapse Supernova Archive \cite{JankaWeb}.). 

In order to demonstrate the consequences of some of our advanced physics ingredients, we also list simulations for the 1.62\,M$_\odot$ models and all EoSs without convection, denoted by an extension ``-c'' of their model names (e.g., 1.62-SFHo-c), and two examples without muons, indicated by an extension ``-m'' of their names (e.g., 1.62-SFHo-m). We picked an average mass for these demonstration cases because of its better compatibility with the SN~1987A neutrino data, although muons are expected to generate larger effects in higher-mass PNSs due to the more extreme densities and temperatures there. In the two simulations without muons we used four-species transport, because in the absence of muons (and tauons) $\nu_\mu$ and $\nu_\tau$ experience the same reaction rates, which differ from those of $\bar\nu_\mu$ and $\bar\nu_\tau$ by the weak-magnetism corrections in the charged-current and neutral-current interactions with nucleons.

In all early CCSN and PNS models computed at the time of SN~1987A and the following decade, but also in many modern SN simulations, heavy-lepton neutrinos are lumped together into a single species $\nu_x$. Collectively treating all species of heavy-lepton neutrinos, the time evolution of $\nu_x$ is followed by solving a single transport problem with opacities averaged for neutrinos and antineutrinos. Because the weak-magnetism corrections imply only relatively small differences in the cross sections, this description is well justified for studying the hydrodynamics problem of CCSNe. It is, however, not sufficient for high-fidelity predictions of the CCSN neutrino emission and for investigating questions connected to three-flavor neutrino oscillations.

\subsubsection{Construction of PNS Models}

Since our simulations are 1D, the SN explosions are triggered artificially by reducing the density in the infall region ahead of the stalled shock by a factor of several 10. This is done at a suitable post-bounce time such that a PNS of the desired mass is left behind when the shock begins to expand outward quickly in response to the decreased ram pressure of the preshock matter. The density is reduced to a level that ensures ---somewhat idealized--- that the late-time PNS cooling signal is not contaminated at any relevant level by neutrino emission from fallback of matter that starts an outward expansion with the SN shock but does not become gravitationally unbound. Although fallback is a common phenomenon in 3D simulations of neutrino-driven explosions~\cite{Stockinger+2020,Bollig+2021,Janka+2022,Wang+2023}, its time dependence cannot be modelled realistically in 1D. Moreover, in our study we are interested in the question whether the SN~1987A neutrino events can be explained by the emission solely originating from PNS post-bounce accretion and Kelvin-Helmholtz cooling. 

\begin{figure*}
    \centering
    \includegraphics[width=1.0\textwidth]{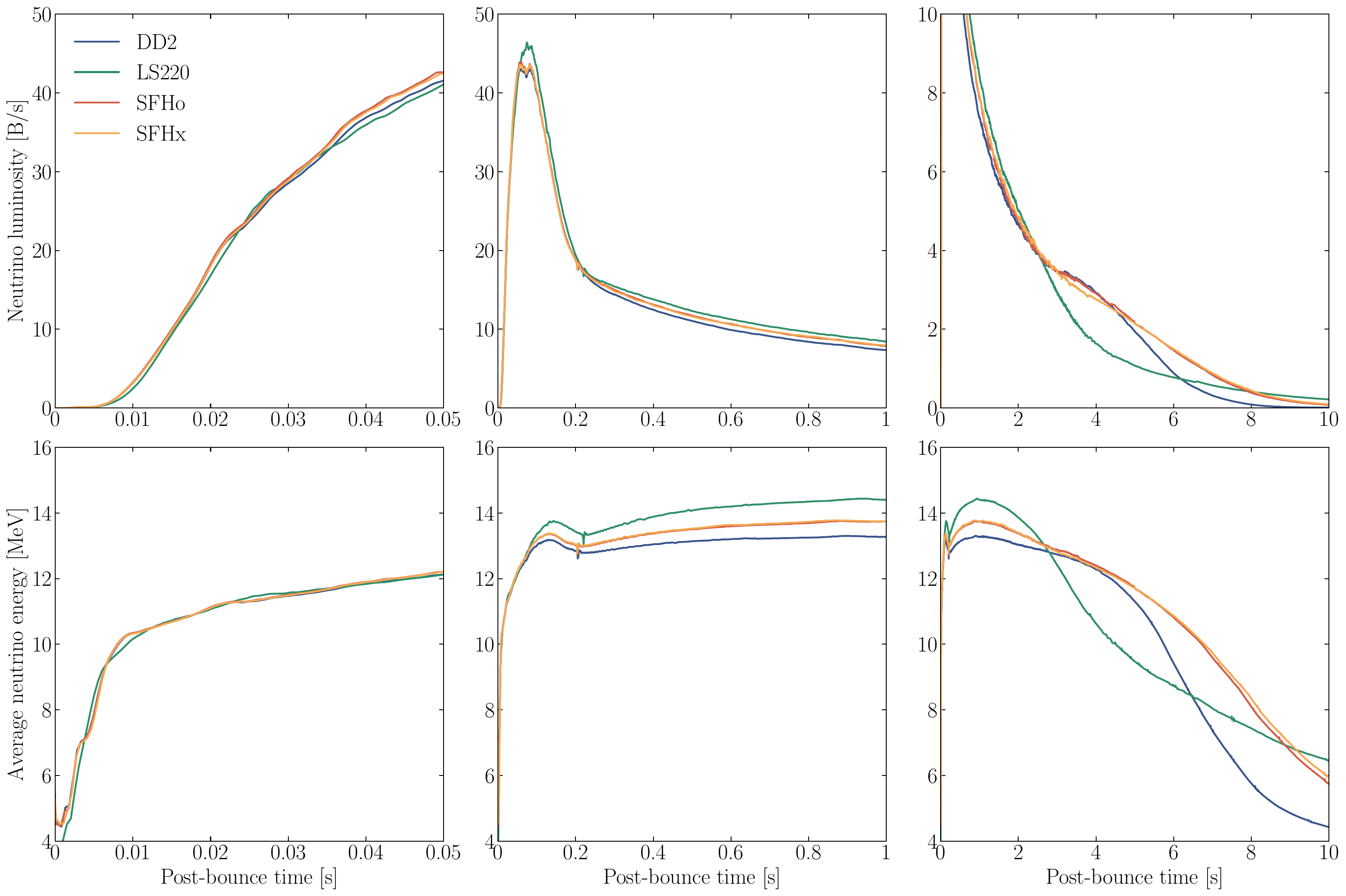}
    \caption{Evolution of the $\bar{\nu}_e$ luminosities (top) and average energies (bottom) during the three main neutrino emission phases of Fig.~\ref{fig:SN-Model-PNSmass} for our 1.44\,M$_\odot$ models with different nuclear EoSs as indicated in the legend.} 
    \label{fig:SN-Model-3}
    \vskip3pt
\end{figure*}

\begin{figure*}
    \centering
    \includegraphics[width=1.0\textwidth]{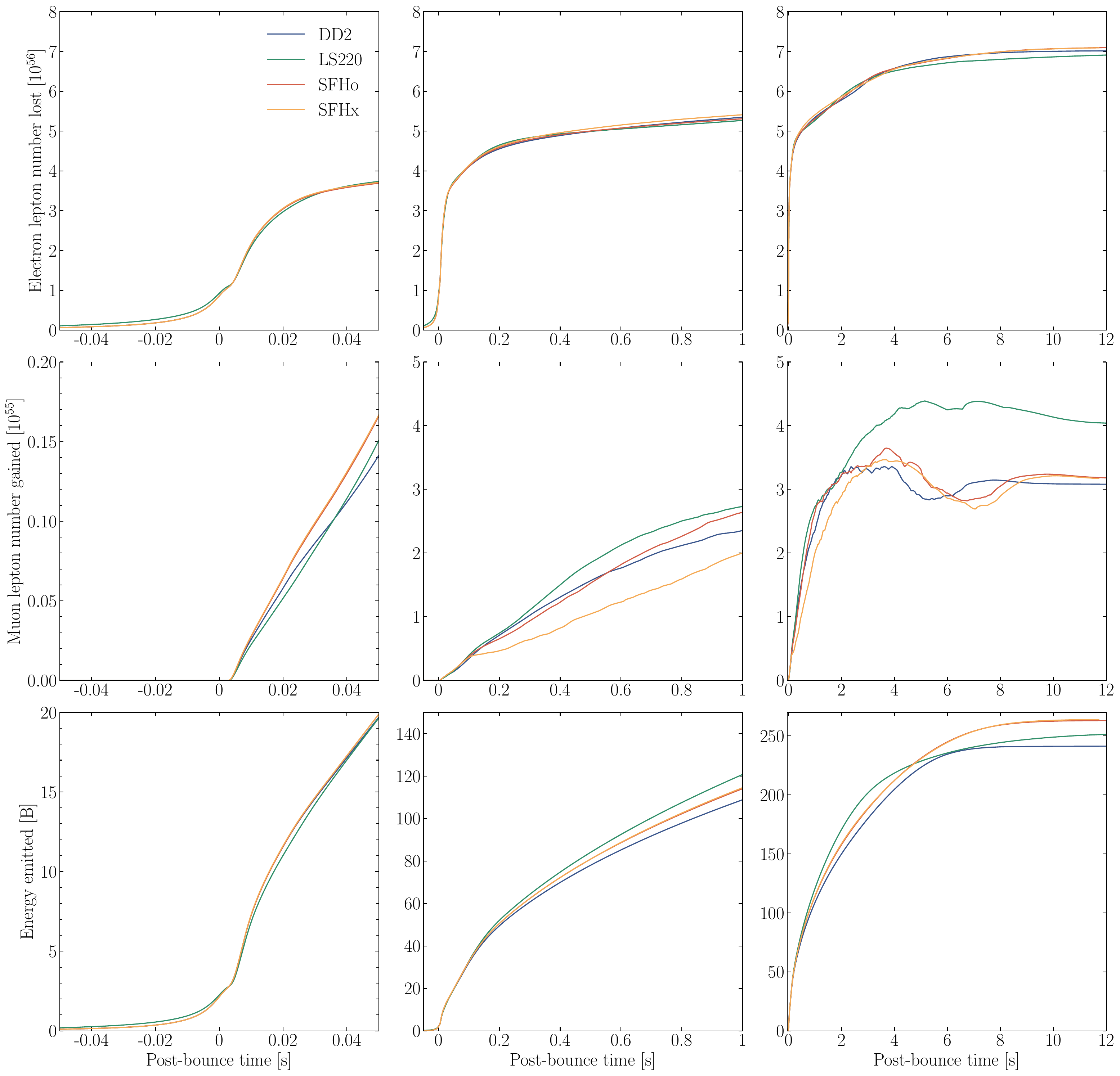}
    \caption{Evolution of the cumulative emitted electron lepton number (top), cumulative gain of muon lepton number in the PNS (middle), and cumulative energy loss (bottom) during the three main neutrino emission phases of Fig.~\ref{fig:SN-Model-PNSmass} for our 1.44\,M$_\odot$ models with different nuclear EoSs as indicated in the legend.}
    \label{fig:SN-Lepton-Numbers}
    \vskip3pt
\end{figure*}

Since the NS mass is expected to correlate in most cases with the iron-core mass or the mass of iron core plus silicon shell (i.e., the mass enclosed by the Si/O composition interface) of the progenitor~\cite{Ugliano+2012,Sukhbold+2016,Wang+2022}, we employ different stellar progenitors to obtain the PNSs of different masses, namely a 9.0\,M$_\odot$ progenitor~\cite{Woosley+2015} for the 1.36\,M$_\odot$ PNS, 18.8\,M$_\odot$ and 18.6\,M$_\odot$ progenitors~\cite{Sukhbold+2018} for the 1.44\,M$_\odot$ and 1.62\,M$_\odot$ PNSs, respectively, a 27\,M$_\odot$ progenitor~\cite{Woosley+2002} for the 1.77\,M$_\odot$ PNS, and a 20\,M$_\odot$ model~\cite{Woosley+2007} for the 1.93\,M$_\odot$ PNS. The choice of these different progenitor models ensures that the post-bounce accretion phase of the PNS is shorter for lower-mass NSs and longer for the more massive ones (Fig.~\ref{fig:SN-Model-PNSmass}; Table~\ref{tab:GlobalProperties}), compatible with the tendency witnessed in self-consistent explosion models. Note that the PNS mass is neither a monotonic function of the pre-collapse mass nor of the ZAMS mass of the progenitor star, as mentioned in Sec.~\ref{sec:intro}.

\subsubsection{Variations of Neutrino Signal with PNS Mass}

Table~\ref{tab:GlobalProperties} lists the values of the post-bounce times $t_\mathrm{acc}$ when the accretion phase ends in our models. PNS accretion in 1D simulations continues as long as the SN shock stagnates and it is visible by the enhanced neutrino (mainly $\nu_e$ and $\bar\nu_e$) emission associated with the settling of the accreted matter onto the PNS surface (Figs.~\ref{fig:SN-Model-PNSmass} and~\ref{fig:SN-Model-1}, middle panels). When the explosion is triggered and the shock begins to run outward, the inflow of matter to the PNS is stopped. This terminates the accretion phase and is reflected in a steep drop of the neutrino luminosities and mean energies from their enhanced values to the level they possess during the Kelvin-Helmholtz cooling of the PNS. The times $t_\mathrm{acc}$ are defined at the base points of this steep decline from the accretion plateau where the $\bar\nu_e$ luminosity transitions into the long-time behavior of the PNS cooling. This is seen in the middle panels of Fig.~\ref{fig:SN-Model-PNSmass}. 

In Fig.~\ref{fig:SN-Model-PNSmass} we also witness a clear correlation of the $\bar\nu_e$ luminosities and after the initial $\sim$50\,ms after bounce also of the mean energies with PNS mass during the different emission phases, namely the phase of shock breakout from the neutrinosphere and corresponding $\nu_e$ burst (left panels; the $\nu_e$ burst is seen in Fig.~\ref{fig:SN-Model-1}), the post-bounce accretion phase of the PNS during shock stagnation (middle panels), and PNS cooling after the onset of the SN explosion (right panels). More massive PNSs are formed as a consequence of higher mass accretion rates, tend to have longer accretion phases and release more gravitational binding energy by neutrino emission (for a given nuclear EoS), all of which explains their higher\break luminosities~\cite{OConnor+2013,Nagakura+2022}. Similarly, higher-mass PNSs posses hotter neutrinospheres, radiating harder spectra~\cite{Mueller+2014,Bruenn+2023}. Due to the structure of its 9\,M$_\odot$ progenitor, the formation of the 1.36\,M$_\odot$ PNS is associated with a slightly longer accretion phase than the 1.44\,M$_\odot$ PNS (Table~\ref{tab:GlobalProperties}), which is reflected by the inversion of their neutrino emission properties between $\sim$0.15\,s and $\sim$0.3\,s after bounce (Fig.~\ref{fig:SN-Model-PNSmass}).

\subsubsection{Neutrino Emission Phases}
\label{sec:nuemphas}

The typical behavior of the neutrino properties (luminosities, mean energies, spectral pinching parameter $\alpha$ of Ref.~\cite{Keil+2003}) of $\nu_e$, $\bar\nu_e$, and heavy-lepton neutrinos in the three emission phases can be seen for the example of model 1.44-SFHo in Fig.~\ref{fig:SN-Model-1} (for a detailed description of the evolution, see, e.g., Refs.~\cite{Mirizzi+2016,Janka2017Handbookb}). The prominent shock-breakout burst of $\nu_e$ reaches a peak luminosity near $4\times 10^{53}$\,erg\,s$^{-1}$ (top left panel). During this phase, the luminosities and mean energies of $\bar\nu_e$ and of all kinds of heavy-lepton neutrinos (collectively denoted as $\nu_x$) begin to rise steeply (left upper two panels), faster for the $\nu_x$ because the $\bar\nu_e$ emission is initially suppressed due to the high degeneracy of electrons (reducing the presence of positrons) and of $\nu_e$ (quenching the pair production of $\nu_e\bar\nu_e$) in the shock-heated layers. Moreover, the escape of $\bar\nu_e$ is also impeded by their frequent absorption on the still abundant protons. We note in this context that the transport of $\bar\nu_e$ and heavy-lepton neutrinos is disregarded before core bounce and switched on only at bounce, because the production of these neutrino species is strongly suppressed by the high electron degeneracy before and during core collapse. Therefore during this pre-bounce evolution the emission of $\nu_e$, created by electron captures, dominates by several orders of magnitude and $\bar\nu_e$ can be expected to be measurable only from future CCSNe in the very close neighborhood of the Sun. 

During the accretion phase, matter falling through the stagnant SN shock settles onto the newly formed PNS in a hot, inflated mantle layer, which radiates mostly $\nu_e$ and $\bar\nu_e$ produced by electron and positron captures onto free protons and neutrons, respectively. Therefore the $\nu_e$ and $\bar\nu_e$ luminosities are up to a factor of 2 higher than the individual ones of the heavy-lepton neutrinos (top middle panel of Fig.~\ref{fig:SN-Model-1}). The $\nu_x$ are produced mainly by nucleon-nucleon bremsstrahlung and by processes involving muons in the higher-density core of the PNS, and they diffuse out from there to escape from deeper but only slightly hotter neutrinospheres (middle panel in the second row). When the onset of the explosion terminates the PNS accretion, the associated enhanced emission of $\nu_e$ and $\bar\nu_e$ comes to an end at around $t_{\rm acc}$. Subsequently, the luminosities of all neutrino species are much more similar and decay quasi-exponentially in the PNS cooling phase. During this long-time evolution and the preceding accretion phase, the mean energies of the radiated $\bar\nu_e$ and of all heavy-lepton neutrinos are a few MeV higher than those of $\nu_e$. The $\bar\nu_x$ spectra are always harder than those of the $\nu_x$, because weak-magnetism corrections reduce (increase) the cross sections for neutral-current neutrino scattering with nucleons for $\bar\nu_x$ ($\nu_x$)~\cite{Horowitz2002}. However, the mean energies of $\bar\nu_e$ and of all heavy-lepton neutrinos differ between each other by less than about 1\,MeV, and partially the mean energies of $\bar\nu_e$ are even higher than those of $\nu_x$. This is in stark contrast to the classical SN and PNS cooling models computed at the time of SN~1987A and is a consequence of including non-conservative (i.e., energy-transfer) effects in neutrino-nucleon scatterings, which softens the spectra of the escaping $\nu_x$~\cite{Raffelt:2001kv,Keil+2003}.   

The emitted spectra are always pinched ($\alpha\agt 2.3$) compared to a Maxwell-Boltzmann spectrum ($\alpha = 2$), decreasing towards later times. Typically (but not always), the spectra of $\nu_e$ and $\bar\nu_e$ are more strongly pinched ($\alpha \sim 3$--5) than those of $\nu_x$ and $\bar\nu_x$ ($\alpha\sim 2.5$--3). The expected detection rates in IMB and Kam-II are shown in Fig.~\ref{fig:SN-Model-1}, bottom panels, under the assumption that the signal is caused by inverse beta-decay (IBD) events of the original $\bar\nu_e$ flux or of the $\bar\nu_x$ flux, assuming a complete flavor swap and defining the $\bar\nu_x$ properties as arithmetic averages of those of $\bar\nu_\mu$ and $\bar\nu_\tau$.

\subsubsection{Variations of Neutrino Signal with EoS}

The employed high-density nuclear EoSs differ considerably in their maximum masses and mass-radius relations for cold NSs, although in all cases these are compatible with astrophysical mass and radius constraints within the current uncertainties. These differences are linked to different nuclear matter properties such as the saturation density, incompressibility, symmetry energy, and the slope of the symmetry energy (all in agreement with current experimental ranges or close to them; for a detailed discussion, see Refs.~\cite{Fischer+2014,Oertel+2017,Tews+2017}). The maximum gravitational masses of cold NSs are 2.05, 2.06, 2.13, and 2.42\,$M_\odot$ for the LS220, SFHo, SFHx, and DD2 EoS, respectively, and the radii of cold 1.4\,$M_\odot$ NSs are 12.67, 11.89, 11.99, and 13.22\,km.\footnote{The mass-radius relations of SFHo and SFHx have a crossing near a gravitational NS mass of 1.25\,$M_\odot$, but the radius differences increase with the distance from this crossing point.} Judged from their small values of this NS radius and the maximum masses, LS220, SFHo, and SFHx are rather ``soft'' EoSs, whereas DD2 is relatively ``stiff.'' It is remarkable that in spite of considerably different nuclear-matter and NS properties of SFHo and SFHx, we generally (for all PNS masses) find only small differences in the neutrino emission properties for simulations with both EoSs, and these differences occur mainly during the very late PNS cooling phase (Figs.~\ref{fig:SN-Model-3} and~\ref{fig:SN-Lepton-Numbers}).

The four considered EoSs also differ in their density dependence of the nuclear symmetry energy. The symmetry energy does not only determine the electron fraction in beta-equilibrium (e.g.,~\cite{Hempel2015,Most+2021,Janka+2023}), but can also have important consequences for PNS convection~\cite{Roberts+2012a}. If the symmetry energy increases sufficiently steeply as a function of baryon density (still staying within current experimental constraints), negative lepton-number gradients can exert a {\em stabilizing} effect on Ledoux convection for high densities and low electron fractions. This counteracts the always destabilizing influence of negative entropy gradients and is in contrast to the situation at low densities and high electron fractions, where both negative entropy and lepton gradients drive convection. Thus regions with negative entropy and lepton gradients in the PNS interior initially develop convection. But as the PNS neutronizes and contracts to lower electron fractions and higher densities, a steep increase of the symmetry energy can lead to a suppression of convective activity. This is the case for the LS220 EoS. In PNS cooling simulations with this EoS convective activity inside the PNS becomes weaker and spatially more restricted after 2--3\,s, which slows down the subsequent loss of electron number and energy. This leads to a characteristic decline of the luminosities and mean energies of the radiated neutrinos at about this time (Fig.~\ref{fig:SN-Model-3}). The decline is steeper and earlier than for all other nuclear EoSs, which permit PNS convection to continue in a larger volume of the PNS. Later on the time dependence of the neutrino emission in the LS220 models flattens and the luminosities and mean energies decrease more slowly and stay on a higher level for longer periods than in simulations with the other EoSs. 

\subsubsection{Properties of Cumulative Signals}
\label{sec:propcumsignals}

Despite these prominent differences in the evolution of the neutrino emission properties, the characteristic cooling times, $\tau_E$ (periods over which 95\% of the total energy loss $E_{\rm tot}^{\rm end}$ happen; Table~\ref{tab:GlobalProperties}), of the simulations with LS220 are not much different from the corresponding times of the SFHo and SFHx models. This is a consequence of a faster initial loss of energy by neutrino emission in the first $\sim$3\,s (Figs.~\ref{fig:SN-Model-3} and~\ref{fig:SN-Lepton-Numbers}) and lower $E_{\rm tot}^{\rm end}$ values than in the cases with SFHo and SFHx EoS (Table~\ref{tab:GlobalProperties}). The electron-deleptonization times, $\tau_e$ (post-bounce times until 95\% of the total emitted electron-lepton number $N_e^{\rm end}$ are carried away), of the simulations with LS220 are usually even shorter than with all other EoSs (except for low-mass PNSs with the DD2 EoS), because $N_e^{\rm end}$ is smallest for LS220 compared to the other EoSs (Table~\ref{tab:GlobalProperties}). The reason is again the higher nuclear symmetry energy of LS220 at high densities. In particular the interaction part of the symmetry energy implies higher electron and muon chemical potentials $\mu_e = \mu_\mu = \mu_n - \mu_p$ in beta-equilibrium~\cite{Hempel2015,Most+2021,Janka+2023} and therefore a lower loss of electron-lepton number and a higher gain of muon number during the PNS evolution (Fig.~\ref{fig:SN-Lepton-Numbers}).   

Since the greater stiffness of the DD2 EoS leads to larger PNS radii, lower values of $E_{\rm tot}^{\rm end}$, smaller central densities and temperatures, and therefore also lower neutrino opacities, the energy loss of the PNSs is generally faster with this EoS than for models with all other EoS cases. This holds true also for the loss of electron-lepton number of low-mass PNSs with the DD2 EoS, though it does not apply for the higher-mass PNSs, where the LS220 models possess the clearly smallest values of $N_e^{\rm end}$ and thus of $\tau_e$, if PNS convection is taken into account.

Figure~\ref{fig:SN-Lepton-Numbers} shows the evolution of the cumulative emission of electron-lepton number, i.e., the excess of $\nu_e$ over the $\bar\nu_e$ emission, for our 1.44\,M$_\odot$ models with all four EoS cases. About half of the total electron-lepton number is emitted during the shock-breakout burst of $\nu_e$ within some 10\,ms after bounce. The electron-deleptonization continues fast until convection reaches the center of the PNS and then proceeds more slowly as the PNS cools further. Notice that, all else being equal, more lepton number can be stored in the form of electrons at higher temperatures. The times $\tau_e$ listed in Table~\ref{tab:GlobalProperties} are typically roughly 1\,s shorter than the energy-loss times $\tau_E$ for DD2, SFHo, and SFHx, but up to $\sim$4\,s shorter for LS220 (when PNS convection is included). 

\begin{figure*}
    \centering
    \includegraphics[width=\textwidth]{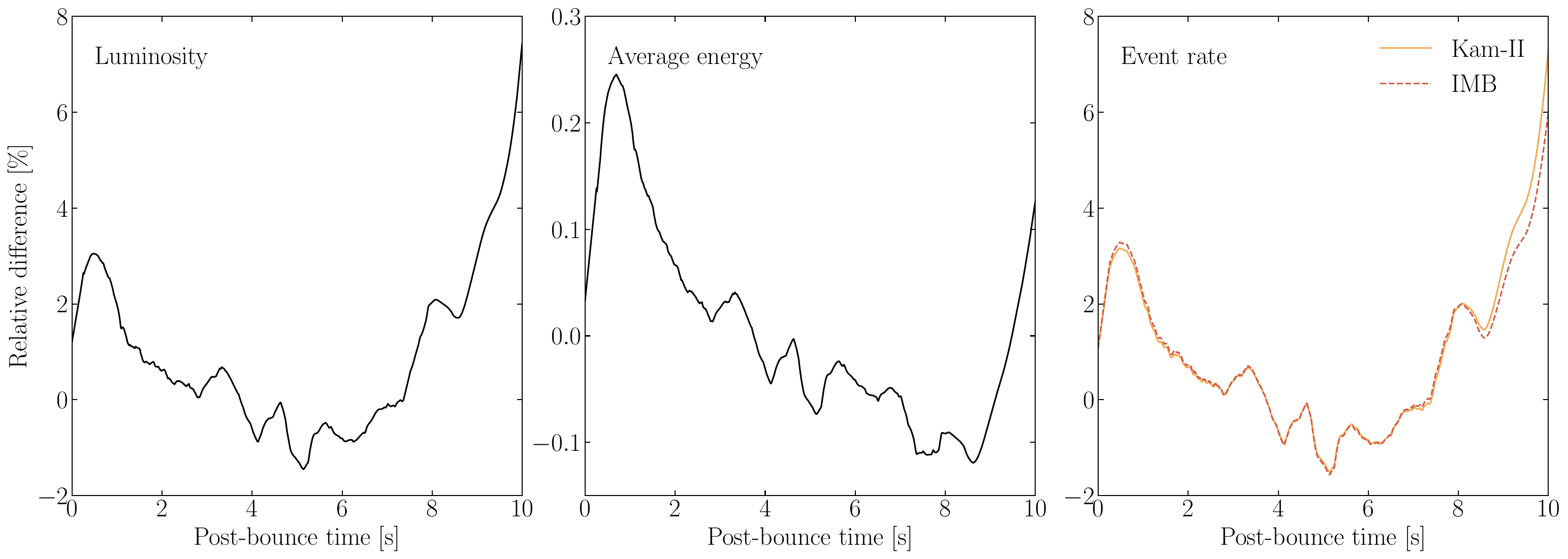}
    \vskip-2pt
    \caption{Relative differences between $\bar\nu_\mu$ and $\bar\nu_\tau$ emission and detection properties according to Eq.~\eqref{eq:diff-mu-tau} for model 1.44-SFHo as in Fig.~\ref{fig:SN-Model-1}. Positive values mean that the quantity for $\bar\nu_\mu$ is larger than the average.
    }
    \label{fig:SN-Model-2}
\end{figure*}

\subsubsection{Signal Variations in the Antineutrino Sector}

Muon formation and the build-up of muon number in the hot PNS is enabled by the high chemical potentials of electrons and the high temperatures, which imply particle energies well above the muon rest-mass energy of 105.66\,MeV. These conditions permit that weak reactions involving muons $\mu^-$ such as $\nu_\mu + e^- \rightleftarrows \nu_e + \mu^-$, $\nu_\mu + n \rightleftarrows p + \mu^-$, and $\bar\nu_e + e^- \rightleftarrows \bar\nu_\mu + \mu^-$ (and similar reactions involving anti-muons $\mu^+$) can take place in the high-density core of the PNS~\cite{Lohs+2014,Bollig:2017lki}.\break
In analogy to electrons and positrons, the thermally excited muons and anti-muons are distributed according to beta equilibrium, fulfilling the chemical potential relation $\mu_\mu - \mu_{\nu_\mu} = \mu_n - \mu_p$, where $\mu_\mu$, $\mu_{\nu_\mu}$, $\mu_n$, and $\mu_p$
are the chemical potentials of muons, muon neutrinos, neutrons, and protons, respectively. In the neutrino-transparent state of the cold NS without any net neutrino number, chemical equilibrium implies $\mu_e = \mu_\mu = \mu_n - \mu_p$. The evolution to this final state is driven by an overall excess emission of $\bar\nu_\mu$ compared to $\nu_\mu$, but it can proceed non-monotonically.
 
Taking the presence of muons into account, our six-species neutrino transport has to make a distinction of $\nu_\mu$, $\bar\nu_\mu$, $\nu_\tau$, and $\bar\nu_\tau$, because the fluxes and spectra of all species are different due to the muon interactions and opposite signs of the weak-magnetism corrections in the neutral-current and charged-current processes of $\nu$ and $\bar\nu$ with nucleons. These weak-magnetism corrections reduce the antineutrino interaction cross sections relative to those of the neutrinos, allowing $\bar\nu_\mu$ to diffuse out of the PNS more readily than $\nu_\mu$.\break
This accelerates the muonization of the PNS medium by the development of an imbalance between the $\nu_\mu$ and $\bar\nu_\mu$ numbers in the PNS interior, leading to transiently positive muon-neutrino chemical potentials $\mu_{\nu_\mu}$. The corresponding net muon number in the neutrino sector is communicated to the charged-lepton sector by the muon-involving weak interactions mentioned above, but it has to decay later again on the way to the neutrino-less state (corresponding to $\mu_{\nu_\mu} = 0$).

The cumulative anti-muon numbers that are emitted until the post-bounce times $t_\mathrm{end}$ when the simulations are stopped, are listed for our models in Table~\ref{tab:GlobalProperties}, and the panels in the middle row of Fig.~\ref{fig:SN-Lepton-Numbers} display the time evolution of the cumulative gain of muon lepton number (i.e., loss of anti-muon number by the excess of $\bar\nu_\mu$ over $\nu_\mu$ emission) for our 1.44\,M$_\odot$ models with all four EoS cases. This muonization process is not monotonic because its evolution depends on the temperature profile in the PNS as well as the competing transport and emission of $\nu_\mu$ and $\bar\nu_\mu$. Since more leptons can be stored at higher temperatures, the change of the net muon number reflects the thermal evolution of the PNS, which heats up transiently in the dense core during its electron-deleptonization~\cite{Burrows:1986ApJ,Pons:1999ApJ}. PNSs deleptonize with respect to their electron number from outside inward and there is a continuous net loss of electron number by an excess number emission of $\nu_e$ over $\bar\nu_e$. In contrast, the build-up of muon number follows the evolution of the temperature maximum, where most of the $\nu_\mu$ and $\bar\nu_e$ are created and facilitate the production of muons~\cite{Bollig:2017lki}. The temperature has an off-center maximum for a few seconds, which moves inward until the temperature peaks at the PNS center and then gradually declines during the cooling evolution. This complex evolution can lead to successive periods of increasing and decreasing total net muon number in the PNS. Finally, in the last phase, the transiently positive muon neutrino chemical potential (which is established by the faster diffusion of $\bar\nu_\mu$) has to settle to zero for the neutrino-less beta-equilibrium in the cold, neutrino-transparent state of the NS. Therefore, in order to come up with an unambiguous definition of the muonization time $\tau_{\bar\mu}$ (Table~\ref{tab:GlobalProperties}), we adopt for this characteristic time scale the \textit{first} instant in the evolution when the emitted anti-muon number reaches $95\%$ of its final value, thus neglecting the later fluctuations.

In the bottom panels of Fig.~\ref{fig:SN-Model-1} we show the predicted detection rates of neutrinos in the Kam-II and IMB detectors (see Sec.~\ref{sec:data}). The inverse beta-decay (IBD) event rate in Kam-II exhibits two crossings at about 50\,ms after bounce and near $t_{\rm acc}$, if we assume the signal to be caused once only by $\bar\nu_e$ (no flavor conversion) and another time only by $\bar\nu_x$ (full flavor conversion). In contrast, in IMB the event rate is always larger if caused by $\bar\nu_x$. In Kam-II the signal therefore reflects the evolution of the $\bar\nu_x$ vs.\ the $\bar\nu_e$ luminosities (upper panels of Fig.~\ref{fig:SN-Model-1}), whereas in IMB the signal is always larger if caused by $\bar\nu_x$ because of the higher neutrino energies needed to trigger this detector (as will be discussed later).

In the context of the SN~1987A neutrino measurements only the antineutrino sector is relevant. In the absence of flavor conversion, the detected signal was caused by the $\bar\nu_e$ emission of the forming PNS, and if this species does not survive after flavor conversion the detection was exclusively caused by heavy-lepton antineutrinos. For the predicted event rates plotted in Fig.~\ref{fig:SN-Model-1} and in our analysis below, we assumed that the $\bar\nu_x$ properties are arithmetic averages of the $\bar\nu_\mu$ and $\bar\nu_\tau$ emission from our source models. Therefore it is interesting how big the differences between the $\bar\nu_\mu$ and $\bar\nu_\tau$ properties are. These are visualized in Fig.~\ref{fig:SN-Model-2}, which displays the relative differences between the luminosities, average energies, and expected event counts, normalized by the sum of the these quantitities for $\bar\nu_\mu$ and $\bar\nu_\tau$, i.e., the plotted relative differences are computed according to
\begin{equation}\label{eq:diff-mu-tau}
    \delta(\bar\nu_\mu,\bar\nu_\tau)=\frac{\bar\nu_\mu-\bar\nu_\tau}{\bar\nu_\mu+\bar\nu_\tau}\,.
\end{equation}
Of course, the right-hand side here is symbolic for the different quantities depending on these flavors. We conclude that despite the effects of muons, the relative differences are fairly small, on the order of a few percent at most, at least for the 1.44\,M$_\odot$ PNS considered in Fig.~\ref{fig:SN-Model-2}. Therefore is it well justified in the context of our analysis to consider a mean species $\bar\nu_x$ with arithmetically averaged properties of $\bar\nu_\mu$ and $\bar\nu_\tau$.

\begin{figure*}
    \centering
    \includegraphics[width=1.0\textwidth]{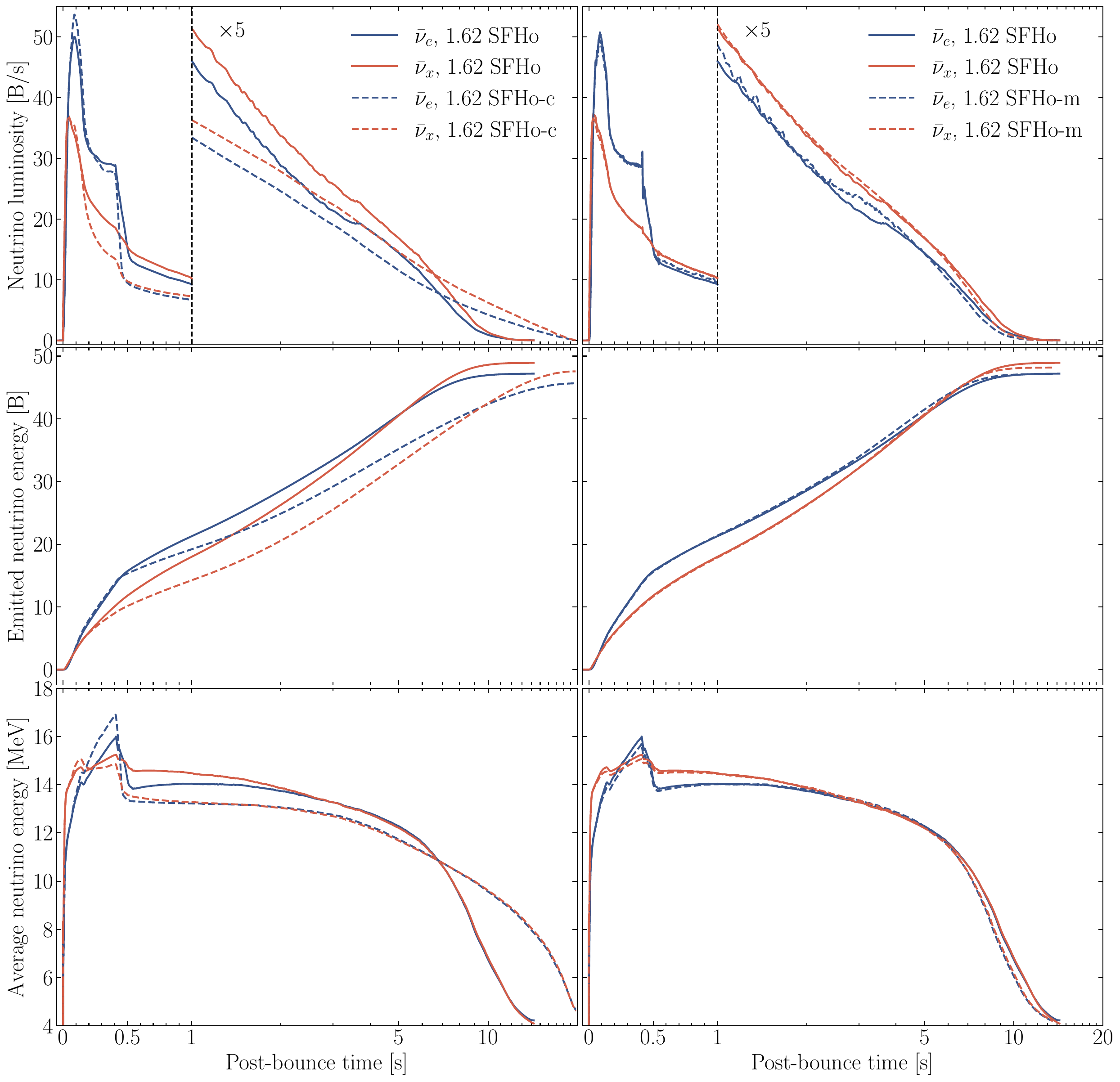}
    \vskip-2pt
    \caption{Post-bounce evolution of the luminosities (top), cumulative emitted energies (middle), and mean energies of $\bar\nu_e$ and $\bar\nu_x$ for our 1.62-SFHo model including PNS convection and muons compared to model 1.62-SFHo-c without convection (but including muons; dashed lines in the left panels) and model 1.62-SFHo-m without muons (but including convection; dashed lines in the right panels). Note the change of the scale on the $x$-axis from linear to logarithmic at 1\,s. The neutrino luminosities emitted at post-bounce times $t_\mathrm{pb}> 1$\,s are scaled by a factor of 5.}
    \label{fig:SN-Model-noconv}
\end{figure*}

\subsubsection{Impact of PNS Convection and Muons}

Finally, we repeat that our state-of-the-art CCSN and PNS-cooling simulations compared to those at the time of SN~1987A include, besides modern nuclear EoS models and a better numerical treatment of the neutrino transport and reaction rates, several major improvements: (1) nucleon-nucleon (RPA) correlations in the charged-current and neutral-current neutrino-nucleon interactions \cite{Horowitz+1991,Sawyer1995,Reddy+1998,Reddy+1999,Burrows+1998,Burrows+1999,Horowitz+2003,Buras+2006,Roberts+2012,Roberts+2017,Oertel+2020,Pascal+2022}, which reduce the neutrino opacities and therefore shorten the PNS cooling times considerably \cite{Huedepohl+2010,Roberts+2012a,Roberts+2017Handbook,Pascal+2022}; (2) PNS convection, which is very efficient in transporting energy and lepton number out of the dense PNS core and thus has an additional impact on shortening the PNS cooling evolution, even dominating the nucleon correlation effects~\cite{Roberts+2012a,Mirizzi+2016,Roberts+2017Handbook,Pascal+2022}; and (3) the presence of muons in the hot PNS medium. 

In order to demonstrate some of these effects, we have also simulated exemplary 1.62\,M$_\odot$ cases with different nuclear EoSs but without convection (model names with extension ``-c'') or without muons (model names with extension ``-m''). 
Figure~\ref{fig:SN-Model-noconv} compares the time evolution of the neutrino emission properties (luminosities, mean energies, cumulative emitted energies) of $\bar\nu_e$ and $\bar\nu_x$ at the example of model 1.62-SFHo, and in Table~\ref{tab:GlobalProperties} the characteristic global parameters of these simulations and a few other cases without convection or without muons are listed, too. 

Convection leads to a sizable enhancement of the $\bar\nu_e$ luminosity mainly after the post-bounce accretion phase and of the $\bar\nu_x$ luminosity, which is generated in deeper PNS layers, already during the accretion phase. Correspondingly, the PNS cooling times $\tau_E$ are significantly longer for the simulations without convection. We stress that the total neutrino energy and electron-lepton number released in models with and without convection should, of course, be identical. The differences seen in Table~\ref{tab:GlobalProperties} are connected to the still incomplete cooling of the non-convective models when their simulations were stopped at $t_\mathrm{end}$ and to small non-conservative effects in the numerical implementation (accumulating errors on the order of 1--2\% over the entire PNS evolution~\cite{Huedepohl+2010}). We anticipate here that the omission of mixing-length convection will not solve the problem that the neutrino emission times in our PNS cooling simulations are too short to account for the last three events measured by Kam-II and the last two events in BUST (Baksan). Even without convection, the opacity reduction by nucleon-nucleon correlations reduces the duration of the signal that could be measured by these experiments to considerably less than 10\,s (see Sec.~\ref{sec:latevents}).

In the relatively low-mass PNS of the displayed results, the effects of muons on the $\bar\nu_e$ and $\bar\nu_x$ (i.e., $\bar\nu_\mu$-$\bar\nu_\tau$-averaged) emission are relatively small. Muons lead to a slight increase of the mean energies during the accretion phase (which can facilitate the shock revival in multi-dimensional simulations~\cite{Bollig:2017lki}) due to a faster contraction of the PNS. Since the final PNS radius is also a bit smaller, PNS cooling including muons also releases slightly higher gravitational binding energies via neutrino emission (despite internal energy being stored in muon rest mass), and the late-time neutrino luminosities and mean energies exhibit a small enhancement for a bit longer cooling times.

\subsubsection{Flavor-dependent Integrated Properties}

In addition to Table~\ref{tab:GlobalProperties}, which provides data for the global properties of the time-integrated neutrino emission of our CCSN models, Table~\ref{tab:Neutrino-flux-properties} contains time-integrated data (total radiated energy $E_\mathrm{tot}^\mathrm{end}$, average neutrino energy $\langle\epsilon_\nu\rangle$, spectral pinching parameter $\alpha$, and typical time scale of the energy emission, $\tau_E$) for all neutrino species and all models individually. For the antineutrinos of the three flavors also characteristic quantities of the predicted signals in each of the three detectors are listed (the total number of expected events $N$, the average detected positron energy $\langle\epsilon_e\rangle$, the signal duration $\tau$ when 95\% of all expected events have accumulated, and the best-fit offset time between bounce and the first detected event, $\delta t$).

The total energies emitted in each neutrino species obey near-equipartition within about 10\%, with $\nu_e$ (or $\bar\nu_\mu$ in some 1.44\,M$_\odot$ cases) carrying away the biggest fraction and $\bar\nu_e$ the smallest. In models with massive NSs ($\ge 1.62$\,M$_\odot$ in our set), which have longer post-bounce accretion phases with associated $\nu_e$ and $\bar\nu_e$ emission, the relative energy share in $\bar\nu_e$ increases and comes close to that of $\bar\nu_\mu$, except for the LS220 models, where the higher nuclear symmetry energy leads to a higher build-up of muon number and thus more energy loss in $\bar\nu_\mu$ (see Sec.~\ref{sec:propcumsignals}). Because of their weak-magnetism-enhanced interaction cross sections with nucleons, $\nu_\mu$ and $\nu_\tau$ always carry away the smallest part of the total binding energy of the massive NSs. 

The average energies of the time-integrated $\nu_e$ signals are the lowest of all neutrino species, and the values increase with the PNS mass from $\sim$9.7\,MeV to over 11\,MeV. The mean energies of all the other neutrinos are nearly the same, in particular those of $\bar\nu_e$, $\bar\nu_\mu$ and $\bar\nu_\tau$, which are 0.3--0.8\,MeV larger than those of $\nu_\mu$ and $\nu_\tau$ (again because of the opacity differences associated with the weak-magnetism corrections). 

The pinching parameters of all time-integrated spectra range between about 2 and 3 with the tendency of lower values (i.e., weaker spectral pinching) for softer EoSs and more massive NSs. These values motivate our choice of $\alpha$-variation in Figs.~\ref{fig:overlap_simulations} and~\ref{fig:banana_plot_flavor_swap}. The time scales $\tau_E$ of 95\% energy loss are shortest for $\nu_e$ because of the shock-breakout burst, but nearly the same for all other neutrino species with differences of at most $\sim$1\,s, leading to very similar 95\% detection times $\tau$ for all antineutrino flavors. Because of the emission differences described above, the expected numbers of neutrino events, $N$, are higher by 1--1.5 in Kam-II and essentially unchanged in IMB and BUST, if complete flavor swap would occur and the $\bar\nu_\mu$ or $\bar\nu_\tau$ instead of the $\bar\nu_e$ were detected by inverse beta decay. The predicted average energies of the produced positrons would differ by at most $\sim$1.5\,MeV in this case.

\onecolumngrid

\vskip30pt

\section{SN 1987A Neutrino Data}

\label{sec:data}

\twocolumngrid

\begin{figure}
    \centering
    \includegraphics[width=1\columnwidth]{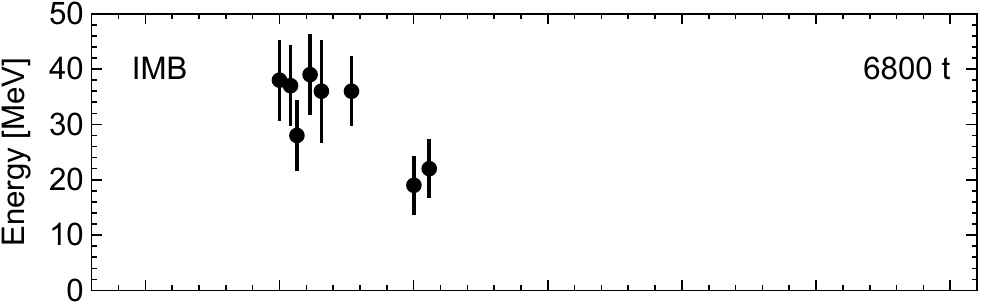}
    \includegraphics[width=1\columnwidth]{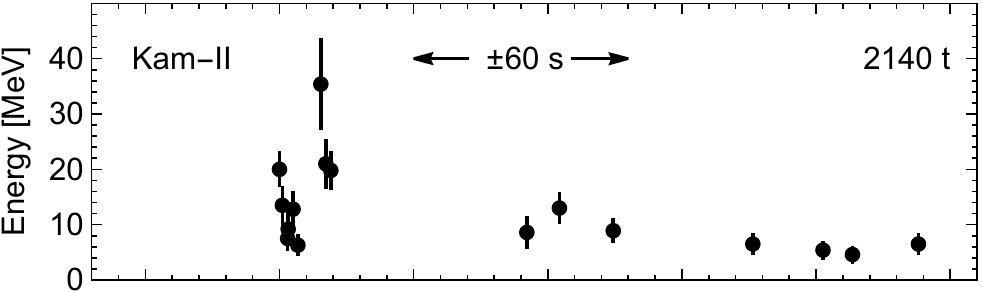}
    \includegraphics[width=1\columnwidth]{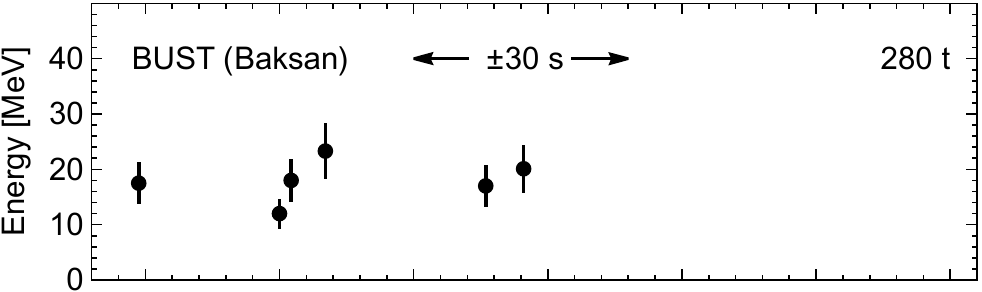}
    \includegraphics[width=1\columnwidth]{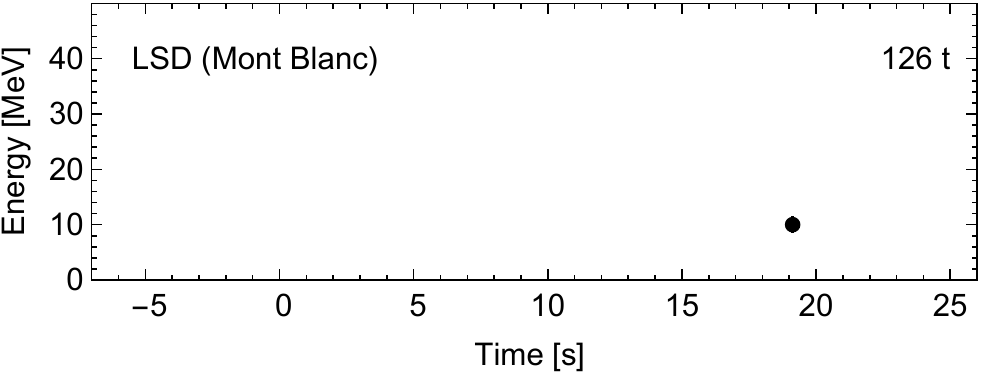}
    \caption{SN~1987A neutrino data at IMB, Kam-II, BUST and LSD. We show the detected positron energy as a function of time relative to the first event at IMB (7:35:41.374 UT on 23 February 1987). The first events of Kam-II and BUST are aligned with that of IMB because both detectors had a significant clock uncertainty. For BUST, we also show an earlier event that was attributed to background. LSD measured an additional burst nearly 5~hours earlier not shown here. In Kam-II, event No.~6 and those after 17~s as well as the late LSD events (another one at 37.5~s) are usually attributed to background. The fiducial masses shown for the scintillator detectors BUST and LSD are water equivalents. The data are listed in Tables~\ref{tab:IMB-data}--\ref{tab:Baksan-data}, the overall detector properties in Table~\ref{tab:Detectors}.
    }
    \label{fig:SNData}
\end{figure}

At the time of SN~1987A, four experiments were sensitive to the neutrino burst as detailed in Appendix~\ref{sec:SN-Observations}. By far the dominant detection process was inverse beta-decay (IBD), $\bar\nu_e+p\to n+e^+$. The largest detector was the Irvine-Michigan-Brookhaven (IMB) \cite{Irvine-Michigan-Brookhaven:1983iap, Bionta:1987qt, IMB:1988suc, Matthews:1987} water Cherenkov detector with a sensitive mass of 6800~t, but relatively sparse photo sensor coverage and concomitant high energy threshold. The time sequence of registered events is given in Table~\ref{tab:IMB-data} and shown in the top panel of Fig.~\ref{fig:SNData}.
We show the trigger efficiency and its uncertainty in the top panel of Fig.~\ref{fig:Detectors} (for more details see Appendix~\ref{sec:IMB} and Table~\ref{tab:Detectors}).  IMB had no low-energy background for the short SN signal duration, although it registered 15 atmospheric muons during the SN burst.

\begin{table}
 \caption{SN~1987A neutrinos in IMB \cite{IMB:1988suc}. Time relative to the first event UT 7:35:41.374 $\pm50$~ms. The energy refers to the detected $e^\pm$, its angle is relative to the opposite LMC direction (i.e.\ the scattering angle).}
 \vskip4pt
    \label{tab:IMB-data}
    \begin{tabular*}{\columnwidth}{@{\extracolsep{\fill}}lllll}
    \hline\hline
     Event& Time     &$N_{\rm hit}$& Energy         & Angle       \\
     No.  & [sec]    &             & [MeV]          & [Degree]    \\
     \hline
     \hp1 ~~[$+1\mu$]$^{a}$  & 0.000 & 47 & $38\pm7$ &\hp$80\pm10$ \\
     \hp2 ~~[$+2\mu$]$^{a}$   & 0.412 & 61 & $37\pm7$ &\hp$44\pm15$ \\
     \hp3    & 0.650 & 49 & $28\pm6$ &\hp$56\pm20$ \\
     \hp4 ~~[$+1\mu$]$^{a}$   & 1.141 & 60 & $39\pm7$ &\hp$65\pm20$ \\
     \hp5 ~~[$+2\mu$]$^{a}$   & 1.562 & 52 & $36\pm9$ &\hp$33\pm15$ \\
     \hp6 ~~[$+5\mu$]$^{a}$  & 2.684 & 61 & $36\pm6$ &\hp$52\pm10$ \\
     \hp7 ~~[$+4\mu$]$^{a}$  & 5.010 & 44 & $19\pm5$ &\hp$42\pm20$ \\
     \hp8    & 5.582 & 45 & $22\pm5$ &  $104\pm20$ \\
     \hline
    \end{tabular*}
    \vskip2pt
    \vbox{\raggedright
     $^a$Number of muons following this event according to the trigger\ numbers in Table~III of Bionta et al.\ \cite{Bionta:1987qt}, with a total of 15~muons.}
\end{table}

\begin{table}
\vskip6pt
 \caption{SN~1987A neutrinos in Kamiokande~II \cite{Hirata:1988ad}. Time relative to the first event at UT 7:35:35 $\pm$1~min.}
 \vskip4pt
    \label{tab:KamII-data}
    \centering
    \begin{tabular*}{\columnwidth}{@{\extracolsep{\fill}}lllll}
    \hline\hline
     Event& Time &$N_{\rm hit}$    & Positron event energy         & Angle       \\
     No.  & [sec]&            & [MeV]          & [Degree]    \\
     \hline
     \hp1         & \hp0.000 & 58 &   $20.0\pm2.9$ &\hp$18\pm18$ \\
     \hp2         & \hp0.107 & 36 &   $13.5\pm3.2$ &\hp$40\pm27$ \\
     \hp3         & \hp0.303 & 25 & \hp$7.5\pm2.0$ &  $108\pm32$ \\
     \hp4         & \hp0.324 & 26 & \hp$9.2\pm2.7$ &\hp$70\pm30$ \\
     \hp5         & \hp0.507 & 39 &   $12.8\pm2.9$ &  $135\pm23$ \\
     \hp6$^{a}$   & \hp0.686 & 16 & \hp$6.3\pm1.7$ &\hp$68\pm77$ \\
     \hp7         & \hp1.541 & 83 &   $35.4\pm8.0$ &\hp$32\pm16$ \\
     \hp8         & \hp1.728 & 54 &   $21.0\pm4.2$ &\hp$30\pm18$ \\
     \hp9         & \hp1.915 & 51 &   $19.8\pm3.2$ &\hp$38\pm22$ \\
       10         & \hp9.219 & 21 & \hp$8.6\pm2.7$ &  $122\pm30$ \\
       11         &   10.433 & 37 &   $13.0\pm2.6$ &\hp$49\pm26$ \\
       12         &   12.439 & 24 & \hp$8.9\pm1.9$ &\hp$91\pm39$ \\
       13$^{a,b}$ &   17.641 &    & \hp$6.5\pm1.6$ &  $103\pm50$ \\
       14$^{a,b}$ &   20.257 &    & \hp$5.4\pm1.4$ &  $110\pm50$ \\
       15$^{a,b}$ &   21.355 &    & \hp$4.6\pm1.3$ &  $120\pm50$ \\
       16$^{a,b}$ &   23.814 &    & \hp$6.5\pm1.6$ &  $112\pm50$ \\
     \hline
    \multicolumn{5}{l}{$^a$Usually attributed to background}    \\
    \multicolumn{5}{l}{$^b$Times and energies from Ref.~\cite{Loredo:2001rx}, angles from Ref.~\cite{Krivoruchenko:1988zg}\kern-15em}   \\
    \end{tabular*}
\end{table}

\begin{table}
 \caption{SN~1987A neutrinos in BUST (Baksan) \cite{Alekseev:1987ej, Alekseev:1988gp}. Time relative to event No.~1 at UT 7:36:11.818 ${+}2/{-}54$~s.}
 \vskip4pt
    \label{tab:Baksan-data}
    \centering
    \begin{tabular*}{\columnwidth}{@{\extracolsep{\fill}}lll}
    \hline\hline
     Event& Time     & Positron event energy         \\
     No.  & [sec]    & [MeV]          \\
     \hline
     \hp0$^a$& $-$5.247 &  17.5$\pm$3.5 \\
     \hp1    &\hm 0.000 &  12.0$\pm$2.4 \\
     \hp2    &\hm 0.435 &  18.0$\pm$3.6 \\
     \hp3    &\hm 1.710 &  23.3$\pm4.7$ \\
     \hp4    &\hm 7.687 &  17.0$\pm3.4$ \\
     \hp5    &\hm 9.099 &  20.1$\pm4.0$ \\
     \hline
    \end{tabular*}
    \vskip2pt
    \vbox{\raggedright
    $^a$Attributed to background by Alexeev et al.\ \cite{Alekseev:1987ej} based on the time structure of the signals in the other detectors and was dropped without mention in subsequent publications.}
\end{table}

To gain intuition for a SN signal, we have defined in Appendix~\ref{sec:Fiducial-Supernova} a ``fiducial SN'' at a distance of 50~kpc that emits a total of $5\times10^{52}$~erg in $\bar\nu_e$ with a Maxwell-Boltzmann spectrum. Assuming $T=4$~MeV, this source causes the production of 53.3 positrons from IBD in the detector volume, of which $6.4^{+3.7}_{-2.8}$ are expected to be seen, to be compared with 8 detected events. The uncertainty is understood as maximum errors derived from the uncertain trigger efficiency. The average observed event energy was 31.9~MeV, to be compared with an expectation of $30.1^{+2.9}_{-2.1}$. Notice that the positron event energy, inferred from the number of PMTs firing in connection with the event, does not directly coincide with the real positron energy due to Poisson fluctuations. Therefore, the event distribution in terms of the positron event energy must include a smearing due to the energy resolution, as we discuss in Appendix~\ref{app:methods_comparison}. In all tables and figures, we will specify whether we are referring to neutrino energy, positron energy, or positron event energy. We summarize these results in Table~\ref{tab:Signals}. 

The second largest of the running experiments was the Kamiokande-II (Kam-II) water Cherenkov detector with 2140~t sensitive mass \cite{Arisaka:1985lki, Hirata:Moriond, Kamiokande-II:1987idp, Hirata:1988ad, Nakahata:1988}, but much lower threshold (Fig.~\ref{fig:Detectors}), making it competitive for SN neutrino detection. For our fiducial SN, a total of 16.7 positrons is produced in the detector, of which 14.0 would be detected with an average energy of 20.0~MeV.

\begin{figure}
    \centering
\includegraphics[width=0.85\columnwidth]{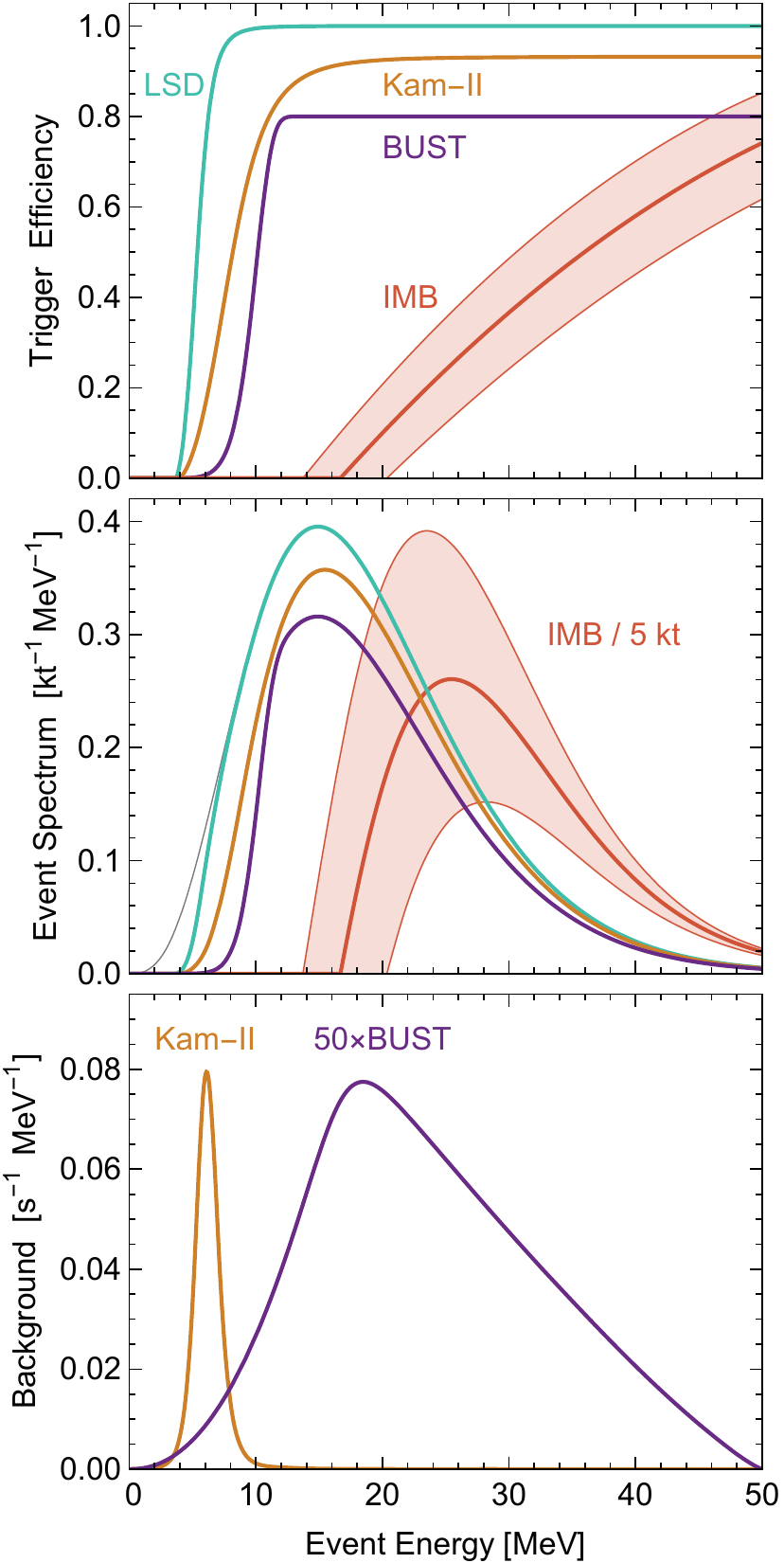}
    \caption{Detector properties. {\em Top:\/} Trigger efficiencies (for IMB with maximum uncertainties). {\em Middle:\/} Event spectrum for our fiducial SN ($T=4$~MeV), normalized to 1~kt water equivalent (5~kt for IMB). As a thin gray line, nearly identical with LSD, we show the event spectrum without threshold. This illustrative plot does not include
    ``smearing'' of the positron event energy by finite energy resolution of the detectors.
    {\em Bottom:\/} Background rate in Kam-II (total 0.187~Hz) and BUST (total 0.034~Hz) as function of reconstructed event energy. These background events have various sources, for example $\beta$ decay of ${}^{214}$Bi in Kam-II.    
    }
    \label{fig:Detectors}
\end{figure}

\begin{table}[ht]
\vskip-10pt
 \caption{Time-integrated signal properties in the detectors for our  ``fiducial'' SN defined in Appendix~\ref{sec:Fiducial-Supernova} and also used in Fig.~\ref{fig:Detectors}. (Total emitted $\bar\nu_e$ energy of 50~B and thermal Maxwell-Boltzmann spectrum with $T=4$~MeV.)
 \label{tab:Signals}}
 \vskip4pt
    \begin{tabular*}{\columnwidth}{@{\extracolsep{\fill}}lllll}
    \hline\hline
           & IMB & Kam-II & BUST & LSD\\
     \hline
     Mass water equivalent & 6800~t & 2140~t & 280~t& 126~t  \\
     Low-energy BKG [Hz]   & ---    & 0.187  & 0.034 & 0.012 \\
     \multicolumn{4}{@{}l}{\bf Total signal (13~s)}  \\
     \quad Events     &\hp8   & 11    &\hp 5   & ---  \\
     \quad + Likely BKG & --- &\hp1   & --- & ---  \\
     \quad Expected BKG  & --- &\hp2.4 &\hp0.44 &\hp 0.16 \\
     \quad $\langle\epsilon_e\rangle$ [MeV]&31.9 & 15.4 &18.1 & --- \\
     \multicolumn{4}{@{}l}{\bf Reduced signal (6~s) without late events}  \\
     \quad Events     &\hp8 &\hp8   &\hp 3   & ---  \\
     \quad + Likely BKG     & --- &\hp1   & --- & ---  \\
     \quad Expected BKG     & --- &\hp1.1 &0.20 &\hp 0.07 \\
     \quad $\langle\epsilon_e\rangle$ [MeV] &31.9 &17.4 & 17.8 & ---\\
     \multicolumn{4}{@{}l}{\bf Late signal alone (7--13~s)}  \\
     \quad Events           & --- &\hp3   &\hp 2   & ---  \\
     \quad Expected BKG     & --- &\hp1.1 &0.20 &\hp 0.07 \\
     \quad $\langle\epsilon_e\rangle$ [MeV] &--- &10.2 & 18.6 & ---\\
     \multicolumn{4}{@{}l}{\bf Expected signal from fiducial SN}  \\
     \quad Events                          &\hp$6.4^{+3.7}_{-2.8}$& 14.0 &\hp1.52&\hp 0.96  \\[0.5ex]
     \quad $\langle\epsilon_e\rangle$ [MeV]& $30.1^{+2.9}_{-2.1}$ & 20.0 & 20.5  & 19.2 \\[0.5ex]
     \hline
    \end{tabular*}
\end{table}

The time sequence of SN~1987A events is shown in the second panel of Fig.~\ref{fig:SNData} such that the first events in IMB and Kam-II are contemporaneous. As explained in more detail in Appendix~\ref{sec:SN-Observations}, IMB had good absolute clock time, whereas that of Kam-II was uncertain within $\pm1$~min so that the relative timing is completely uncertain within the burst duration of a few seconds.

Besides atmospheric muons at large energies, \hbox{Kam-II} had a low-energy background mostly from the beta-decay of $^{214}$Bi with a rate of 0.187~Hz. The spectrum is shown in the bottom panel of Fig.~\ref{fig:Detectors}, where we see that there is a fairly clean energy cut between signal and background. Usually event No.~6 (Table~\ref{tab:KamII-data}) is attributed to background as well as the very late events after 17~s. According to these assumptions, there are 11 SN-related events with an average energy of 15.4~MeV.

We will later see that the three late Kam-II events at around 10~s are difficult to explain by PNS cooling. If indeed they have a different origin as discussed later, then we should only consider the first 9 events minus No. 6 if interpreted as background. In this case there are 8 SN-related events with an average energy of 17.4 MeV. It was often mentioned that there is significant tension
between the average IMB and Kam-II event energies if interpreted as coming from a common quasi-thermal $\bar\nu_e$ flux. This tension is much smaller if the late Kam-II events are not part of the cooling signal.

The third yet smaller detector was the Baksan Underground Scintillator Telescope (BUST) \cite{Pomansky:1978xc, Alekseev:1979pn, Chudakov:1979vh, Alekseev:1987ej, Alekseev:1988gp, Alekseev:1993dy, Kuzminov:2017, Novoseltsev:2019gdt, Novoseltseva:2022qic} with a water-equivalent mass of 280~t, meaning the mass holding the same number of protons relevant for IBD. The sensitive mass is only around 13\% that of Kam-II and the trigger efficiency is somewhat smaller, so from our fiducial SN one expects 1.52 events with average energy 20.5~MeV. BUST had a low-energy background of 0.034~Hz with a much broader spectrum (see Fig.~\ref{fig:Detectors}) so that there is no energy cut between signal and background. On the contrary, they peak around the same energy of around 20~MeV. However, over the full Kam-II duration of 13~s one would expect only 0.44 background events. BUST observed 6 events, of which the one called No. 0 in Table~\ref{tab:Baksan-data} was dismissed as background based on the time structure of the overall signal. BUST also had a significant clock uncertainty and in Fig.~\ref{fig:SNData} event No.~1 is aligned with No.~1 in IMB. In our time-dependent analysis we will find that it is indeed likely that this event was background. Considering only the 5 remaining ones, the average observed energy is 18.1~MeV, compatible both with a SN or background origin. However, if the BUST events were actually background, they would represent a
rare upward Poisson fluctuation.

If we interpret the late Kam-II events as not coming from PNS cooling, the same cut should be applied to the BUST data, reducing the signal to 3~events with an average energy of 17.8~MeV (Table~\ref{tab:Signals}). Either way, there is no objective justification for not including the BUST signal in our analysis.

The smallest relevant detector was the Liquid Scintillator Detector (LSD) \cite{Badino:1984ww, Aglietta:1987it, Aglietta:1987we, Dadykin:1987ek, Saavedra:1987tz} in the Mont Blanc tunnel with a w.e.\ mass of 126~t, but better trigger efficiency than BUST, so the expected SN signal is about 2/3 that of BUST. This detector observed a burst almost five hours earlier than the other detectors, which subsequently has been dismissed (see Appendix~\ref{sec:LSD} for a more detailed discussion). On the other hand, LSD was of course also sensitive at the time of the other detectors and it had good absolute clock time. As shown in the bottom panel of Fig.~\ref{fig:SNData}, it observed an event at around 19~s (and another one at 37.5~s) that should be attributed to background. LSD had a low-energy background rate of 0.012~Hz, so over 13~s one expects 0.16 events. Its non-observation during the signal at the other detectors provides a constraint on the overall neutrino flux. This constraint is not very restrictive, but somewhat counteracts the upward fluctuation at BUST. In any case, there is no reason to exclude LSD from the analysis.

\onecolumngrid


\section{Fit of time-integrated flux}
\label{sec:Time-Integrated-Analysis}
\vskip8pt
\twocolumngrid

\subsection{Maxwell-Boltzmann Spectrum}

As a first step to interpret the SN~1987A data beyond the simple averages of the previous section, we perform a maximum-likelihood analysis similar to previous works and detailed in Appendix~\ref{app:methods_comparison}. The main difference to previous studies is including an upper limit from the non-observation at LSD and the uncertainty of the IMB trigger efficiency, which has the main effect of increasing the uncertainty of the measured event rate, whereas the average event energies remain similar (see Appendix~\ref{sec:IMB} for more details). We treat the uncertainty of $\pm0.05$ and uncertain energy calibration of $\pm10\%$ as separate nuisance parameters, assuming a top-hat distribution between the extreme values. Together these effects lead to the maximum range shown in Fig.~\ref{fig:Detectors}. Moreover, we model the background spectrum rather than censoring individual events, where for the signal duration we use 13~s as suggested by the full Kam-II signal. 

We follow the procedure outlined in Appendix~\ref{app:methods_comparison} to obtain the time-integrated, energy-dependent event rate at each of the experiments IMB, Kam-II,  BUST, and LSD.  For each of them, we obtain a likelihood $\mathcal{L}_E$ based on the time-integrated, energy-dependent analysis according to Eq.~\eqref{eq:energy_likelihood}. As in the previous section, we assume that each detector observes a $\bar\nu_e$ flux with a Maxwell-Boltzmann spectrum that is given in terms of a temperature $T$. Instead, we use the average neutrino energy $\bar\epsilon=3T$ as our spectral parameter as it lends itself to generalization when we later consider quasi-thermal (pinched) spectra. The total flux is expressed in terms of the total energy $E^{\bar{\nu}_e}_\mathrm{tot}$ emitted at a distance of 50~kpc, for now ignoring the issue of flavor conversion. The IMB likelihood in addition depends on the two nuisance parameters $\zeta$ and $\xi$, describing the uncertainty in the energy scale and normalization of the efficiency curves as described in Appendix~\ref{app:methods_comparison}. 

We then proceed to fit separately each experiment, using our maximum-likelihood definition. In the case of IMB, for each choice of $\bar{\epsilon}$ and $E^{\bar{\nu}_e}_\mathrm{tot}$ we marginalize over the nuisance parameters by maximizing the likelihood with respect to them. For BUST, we use the detected events from Table~\ref{tab:Baksan-data}, including event number $0$ conventionally attributed to background, since we are anyway modeling also the background spectrum for the detector following Ref.~\cite{Loredo:2001rx}; we have verified that adding or removing this event leads to negligible changes in the results. For LSD we consider no detected event, i.e., we do not worry about their very late events.

\begin{figure}
\vskip-6pt
    \centering
    \includegraphics[width=0.46\textwidth]{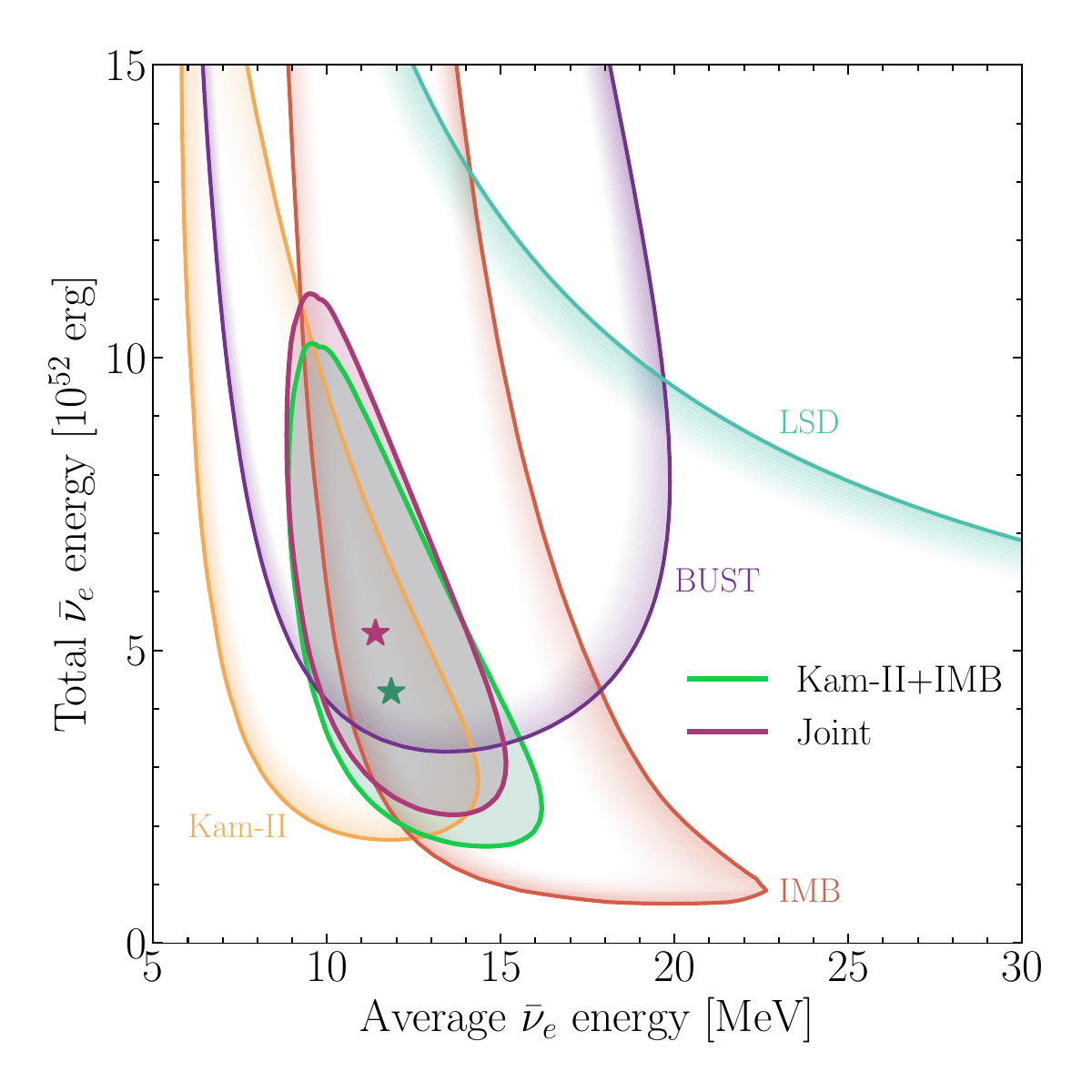}
    \vskip-6pt
    \caption{Allowed regions (95\% CL) for the total and average $\bar\nu_e$ energy measured at different detectors, assuming a Maxwell-Boltzmann spectrum with $T=\bar\epsilon/3$. Joint region and best-fit value from IMB+Kam-II in green, from all experiments in magenta.}
    \label{fig:pinched_all}
    \vskip-6pt
\end{figure}

Figure~\ref{fig:pinched_all} shows the $95\%$ confidence contours in the parameter space of $\bar{\epsilon}$ and $E^{\bar{\nu}_e}_\mathrm{tot}$ for all experiments separately. We use the test statistic $\Lambda$ as defined in Eq.~\eqref{eq:TS_definition}, asymptotically distributed as a chi-squared variable with two degrees of freedom under the null hypothesis. Therefore, we obtain $68\%$ and $95\%$ confidence level (CL) contours by setting $\Lambda=2.3$ and $\Lambda=6$ respectively.

The confidence regions for IMB and Kam-II are the usual banana-shaped contours that allow for a large flux at low energies due to the vanishing trigger efficiency. Even a huge flux at very low energies would not be seen. If we were to combine IMB with LSD, the non-observation at the small LSD detector would chop off the upper part of the IMB banana because of the low LSD threshold. 

However, if we combine IMB with Kam-II as in many previous studies, we obtain the closed 95\% confidence region shown in green in Fig.~\ref{fig:pinched_all} with a best-fit value close to the parameters of our earlier fiducial SN. For such a joint analysis of different experiments, we define as a likelihood the product of the individual likelihoods, marginalized over the nuisance parameters of IMB. 

Similar to previous studies, we find a tension between the Kam-II and IMB confidence regions. While at $2\sigma$ these overlap generously, Kam-II points to lower average $\bar{\nu}_e$ energies than IMB, an effect that we already saw in 
Table~\ref{tab:Signals}. We find a similar tension with BUST, which also points to somewhat larger $\langle\epsilon_{\bar{\nu}_e}\rangle$ and significantly larger total flux. Kam-II and BUST have similar trigger response, but the latter has only 13\% of the Kam-II mass, yet observed half as many events. For LSD, since no event was observed in the time frame of the other measurements, only an upper bound on $E^{\bar{\nu}_e}_\mathrm{tot}$ is found. Notice that for our earlier fiducial SN, LSD should have seen around 2/3 as many events as BUST, so it is not ridiculously small by comparison. The non-observation at LSD is somewhat of a downward fluctuation, the events at BUST a large upward fluctuation of signal, background, or both.
The impact of adding BUST and LSD to the joint analysis of all the experiments is to favor a region of slightly \textit{lower} $\langle\epsilon_{\bar\nu_e}\rangle$ and \textit{higher} $E_{\rm tot}^{\bar\nu_e}$, due to the large flux needed to explain the BUST observations.

\subsection{Excluding Late Events}

\begin{figure}
\vskip-12pt
    \centering
    \includegraphics[width=0.46\textwidth]{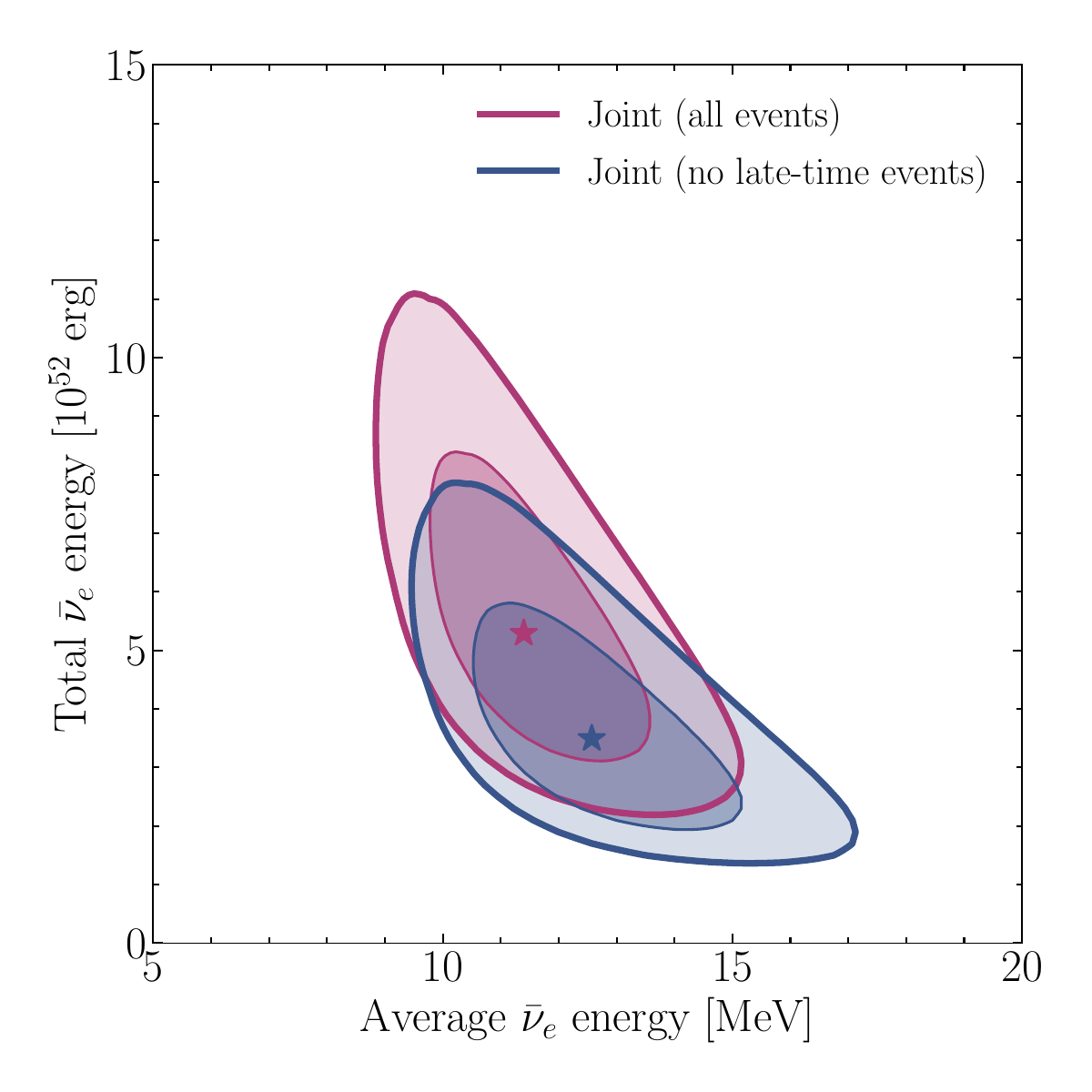}
    \vskip-10pt
    \caption{Confidence regions for all experiments combined, with (magenta) and without (blue) the late-time events, assuming a Maxwell-Boltzmann spectrum. We show $68\%$ and $95\%$ confidence regions as well as best-fit points.}
    \label{fig:pinched_no_late}
\end{figure}

In our time-dependent comparison between numerical models and SN~1987A data we will see that the late-time Kam-II events as well as those at BUST are difficult to explain by PNS cooling.
We will discuss this topic in more detail in Sec.~\ref{sec:speculations} and here only ask how the confidence contours for the PNS cooling signal change under this assumption. In other words, we assume that the PNS cooling signal only lasts until 6~s, including all IMB events, while leaving out Nos.~10--12 in Kam-II as well as Nos.~4--5 in BUST. As we have already seen in Table~\ref{tab:Signals}, the remaining signals are more compatible (lesser tension of average energies). 

In Fig.~\ref{fig:pinched_no_late} we show the shift of the confidence region and best-fit parameters when we ignore the late \hbox{Kam-II} and BUST events. As anticipated in Table~\ref{tab:Signals}, the implied $\langle\epsilon_{\bar\nu_e}\rangle$ is larger, whereas the required $\bar\nu_e$ flux to explain the data is naturally smaller because there are fewer events to account for and higher energies imply that a smaller flux is enough to explain the same number of events.

\begin{figure}
 \vskip-10pt
    \centering
    \includegraphics[width=0.46\textwidth]{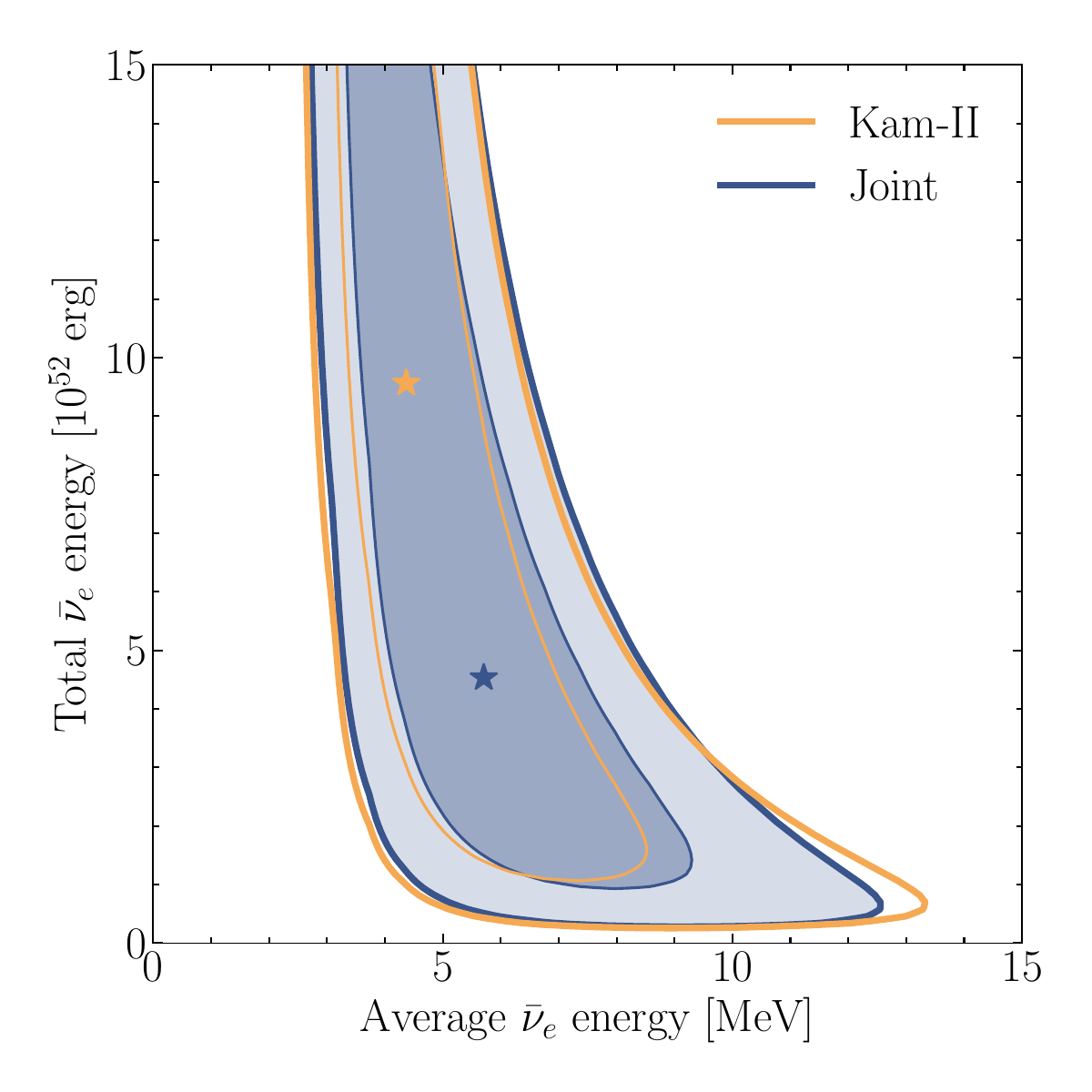}
     \vskip-10pt
    \caption{Confidence regions (68\% and 95\%) and best-fit point (golden) if the three late Kam-II events are attributed to a Maxwell-Boltzmann flux originating from a SN fallback signal. Joint analysis from all experiments (notably two late BUST events) in blue.}
    \label{fig:pinched_only_late}
    \vskip-4pt
\end{figure}

\begin{figure*}
    \centering
    \includegraphics[width=0.76\textwidth]{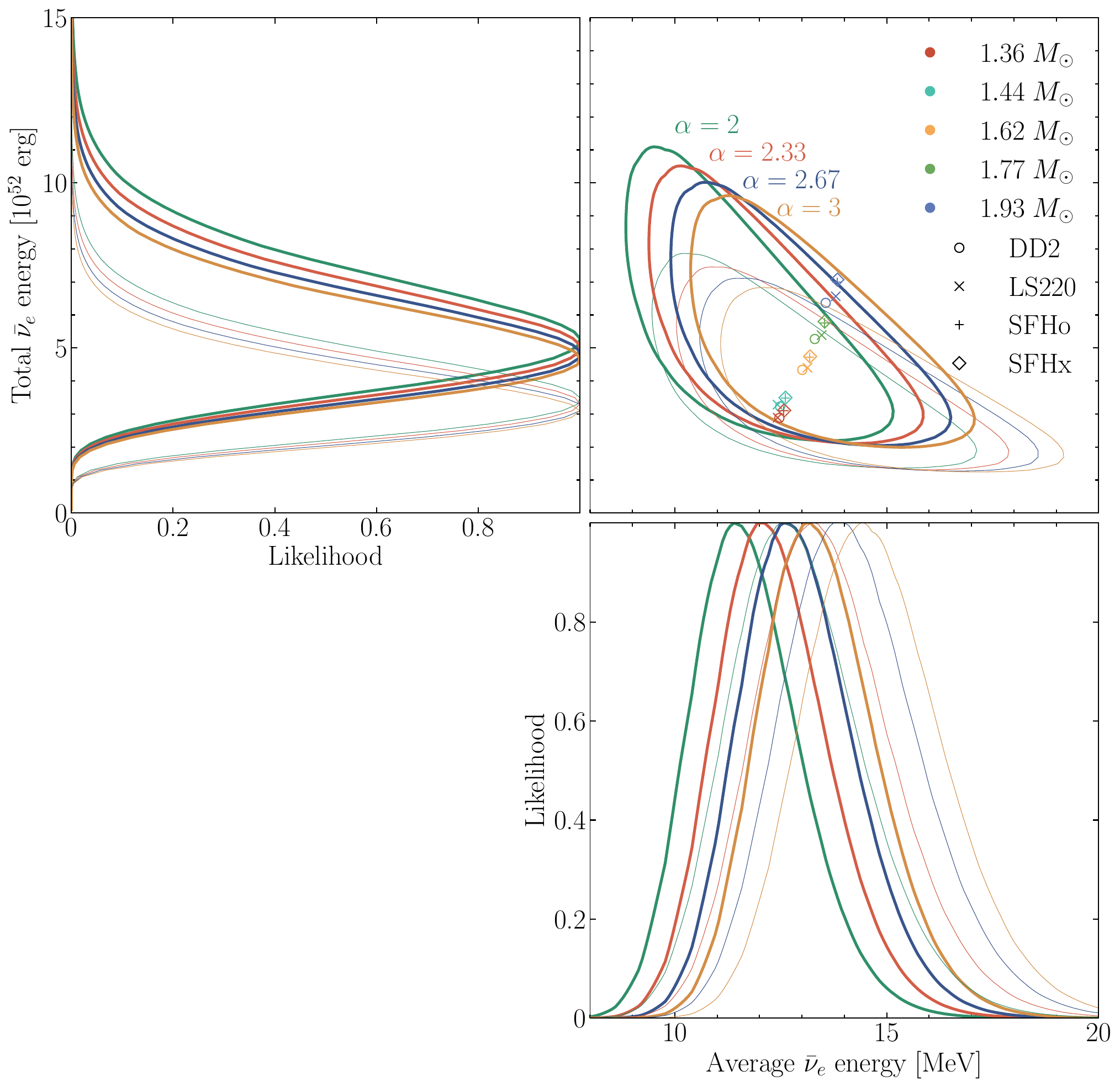}
    \caption{Joint $95\%$ confidence contours from all experiments for the shown values of the pinching parameter. Thick contours include all data, thin ones exclude the late-time events. Our model predictions are superimposed, ignoring flavor conversion, with different colors indicating the different PNS masses. The four groups with different symbols correspond to the different EoS models. The one-dimensional likelihoods marginalized over the complementary parameters are shown as a corner plot.}
    \label{fig:overlap_simulations}
\end{figure*}

\begin{figure*}
    \centering
    \includegraphics[width=0.76\textwidth]{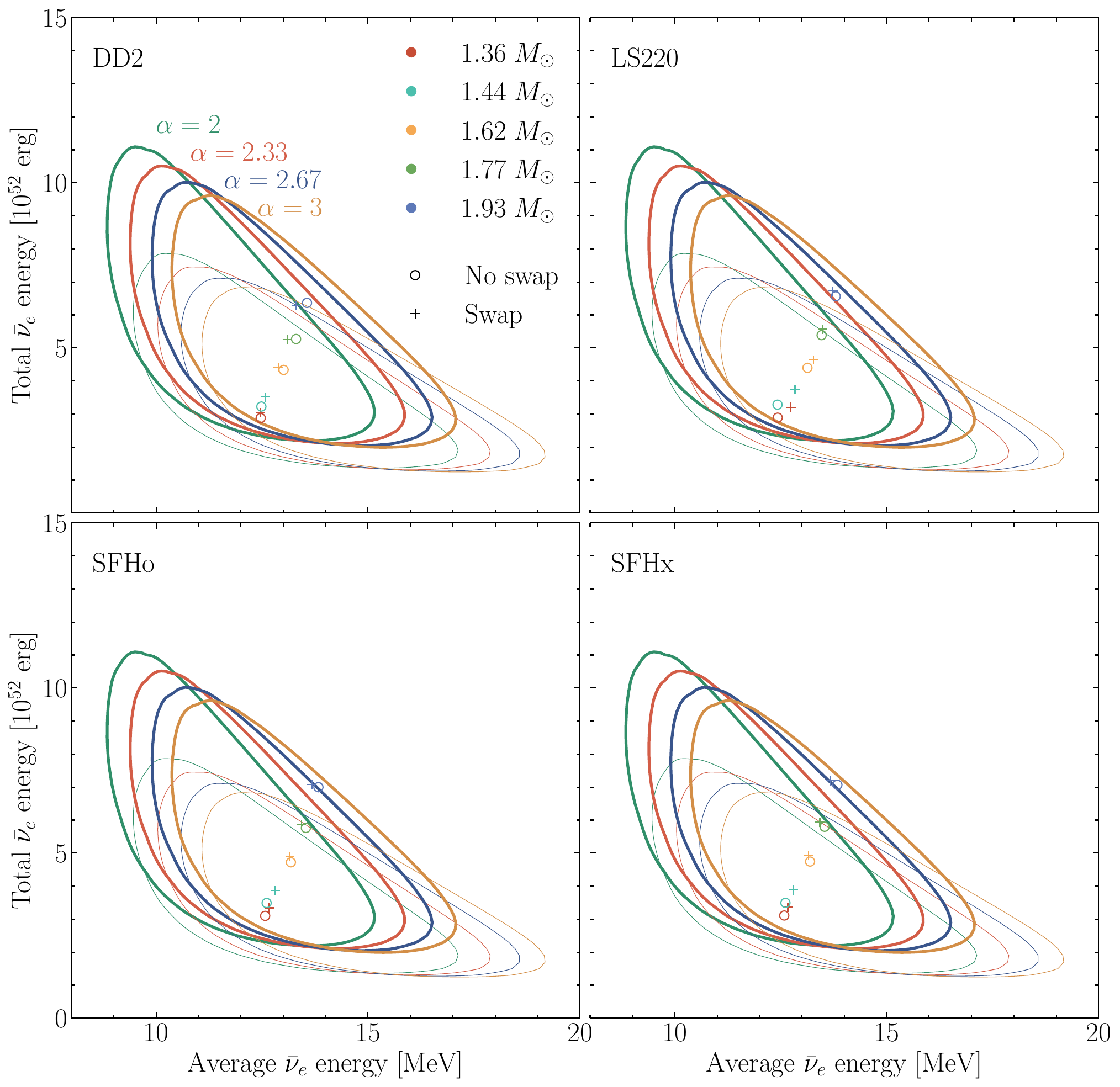}
    \caption{Impact of flavor conversion. Same confidence contours as in Fig.~\ref{fig:overlap_simulations} in each panel. For each EoS we show the model predictions in a different panel for the different progenitor masses without (open circles) and with complete (plus signs) $\bar\nu_e$--$\bar\nu_x$ flavor swap.}
    \label{fig:banana_plot_flavor_swap}
\end{figure*}

\begin{figure*}
\includegraphics[width=0.99\textwidth]{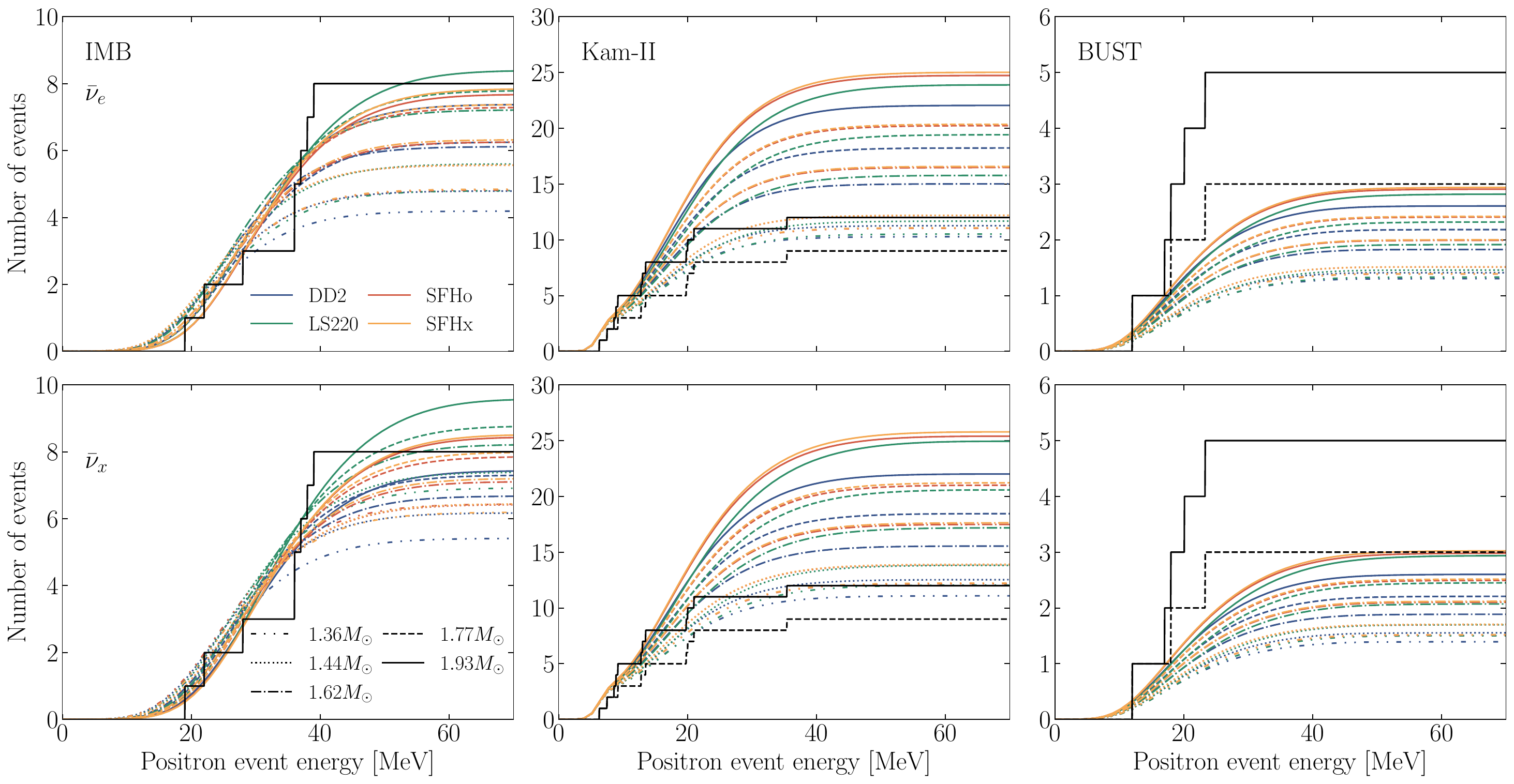}
\caption{Cumulative energy distributions in IMB, Kam-II, and BUST together with the expectations from our models for no flavor swap ({\em top row}) and full flavor swap ({\em bottom row}). The black dashed lines for Kam-II and BUST leave out the late events.
}\label{fig:cumulative_spectrum}
\end{figure*}

\subsection{Late Events Alone}

As one speculation one may ask what is required to explain the late Kam-II and BUST events by a SN fallback signal that produces another $\bar\nu_e$ burst with a Maxwell-Boltzmann spectrum? In Fig.~\ref{fig:pinched_only_late} we show the corresponding confidence contour and best fit point
for \hbox{Kam-II} alone (golden) and joint for all experiments (blue). For IMB and LSD, this means no observation; for IMB, we fix the efficiency parameters to their nominal values. The required flux is very large, comparable to the initial burst, because the small energies of the late events together with the smaller cross section and smaller trigger efficiency require a large flux to cause three events, which themselves on the other hand are subject to huge Poisson fluctuations. The total required flux is poorly constrained. Including all detectors shifts the contours to larger $\bar\nu_e$ energies and the best-fit value to a smaller total flux, simply because the late events in BUST have relatively high energies.

Concerning the total SN emitted energy, PNS and fallback burst have different interpretations. We here only fit parameters assuming a Maxwell-Boltzmann $\bar\nu_e$ flux arrives at the detectors. However, the PNS cooling signal produces fluxes of all flavors and we will see that flavor conversion has only a small impact. On the other hand, if the fallback signal stems from an accretion disk, it will produce primarily $\bar\nu_e$ and $\nu_e$, and the detected and emitted fluxes can be more strongly modified by flavor conversion. For a large $\bar\nu_e$ survival probability, the source will need to emit much less total energy than during the PNS cooling phase to explain the same number of events.

\subsection{Pinched Spectra}

The neutrino fluxes emerging from a SN are not exactly thermal, but rather tend to be ``pinched,'' meaning that they are more narrowly peaked than a Maxwell-Boltzmann spectrum. Of course, if neutrinos would emerge as truly thermal radiation, their spectrum would be of Fermi-Dirac type. Phenomenologically, the numerical quasi-thermal spectra are well represented as a Gamma distribution (``alpha fit'') of the form, here for the fluence arriving at Earth,
\begin{equation}
    \frac{d\mathcal{F}_{\bar{\nu}_e}}{d\epsilon_\nu}=
    \frac{E^{\bar{\nu}_e}_\mathrm{tot}}{\Gamma_{1+\alpha} \bar{\epsilon}^{\,2}}
    \frac{(1+\alpha)^{1+\alpha}}{4\pi d_\mathrm{SN}^2} \left(\frac{\epsilon_\nu}{\bar{\epsilon}}\right)^\alpha 
    e^{-(1+\alpha)\epsilon_\nu/\bar{\epsilon}}.
\end{equation}
$E^{\bar{\nu}_e}_\mathrm{tot}$ is the total energy injected in $\bar{\nu}_e$ and $\Gamma_x$ is the Gamma function at argument $x$, not to be confused with the Gamma distribution itself. The normalization is arranged such that the average energy is $\langle\epsilon_{\bar\nu_e}\rangle=\bar\epsilon$, the latter being a fit parameter. A Maxwell-Boltzmann spectrum has $\alpha=2$ and then $\bar\epsilon=3T$. For larger pinching parameter ($\alpha>2$), the spectrum is more narrowly peaked for the same average energy. Instantaneous fluxes can be strongly pinched, depending on flavor and SN emission phase, whereas the time-integrated numerical fluxes tend to have $\alpha=2$--3 as shown for our models in Table~\ref{tab:Neutrino-flux-properties}. For more technical details about the Gamma distribution see Appendix~\ref{sec:Gamma}.

Figure~\ref{fig:overlap_simulations} shows the joint $95\%$ confidence contours from all experiments for the indicated values of the pinching parameter $\alpha$ in the range 2--3, both including (thick lines) and excluding (thin lines) the late-time events. We also show the one-dimensional likelihoods marginalized over the complementary parameters as a corner plot. We see that the best-fit overall neutrino flux depends only minimally on the assumed pinching, whereas the implied $\langle\epsilon_{\bar\nu_e}\rangle$ is shifted from around 11.3~MeV to around 13.1~MeV. The IMB events, which are always in the tail of the distribution, pull the implied average to higher values if the spectrum is more narrowly pinched.

\subsection{Model Comparison}

For comparison, we superimpose on the figure the parameters derived from our suite of models, for now ignoring flavor conversion. We find that the points lie more or less within the range identified by the data, with a relatively large tension with the heavy progenitors $M=1.93\;M_\odot$; we will see that this is indeed confirmed by our more detailed analysis in the next section. Excluding the late-time events does not strongly modify this conclusion. In principle, of course, each model also has its own value of $\alpha$ so that each model would have to be compared with only one of the contours. However, it is clear that one cannot discriminate between the models in a meaningful way based on this analysis. On the other hand, the $1.62\,M_\odot$ models cluster around the best-fit values and in this regard the exact pinching parameter and the late-time events make no big difference.

\subsection{Flavor Conversion}

We have already seen that the integrated $\bar\nu_e$ and $\bar\nu_x$ spectra are not very different so that flavor conversion will not have a large impact on the signal. We here return to this question and show in Fig.~\ref{fig:banana_plot_flavor_swap} the model predictions without flavor swap (the same as in the previous figure) and with complete swap of $\bar{\nu}_e$ with $\bar{\nu}_x$, i.e., the signal is assumed to be caused by the  $\bar{\nu}_x$ flux from the SN.

The shifts of the loci in this figure are so small that we show them in different panels for each EoS, and then for each progenitor mass. On the scale of the confidence regions provided by the sparse SN~1987A data, these shifts are minimal and certainly one cannot discriminate between flavor conversion scenarios based on these data.

\subsection{Cumulative energy spectrum}

Figure~\ref{fig:cumulative_spectrum} displays the cumulative energy distributions for the neutrino events detected by all experiments compared to the predicted signals from our SN and PNS models. We include it here in our discussion of the time-integrated signal properties although we need to anticipate some choices of parameters that will play a role for the time-dependent analysis presented in Sec~\ref{sec:Time-Dependent-Analysis}. Specifically, for IMB, we use the efficiency parameters $\xi$ and $\zeta$ that best fit the time and energy structure according to the procedure described in Section~\ref{sec:cumdis}. The signal is time-integrated; for Kam-II and BUST, the background is integrated over a nominal signal duration of 13\,s.

The cumulative energy distributions of Fig.~\ref{fig:cumulative_spectrum} again visualize, from a different perspective, the tension between the three experiments that we already concluded from the confidence contours in the parameter space of $\bar{\epsilon}$ and $E^{\bar{\nu}_e}_\mathrm{tot}$ and that we will also find in our time-dependent analysis. The plots show that the cumulative energy distribution in IMB is matched quite well by the most massive (1.77 and 1.93\,M$_\odot$) of our PNS models, in particular under the assumption of a full flavor swap between $\bar\nu_e$ and $\bar\nu_x$. In contrast, the cumulative energy distribution in Kam-II (also including the latest events) is best compatible with the model signals of our lightest (1.36 and 1.44\,M$_\odot$) PNS cases. The BUST detection can be matched only by some of our most massive PNS models and if the last two BUST events are omitted from the cumulative energy distribution. These results will be confirmed by our comparison of the cumulative time distributions (Fig.~\ref{fig:cumulative_distributions}; Sec.~\ref{sec:Time-Dependent-Analysis}) and the test statistic $\Lambda$ measuring the quality of the fits (Fig.~\ref{fig:likelihood_exps}; Sec~\ref{sec:Overall-Comparison}).  

Interestingly, however, the agreement between a subset of our models and the Kam-II data cannot be achieved equally well when the last three Kam-II events are excluded. This is puzzling and a strange aspect of the Kam-II measurement, whose time structure with the $\sim$7\,s gap, followed by the three events more than 9\,s after the first one, poses many more challenges for an explanation by PNS cooling signals than the IMB data. We will investigate this aspect in more detail in the following section.

\onecolumngrid

\section{Time-Dependent Analysis}
\label{sec:Time-Dependent-Analysis}
\vskip8pt
\twocolumngrid

\subsection{Signal duration}
\begin{figure}[ht!]
    \vskip4pt
    \includegraphics[width=1\columnwidth]{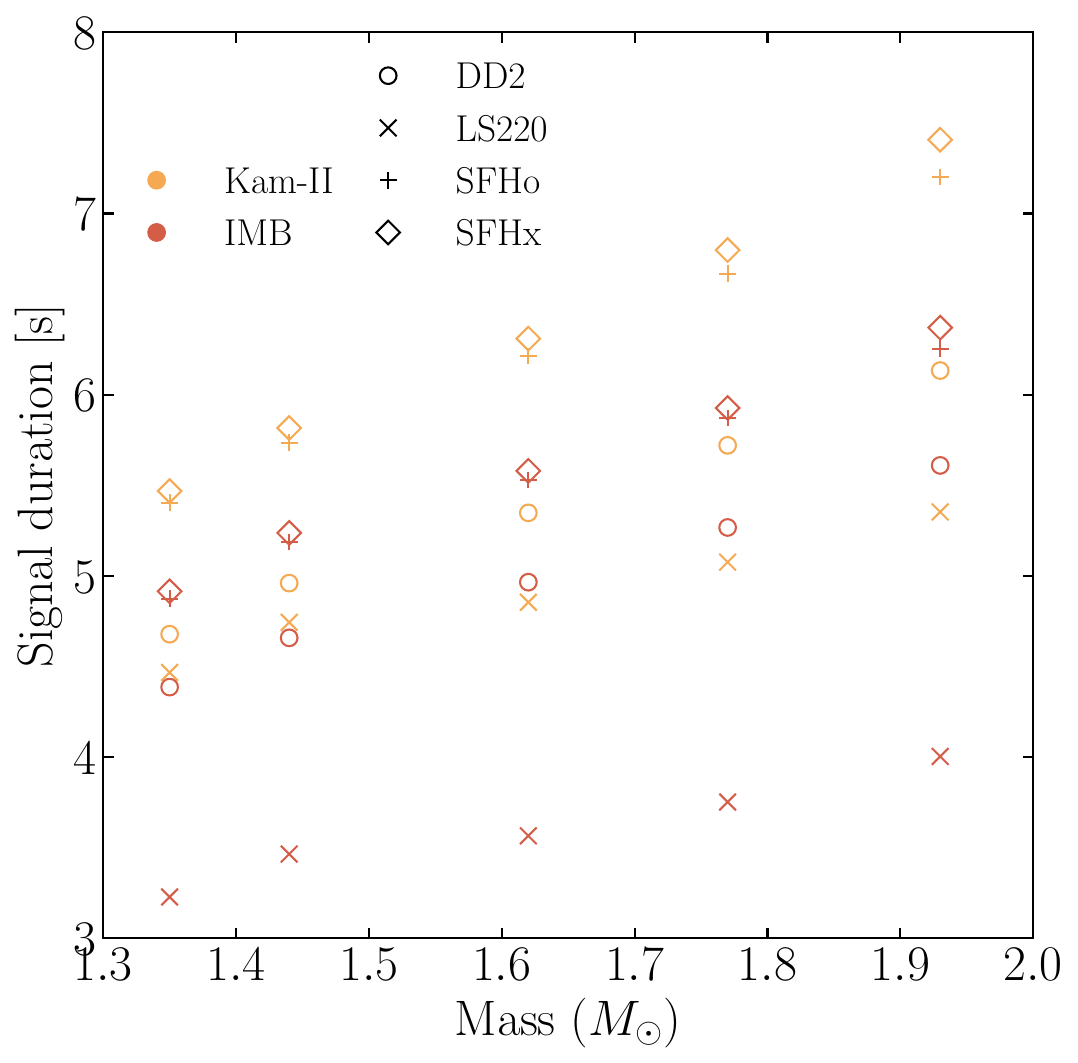}
    \caption{Expected signal durations in IMB and Kam-II, i.e., the time when 95\% of the expected events have accrued, for all our SN models based on the $\bar\nu_e$ signal without flavor conversion. These results are also shown in Table~\ref{tab:Neutrino-flux-properties}, where the signal duration based on full flavor swap is provided as well.
    \label{fig:SignalDuration}}
    \vskip2pt
\end{figure}

\vskip-4pt
We have seen that the time-integrated neutrino fluxes produced by our models agree well with the data, given that the sparse data themselves show many fluctuations or tensions. However, when we turn to the time structure of the model predictions, the picture changes because generally the signal duration implied by the models is too short. As a simple definition of ``signal duration'' $\tau$ we use the time it takes for 95\% of the expected events in a given detector from a given SN model to build up. In Table~\ref{tab:Neutrino-flux-properties} we tabulate $\tau$ for all our models and the detectors IMB, Kam-II, and BUST, assuming either that all events are caused by the original $\bar\nu_e$ flux or by $\bar\nu_\mu$ or $\bar\nu_\tau$, assuming a complete flavor swap. We show the same information graphically in Fig.~\ref{fig:SignalDuration} for the no-swap case and only for IMB and Kam-II. BUST and LSD predict similar signal durations as Kam-II because of their similar detector response.

The signal durations vary, but typically are 5--7~s, where the LS220 models are exceptional in that they predict particularly short durations at IMB. Generally, IMB yields a significantly shorter duration than Kam-II and BUST, because it is mostly sensitive to the higher-energy neutrinos emitted in the early phase of the cooling. The signal duration increases with NS mass, but is always much shorter than suggested by the three late Kam-II events and the two late BUST events.

\subsection{Cumulative Distributions}
\label{sec:cumdis}

Another way of illustrating this point is to show the observed cumulative event distribution as a function of time for the different detectors together with expectations. However, to compare data with models, we need the relative offset time $\delta t$ between the SN bounce time (the zero model time) relative to the first event in each detector. Because of the clock uncertainties in Kam-II and BUST, there is an independent $\delta t$ in each of them. It is determined for each SN model and assumed source signal ($\bar\nu_e$, $\bar\nu_\mu$ or~$\bar\nu_\tau$) and shown in Table~\ref{tab:Neutrino-flux-properties}. We find $\delta t$ with a maximum-likelihood fit, however leaving out the late events in Kam-II and BUST, which may not be part of the PNS cooling signal. In any case, the shifts are in the range of 10--100~ms and the late-time events do not have any noticeable impact, as we have verified.

One corollary of this exercise is that the maximum likelihood never tries to pull event No.~0 at BUST into the SN signal range. As expected in the original BUST publication \cite{Alekseev:1987ej}, 
the time structure strongly suggests this event to be background.

\begin{figure*}
\includegraphics[width=\textwidth]{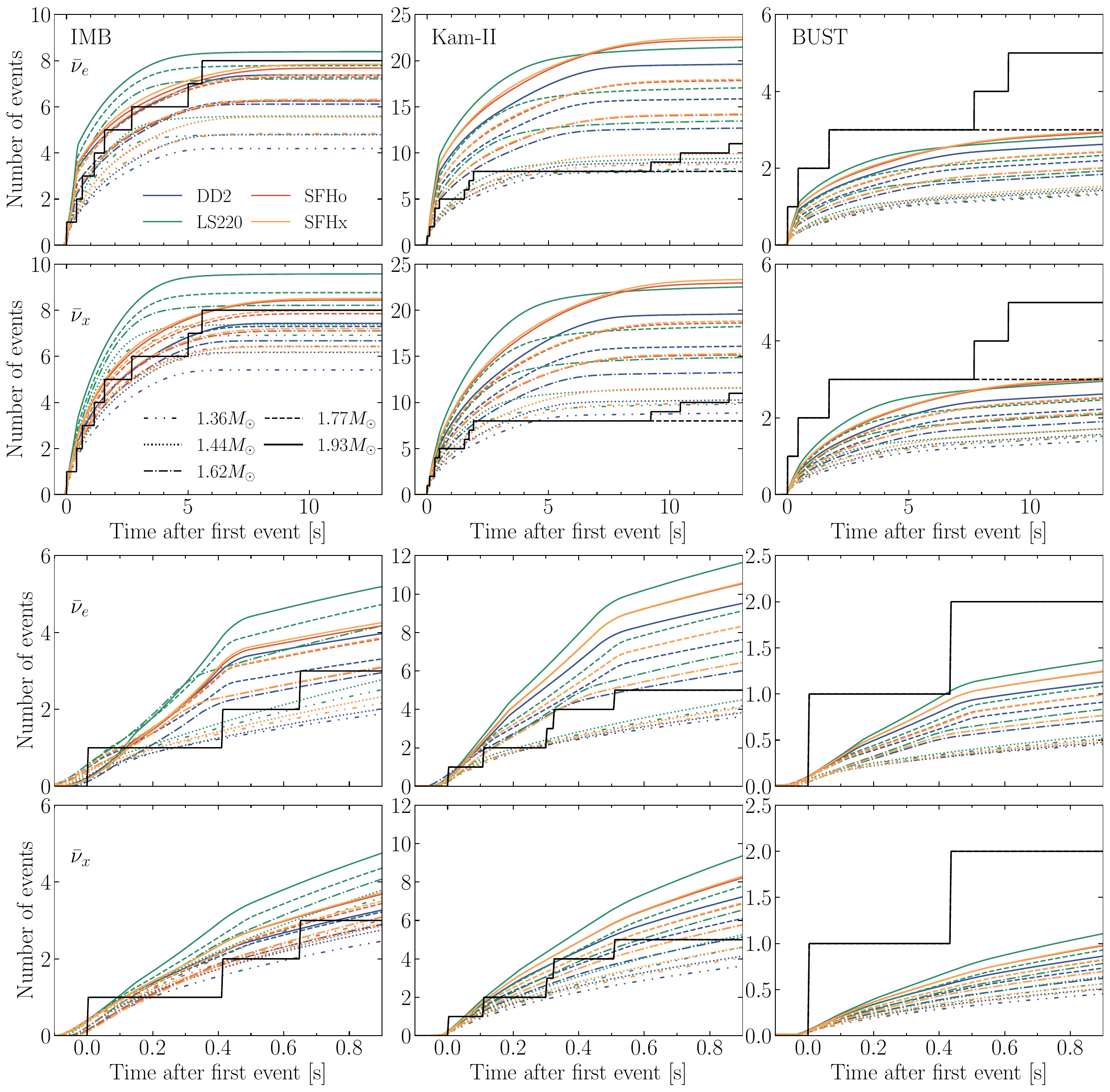}
\caption{Cumulative event distributions in IMB, Kam-II, and BUST together with the expectations from our models. Time is measured relative to event No.~1 in each experiment, whereas the SN bounce time is shifted by $\delta t$ according to the best-fit values provided in Table~\ref{tab:Neutrino-flux-properties} for each model. In BUST we include the background in the signal prediction, explaining the positive slope in the signal prediction even at late times. {\em Upper two rows}: Full signal duration, using the $\bar\nu_e$ or $\bar\nu_x$ flux in the rows as indicated (no or full flavor swap). In IMB, we use the best-fit trigger efficiency, explaining the smaller spread in total signal prediction. {\em Bottom two rows}: Same for the first second. }\label{fig:cumulative_distributions}
\end{figure*}

With these insights we show the cumulative event distribution for each detector in Fig.~\ref{fig:cumulative_distributions}, where $t=0$ is the time of event
No.1 in each detector, whereas the curves from the models are shifted by the best-fit offset times $\delta t$ provided in Table~\ref{tab:Neutrino-flux-properties}. IMB has no background and in Kam-II we show only the events attributed to the SN signal, leaving out No.~6 that is identified as background by an energy cut. In turn, for the model signal at Kam-II we only integrate for reconstructed energies larger than $7.5$~MeV, in the range in which non-background events were detected. One can easily see that the predicted cumulative distributions become practically flat at 5--7~s, reflecting the short signal duration discussed earlier.

For BUST, one cannot make such an energy cut and signal and background cannot be separated. In this case we include the expected background that builds up linearly in time, explaining the slope of the predictions at late times. Without this effect, the predictions are as flat as in Kam-II after 5--7~s.

For the case of IMB, the prediction itself is uncertain due to the significant uncertainty of its trigger efficiency. For each model, we use the trigger efficiency that maximizes agreement with the data, i.e., we maximize the likelihood over the two nuisance parameters $\xi$ and $\zeta$ discussed in Appendix~\ref{eq:Likelihood} around Eq.~\eqref{eq:IMB-nuisance}. This optimization explains the smaller spread in the IMB final event numbers relative to Kam-II, where the more massive SN models are disfavored based on the vast overprediction of total Kam-II event numbers.

In the upper two rows we show the signal for the full duration determined by the last Kam-II event. For the top row, it is assumed to be caused by the original $\bar\nu_e$ flux (no flavor conversion), in the second row by $\bar\nu_x$ (complete flavor swap). There is no big difference except that for a full flavor swap, the total event number (at the end time) is slightly larger.

\begin{figure*}
\includegraphics[width=\textwidth]{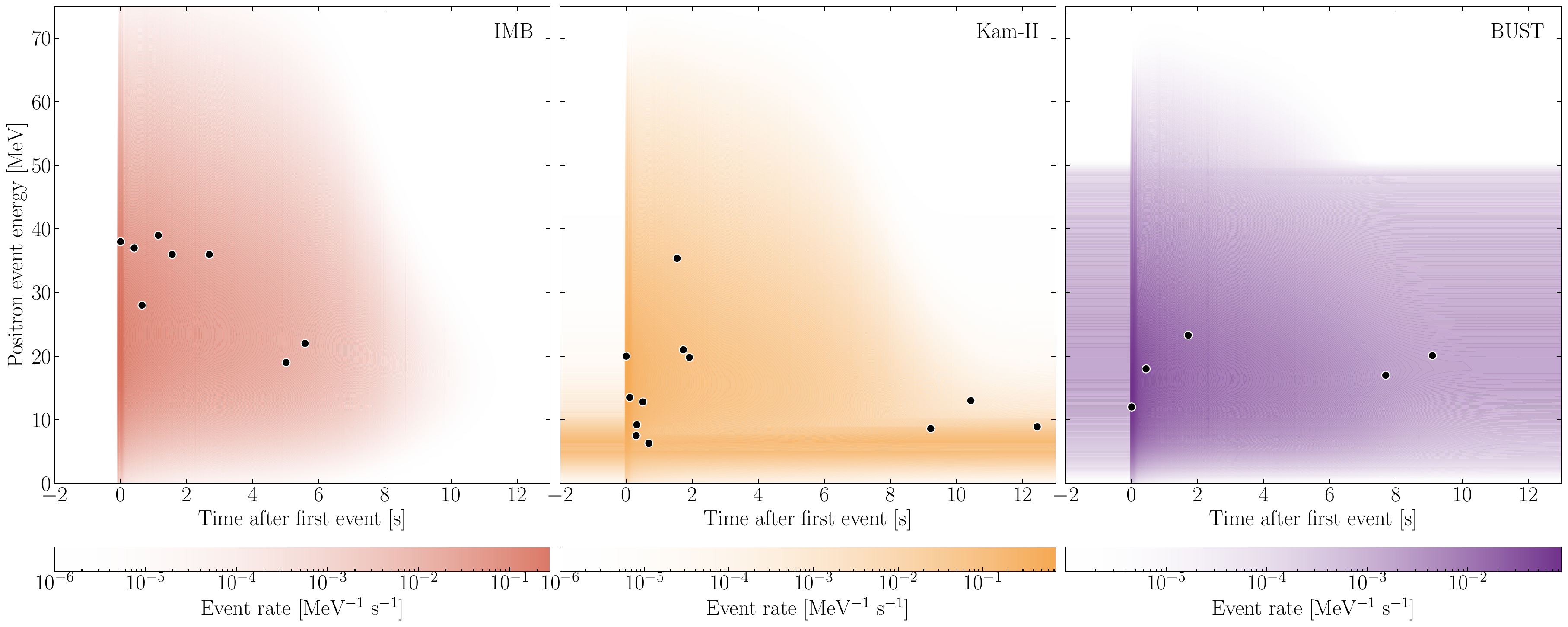}
\caption{Differential event distribution (signal and background) at each experiment, compared with the observations. Results are shown for model 1.44-SFHo without flavor swap; the offset time for each experiment is chosen as the best-fit value reported in Table~\ref{tab:Neutrino-flux-properties}.
}\label{fig:differential_distributions}
\end{figure*}

The effect of flavor swap is more pronounced in the early signal that we show, for the first second, in the two bottom rows. The conspicuous kink in the slope of the expectation corresponds to the end of the accretion period $t_{\rm acc}$ that is listed in Table~\ref{tab:GlobalProperties}. For the heavier models, $t_{\rm acc}\simeq0.54$--0.67~s, whereas for the 1.44\,M$_\odot$ models it is around 0.2~s post bounce. As expected from the behavior of the models illustrated in Fig.~\ref{fig:SN-Model-1}, after a full flavor swap, the event rate is smaller until the cross-over at $t_{\rm acc}$ and larger afterwards. For most models, the early signal would prefer a full flavor swap, except of course BUST which has seen a large upward fluctuation of events and therefore prefers the largest predicted signal.

As discussed earlier, the flavor dependence is a thorny issue that may not have the same answer for the full signal duration. Flavor dependence may differ in time and model-dependent ways from what is produced by our neutrino transport scheme under the assumption of no conversions, and flavor conversions may depend on these factors as well. Moreover, because of large-scale neutrino emission anisotropies connected to global asymmetries in the SN core such as the LESA and SASI (standing accretion shock instability) phenomena \cite{Tamborra+2014,Tamborra+2014a,Nagakura+2021a},
the early signal depends on the observer direction in a 3D world. Therefore, juxtaposing our no-swap and full-swap cases merely illustrates the extreme effects of flavor that can be extracted from our 1D models.

Generally we do not find the early $\bar\nu_e$ signal to be strongly over-predicted by the models, although averaged over models, a reduced early signal (by flavor swap) fits better in Kam-II in agreement with the findings of Ref.~\cite{Li:2023ulf}, in which, however, BUST was left out from the analysis, where the trend is the other way around. Moreover, it is not necessary that all models fit the data well, since SN~1987A must have had one specific but unknown final NS mass. 
Therefore, in view of the huge Poisson uncertainty of the measured event number during the first second, we do not find that there is a particular problem between models and data for the early signal, although, of course, the heavier-mass models strongly overpredict the overall Kam-II event rate.

\subsection{Late Events}
\label{sec:latevents}

The three late events in Kam-II beginning at 9.2~s after their first event
and the two late ones in BUST, beginning at 7.7~s after their No.~1, are typically not consistent with our models because of our short cooling times and concomitant predicted signal durations. One way to illustrate this point is to show the predicted detection rate, differential with regard to detected energy and time, overlaid with the actual data. For the case of model 1.44-SFHo we show such a plot in Fig.~\ref{fig:differential_distributions}, where detector backgrounds are included. 

In BUST, the entire signal is a strong upward fluctuation in that over 13~s,  as defined by the overall Kam-II duration, only 0.44 background events are expected. Assuming No.~0 in BUST to be background and No.~1 to indicate the beginning of the SN signal, and if the first 3 events are attributed to the SN, the Poisson probability for two additional background events anytime in 13~s is 6.2\%, and for two to appear in the period 7--13~s is 1.7\%. This probability is not extremely small and so it is not entirely implausible that these two events could be attributed to background.

\begin{figure*}
    \hbox{\includegraphics[width=0.99\textwidth]{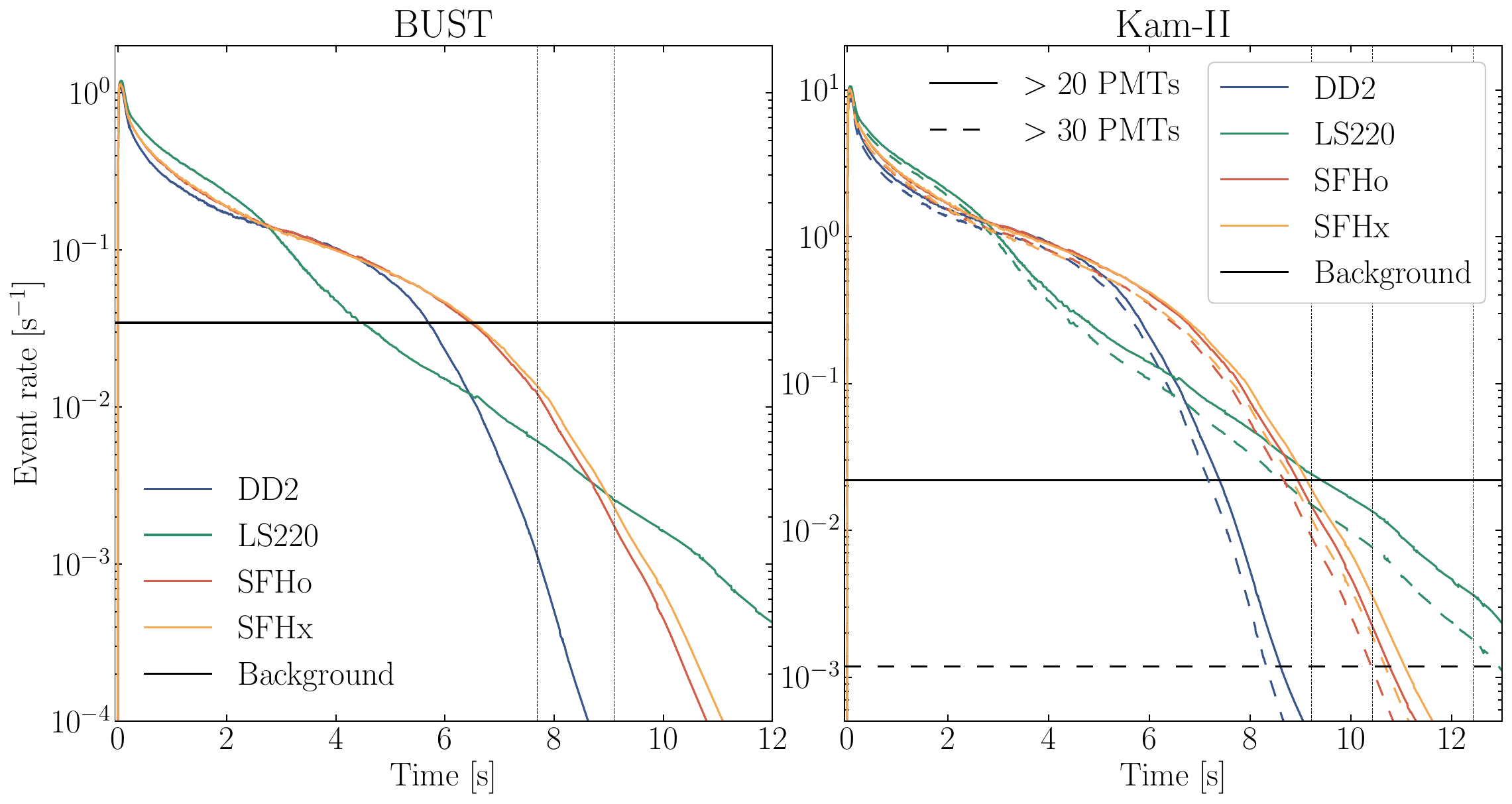}}
    \caption{Expected late signal for our 1.44\,M$_\odot$ models compared with background. We have assumed full flavor swap (signal caused by $\bar\nu_x$) to maximize the predicted signal.
    {\em Left:} BUST. The vertical lines indicate the timing of the two late events. {\em Right:} Kam-II. We show the energy-integrated signal with $E_{\bar\nu_\mu}^{\rm det}\geq7.6$~MeV, corresponding to $N_{\rm hit}\geq20$, as solid lines and \smash{$E_{\bar\nu_\mu}^{\rm det}\ge11.4$~MeV}, corresponding to $N_{\rm hit}\geq30$, as dashed lines. The vertical lines indicate the timing of the three late events.
    }
    \label{fig:late}
\end{figure*}

Could these late events also come from an upward fluctuation of the signal rather than of the background? To answer this question we show in Fig.~\ref{fig:late} the expected rate at late times for a baryonic NS mass of 1.44\,M$_\odot$ compared with the background rate. The expected signal, integrated after 7~s for the shown EoS cases, is $2.24\times 10^{-3}$ (DD2), $1.51\times 10^{-2}$ (LS220), $2.06\times 10^{-2}$ (SFHo), and $2.38\times 10^{-2}$ (SFHx). We have assumed complete flavor swap to maximize the predicted signal at late times. For these models, clearly it is much more likely that the late events are an upward background fluctuation than an upward signal fluctuation. These numbers confirm the visual impression of Fig.~\ref{fig:differential_distributions}.

The question of a background attribution of the three late Kam-II events is more difficult to address because the dominant background from radioactive decays strongly peaks at 6~MeV and can be removed by an energy cut. For the remaining background, with energies above 7.5~MeV or so, no good spectral information is available. Moreover, such events tend to be clustered near the surface of the detector because they tend to be caused by radioactivity in the surrounding rock or the PMTs themselves. Therefore, the vertex location in the detector provides some important clues.

The Kam-II collaboration has provided the vertex coordinates and directional cosines of all events \cite{Hirata:1988ad}. Event No.~10 has $(x^2+y^2)^{1/2}=7.18~{\rm m}$ and $z=7.01$~m, where the cylindrical detector has the PMT surface at radius 7.2~m and overall height of 13.1~m. The coordinate $z=0$ is offset by 1~m in the positive (upward) vertical direction, so the upper PMT plane is at $z=7.05$~m. In other words, event No.~10 is both close to the cylindrical walls and the top layer in the detector corner, where most of the external background shows up. From the directional cosines reported in Table~I of Ref.~\cite{Hirata:1988ad},
we gather that the electron or gamma was indeed pointing inward relative to the cylindrical surface, but toward the top layer, so a Cherenkov ring could not have been seen. Actually, we believe that this is a typographical error in their Table~I and that $\cos\gamma$ of this event should be reversed, making it point inside the detector so that the Cherenkov ring would have been visible on the opposite walls. Indeed, one needs to reverse this sign to obtain the scattering angle reported in their Table~II. We conclude that this event came from the outside direction and judging by its vertex position, it is consistent with an external background entering the detector.

Event No.~11 has $(x^2+y^2)^{1/2}=4.24~{\rm m}$ and $z=-1.73$~m, so it is far from the walls, and thus much less likely to come from the outside, yet has relatively high energy of $13.0\pm2.6$~MeV.

Finally, event No.~12 has $(x^2+y^2)^{1/2}=4.29~{\rm m}$ and $z=-0.97$~m and thus is also far from the walls, but with its relatively low energy of $8.9\pm1.9$~MeV more plausibly could come from low-energy radioactive background.

The event energies in Kam-II are primarily derived from the number $N_{\rm hit}$ of PMTs hit by these events. We have the information that the background around the time of the SN was 21.9~mHz for $N_{\rm hit}\geq20$ and 1.18~mHz for $N_{\rm hit}\geq37$ (see Sec.~\ref{sec:Kam}). We identify $N_{\rm hit}=20$ with the reconstructed energy of 7.6~MeV and $N_{\rm hit}=30$ with 11.4~MeV. If we attribute event No.~6 to background because of its low energy, and the other of the early nine events to the SN, the probability for {\em three or more} additional events with $N_{\rm hit}\ge20$ during the full 13~s period is 0.3\%. To find three such background events in the 7--13~s period is 0.03\%. To find {\em one or more} background events with $N_{\rm hit}\geq30$ in the 13~s period is 1.5\%, and to get one or more in the 7--13~s interval is 0.7\%. 

In the right panel of Fig.~\ref{fig:late} we display the Kam-II signal for the two conditions $N_{\rm hit}\geq20$ or 30 compared with the cited background information. For the models shown, the background rate with $N_{\rm hit}\ge20$ is larger than the SN signal, so from this perspective, these events are more likely background than SN signal. On the other hand No.~11, which has $N_{\rm hit}=37$, is more likely an upward signal fluctuation for all EoS other than DD2. 

In summary, while it is not impossible to attribute the late events in Kam-II and BUST to a combination of upward background and signal fluctuations, these events do not fit well to our signal predictions, also in view of the short cooling timescales discussed earlier. Therefore, in Sec.~\ref{sec:speculations} we will speculate about other options.

\onecolumngrid

\begin{figure*}
\includegraphics[width=\textwidth]{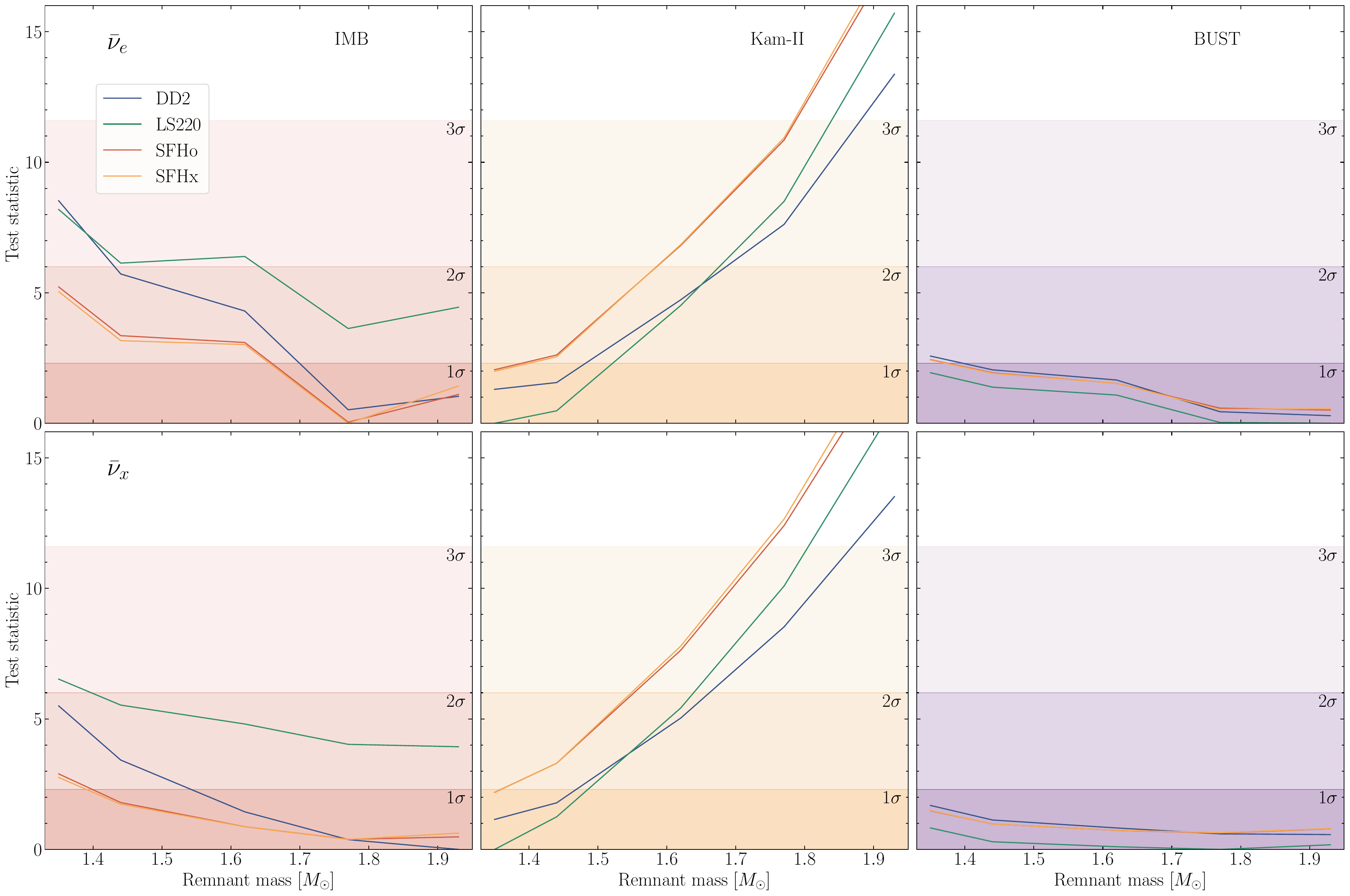}
\caption{Test statistic $\Lambda$ for the different models, separately for each experiment, i.e., the log-likelihood relative to the maximum as defined in Eq.~\eqref{eq:TS_definition}. Late-time events are excluded. We show in gradual shades of colors the 68.3\%, 95\%, and 99.7\% confidence levels. Signal caused by $\bar\nu_e$ ({\em top row}) or $\bar\nu_x$ ({\em bottom row}). One can only compare the models within one panel relative to each other, not between panels, because $\Lambda$ is normalized to zero in each panel for the best case. }\label{fig:likelihood_exps}
\vskip30pt
\includegraphics[width=0.7\textwidth]{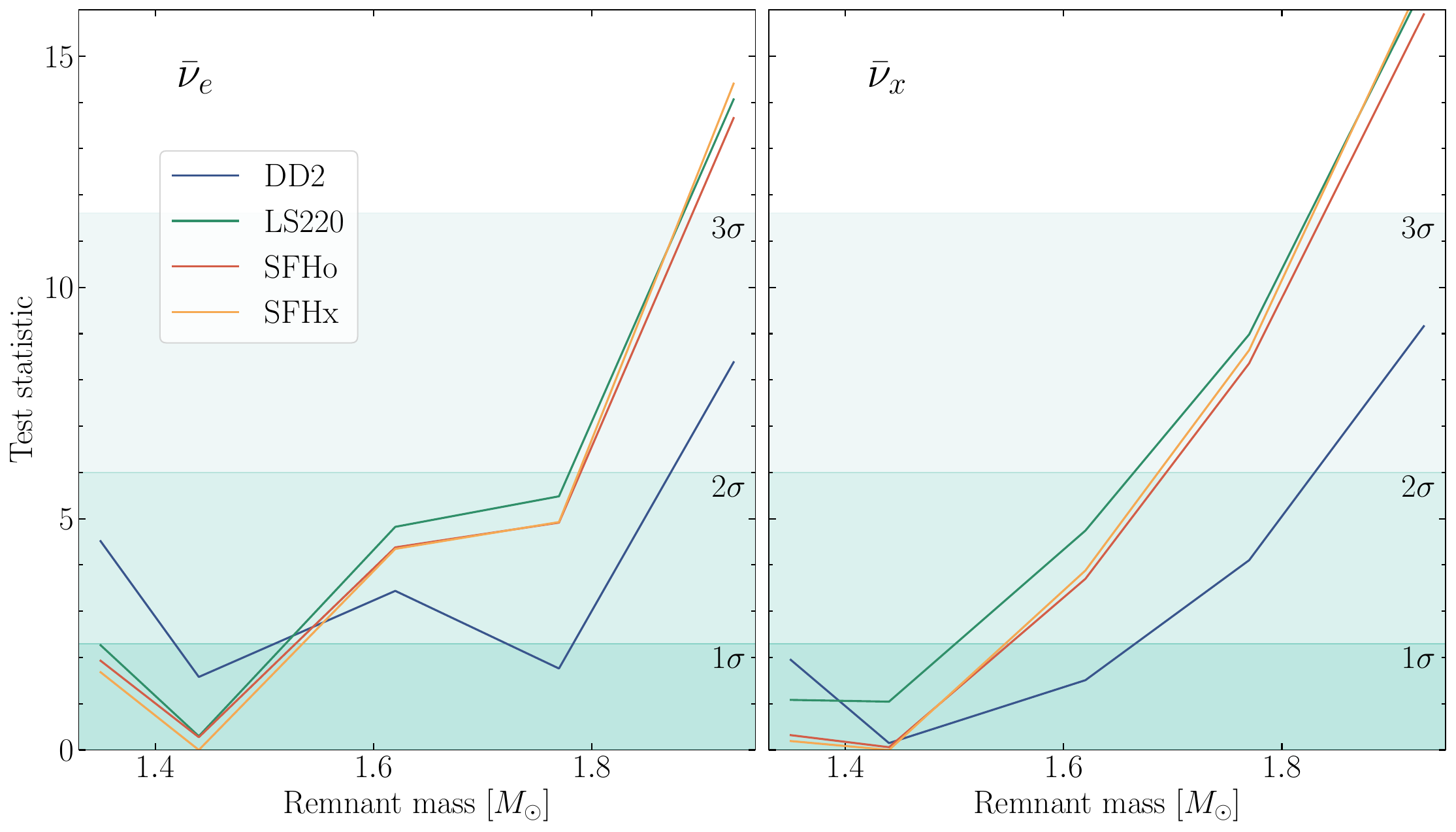}
\caption{Same as Fig.~\ref{fig:likelihood_exps} for all experiments combined, i.e., common likelihood as a product of the individual ones and also including the nonobservation at LSD. Results are shown both for the original $\bar\nu_e$ spectrum and for complete flavor swap.
}\label{fig:likelihood_plots}
\end{figure*}

\section{Overall Model Comparison}
\label{sec:Overall-Comparison}
\twocolumngrid

As a final step, we compare the likelihood of our models with the SN~1987A data. Backgrounds are now always included and on the basis of energy and time, the likelihood interpolates between the cases that a given event is signal or background. As an indicator of the relative goodness of fit among different models we use the test statistic (TS) $\Lambda$ defined in Eq.~\eqref{eq:TS_definition}, i.e., the log-likelihood difference.

We do not think that attempting to discriminate between our models on the basis of the late events provides meaningful information because within our model space, the late events are a rare signal and/or background fluctuation. One would favor a model that achieves the best compromise between the least underprediction of the late events and smallest overprediction of the total event number, without a believable result on either end. Therefore, in this analysis we assume that the late events in Kam-II and BUST have a different explanation than the cooling signal of our convective PNS models.

Our models are defined by their EoS and the final NS mass. We can give a rough indication of the level by which a model is preferred over another by assuming that the TS follows a chi-squared distribution with two degrees of freedom, by qualitatively considering the mass and the EoS as if they were two parameters. Of course we do not expect this to be strictly valid, since the EoS is not really a continuous parameter, and furthermore because the sparse data do not allow us to really expect a chi-squared distribution for the TS -- as would be the case in the asymptotic limit of many events -- but this simplification allows us to show indicative values for the confidence levels with which the models are preferred relative to each other.

For each experiment we show in each panel of Fig.~\ref{fig:likelihood_exps} the TS for the different models as a function of NS mass and EoS given by the line colors as indicated. In the top row, the signal is caused by the original $\bar\nu_e$ (no flavor swap) and in the second row for $\bar\nu_x$ (full swap).
With this method, we can only compare models within a given panel, not between panels, because the TS is normalized in each panel to the best case, allowing only for a relative comparison.

In the left column we see that IMB alone would favor larger-mass models to explain the relatively large number of events. We have already included the uncertain trigger efficiency as a nuisance parameter, so already the best compromise between trigger efficiency and event number is achieved. On the other hand, BUST has practically no discriminatory power. If we attribute only the first three events to the SN, the total event number is not strongly overpredicted relative to the large Poisson uncertainty.

The model discrimination is most pronounced for Kam-II with little difference between flavor-swap scenarios. The considerably overpredicted event numbers in the higher-mass models strongly favor the lower-mass ones for all cases of the EoS. 

This effect is so strong that it dominates also in a combined analysis of all experiments shown in Fig.~\ref{fig:likelihood_plots}, where the global likelihood is a product of the individual ones. The smaller-mass models are significantly favored. The only exception is DD2, where the 1.44 and 1.77\,M$_\odot$ models are equivalent for the $\bar\nu_e$ case, but the lower mass is strongly favored for $\bar\nu_x$. This ``anomaly'' appears to be driven by 1.44-DD2-$\bar\nu_e$ being a particularly poor fit to IMB. 
Broadly, the 1.44\,M$_\odot$ models are somewhat favored, and the 1.93\,M$_\odot$ ones are clearly disfavored, but without a better grasp of flavor conversion physics, one could not easily dismiss any of the models with masses of 1.77\,M$_\odot$ or below.

\onecolumngrid

\section{Speculations about Late Events}
\label{sec:speculations}
\vskip4pt
\twocolumngrid

The late three Kam-II events as well as the last two events in BUST are hard to explain by any of our models that fit the IMB signal and that account for the rise of the cumulative event numbers in Kam-II and IMB during the first second. Convection and nucleon correlations in the neutrino interactions included, the PNS cooling becomes so short that the predicted event rates fall below $\sim$0.5\,s$^{-1}$ in Kam-II at $t_\mathrm{pb}\sim 6$\,s and are orders of magnitude below the detected event rates at $t_\mathrm{pb}\gtrsim 9$\,s (Fig.~\ref{fig:SN-Model-1}). Another feature of the long Kam-II and BUST signals are the conspicuous gaps of 7.3\,s in Kam-II and 6.0\,s in BUST preceding the last events. Their low statistical probability ($\lesssim\,$2\% in Kam-II \cite{Suzuki+1988}) inspired
early speculations that the signal consists of several separate bursts and the last one might be connected to a delayed nuclear phase transition in the new-born NS~\cite{Takatsuka1987, Takahara+1988}. Here we will briefly discuss such a possibility as well as other physical effects that could extend the measurable neutrino emission from the core of SN~1987A beyond 10\,s. 

\begin{figure*}[ht]
\includegraphics[width=\textwidth]{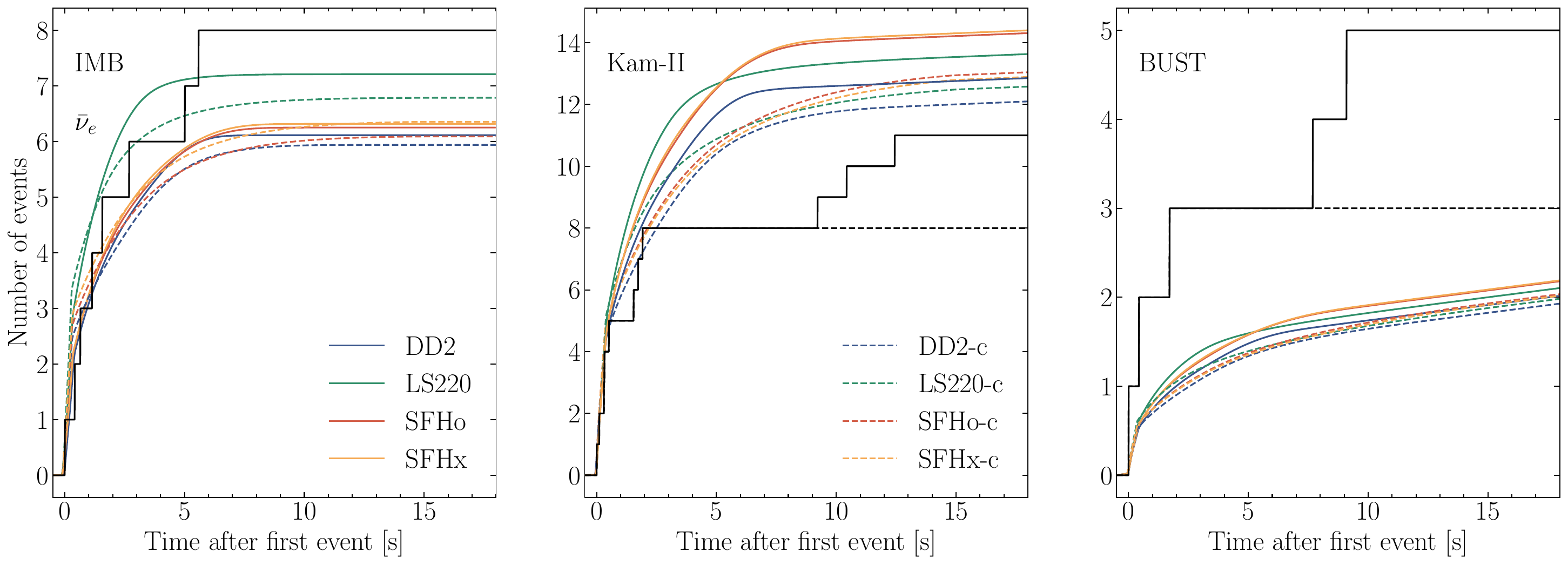}
\caption{Cumulative $\bar\nu_e$ event distributions at all experiments. The predictions for our 1.62\,M$_\odot$ models are shown with (solid) and without (dashed) PNS convection, the experimental results of Kam-II and BUST with the late events included (solid) and excluded (dashed). For IMB, we choose for each model both the efficiency parameters and the time offset of the first event with respect to bounce that best fit the data. For Kam-II, we consider events integrated above $7.5$~MeV to minimize the impact of the background; nevertheless, for a fair comparison, we do include the background integrated above this threshold energy. For BUST, we include the background spectrum without any threshold, explaining the increase of event number at late times. 
}\label{fig:convection_cumulative}
\end{figure*}

\subsection{Reduced PNS convection}
\label{sec:reducedconvection}

PNS convection is the primary cause for reducing the cooling times in our simulations to below $\sim$6\,s for the main period of neutrino emission with detection rates of at least several 0.1\,Hz (Figs.~\ref{fig:SN-Model-1} and~\ref{fig:late}). Even for the LS220 EoS, which facilitates damping of convective activity after a few seconds (due to the steeper density dependence of the nuclear symmetry energy) and thus a slower decay of late neutrino emission, the neutrino luminosities and mean energies decline to a low level with detection rates below 0.1\,Hz after only $\sim$5\,s (Figs.~\ref{fig:SN-Model-3}, \ref{fig:late}, and Table~\ref{tab:Neutrino-flux-properties}). Even in our simulations without convection, the luminosities and mean neutrino energies reach the low values that correspond to detection rates of $<$0.1\,Hz well before 10\,s after bounce (Fig.~\ref{fig:SN-Model-noconv}), mainly because nucleon-nucleon correlations in the neutrino opacities alone shorten the cooling times considerably \cite{Huedepohl+2010, Roberts+2017Handbook, Pascal+2022}. Therefore, it is not unexpected that switching off convection by itself cannot solve the problem of the long observed signal durations.

This expectation is confirmed by Fig.~\ref{fig:convection_cumulative} and Table~\ref{tab:Neutrino-flux-properties} and, in particular, Table~\ref{tab:with-without-convection}, where we extract some of the numbers from the long Table~\ref{tab:Neutrino-flux-properties} in order to directly compare convective and non-convective models for our 1.62\,M$_\odot$ simulations with different EoSs. Of course, as discussed before, all of the 1.62\,M$_\odot$ models overpredict the total number of events in Kam-II and underpredict those in BUST. The middle panel of Fig.~\ref{fig:convection_cumulative} also shows that even in this extreme case the detection of three events after 8\,s by Kam-II cannot be accounted for, but almost all of the extra events are expected in the time interval between 2\,s and 9\,s, where Kam-II had its long detection gap; see also the numbers for the detection periods $\tau$ in Table~\ref{tab:with-without-convection}. In contrast, for IMB both the convective and non-convective models are able to provide a reasonable match with the data and the effects of PNS convection have less impact.

While our PNS cooling times of 5--8\,s are fully compatible with other recent 1D calculations that include PNS convection via a mixing-length treatment~\cite{Roberts+2012a, Roberts+2017Handbook, Pascal+2022}, one needs to keep in mind that this description is an approximation of the multi-dimensional hydrodynamic transport, which involves free parameters~\cite{Mirizzi+2016} and ultimately requires validation by 3D simulations of the long-time evolution of convective PNSs with different masses and nuclear EoS cases (for a first step, see Ref.~\cite{Nagakura+2020}). It is, however, unlikely that 3D models will yield much longer time scales for the internal PNS cooling, because over- and undershooting at the base and top of the convective layer in the PNS are not accounted for by the 1D mixing-length approximation but could increase the efficiency of convective transport in 3D. 

\subsection{Fallback}

In 1D models, accretion ends as soon as the explosion sets in and the shock begins to expand. A high-entropy ``bubble'' develops between the neutrino-cooling PNS on the one side and a thick, dense shell of stellar material moving in the wake of the outgoing shock on the other side. This bubble is filled by a low-density, spherical wind of baryonic matter that is blown off the PNS surface by energy transfer through neutrinos (e.g., Refs.~\cite{Duncan+1986, Witti+1994, Qian+1996, Otsuki+2000, Thompson+2001}). 

Recent 3D simulations suggest, however, that the volume between PNS and shock is filled by downflows of postshock matter that falls inward to the close vicinity of the PNS, where it absorbs energy from neutrinos and from where it expands outward again, adding considerable amounts of energy to the explosion of the SN~\cite{Mueller+2017, Bollig+2021, Wang+2023}. This generically multi-dimensional hydrodynamic phenomenon of long-lasting simultaneous inflow and outflow activity can be reliably modeled only in 3D, because the assumed axisymmetry in 2D simulations imposes artificial constraints on the flow geometry and dynamics~\cite{Mueller2015}. In 3D simulations of successful SN explosions, the accretion luminosity from the downflows is found to contribute significantly to the neutrino emission only for post-bounce periods of less than a second~\cite{Mueller+2017, Bollig+2021, Vartanyan+2023}. Therefore we have confidence that our 1D models can capture the essential effects of additional neutrino emission by post-bounce PNS accretion sufficiently well for comparisons with the SN~1987A neutrinos, even if the time profile of this early signal phase differs from the 3D predictions. 

However, it is a viable possibility that the late neutrino emission beyond 5\,s after bounce, when the luminosity from the PNS interior declines to low values, is affected by an accretion component associated with the sustained downflows of matter towards the PNS. When the neutrino energy deposition in the vicinity of the PNS declines and thus the driving force of outflows gradually abates, the downflows will gradually transition to fallback of matter that is initially swept out by the SN shock but ultimately cannot escape the gravitational pull of the new-born NS. This fallback material can end up on the NS surface and release its gravitational binding energy by radiating neutrinos.    

The 7.3\,s and 6\,s gaps in the Kam-II and BUST signals and the subsequent detection of a larger bunch of late events with relatively low neutrino energies in both experiments suggest a more quiescent period, followed by an eruptive episode of neutrino emission rather than a continuous accretion luminosity. The gaps could be plausibly explained by a pause or relatively inactive phase of the neutrino production, whose beginning is marked by the last IMB event, at latest. 

Figure~\ref{fig:pinched_only_late} implies that an energy of a few $10^{51}$\,erg emitted in $\bar\nu_e$ could be sufficient to account for the late events in Kam-II and BUST. If the energy requirement is as low as this value, it corresponds to an accreted mass of only some $10^{-2}$\,M$_\odot$ for a PNS with a mass of 1.5\,M$_\odot$ and a radius of about 12\,km (Eq.~(8) in~\cite{Janka1996}), provided the gravitational binding energy of this mass is radiated equally in $\nu_e$ and $\bar\nu_e$ because the $\nu_x$ emission can be assumed to be of minor importance. Such fallback masses, accreted with rates around $\sim$\,$10^{-2}$\,M$_\odot$\,s$^{-1}$ over several seconds, do not appear to be implausible~\cite{Janka1996,Ertl+2016,Janka+2022}. Since the fallback matter is likely to carry high angular momentum~\cite{Bollig+2021,Janka+2022}, it might form a centrifugally supported accretion torus or accretion belt around the effectively cold NS. The neutrino release from such a low-density accretion belt will be dominated by $\nu_e$ and $\bar\nu_e$, in contrast to the production of significant additional $\nu_x$ emission at the high-density conditions for radial accretion of spherically symmetric flows~\cite{Akaho:2023alv}. Moreover, because the thermodynamic conditions of the emission region envisioned here could be similar to those in accretion tori around black holes \cite{Ruffert+1999,Setiawan+2006,Blum+2016,Just+2022}, the escaping neutrinos will be cooler than estimated for the spherical situation. 

For these reasons, accretion of fallback matter with dynamically relevant angular momentum seems to be an interesting scenario to explain the latest events measured by Kam-II and BUST, if energies of the $\bar\nu_e$ burst in the lower part of the banana-shaped volumes in Fig.~\ref{fig:pinched_only_late} are considered. The opposite conclusion, however, applies if the energy requirement is set by the most likely values in these confidence regions, because these energies are comparable to the energy carried by the previous PNS cooling signal~\cite{Janka1996}. Fallback accretion must be expected to be a highly variable phenomenon, which depends sensitively on the explosion properties and the core structure of the progenitor star, and which deserves closer investigation by 3D simulations.

\subsection{Nuclear Phase Transition}

Phase transitions (for example between hadrons and quarks in QCD or in the form of pion or kaon condensation) in the high-density nuclear matter of the hot PNS were discussed as a possible mechanism to trigger and strengthen CCSN explosions even before SN~1987A \hbox{\cite{Migdal+1979, Kaempfer1985, Takahara+1985, Takahara+1986}}. They garnered attention also in later years for the same  reason~\cite{Takahara+1988a, Gentile+1993}, but also to explain the 7\,s gap and following three neutrino events in the \hbox{Kam-II} measurement~\cite{Takatsuka1987, Takahara+1988} and as a path to black-hole formation~\cite{Brown+1994}. More recently the topic has received new attention not only as an alternative mechanism to drive CCSN explosions~\cite{Sagert+2009, Fischer+2018}, but also as a source of characteristic multi-messenger signals (neutrinos, gravitational waves, trans-iron elements; e.g., Refs.~\cite{Zha+2021, Kuroda+2022, Jakobus+2022}). 

The exact physical conditions under which the phase transition takes place, the PNS evolution time until it happens, and the associated energy release depend strongly on its physical nature (first or second order; \cite{Takahara+1985, Takahara+1986}). Also the detailed assumptions about the nuclear medium play a crucial role~\cite{Fischer+2018, Zha+2021, Jakobus+2022}. PNS cooling times of several seconds prior to the phase change in the EoS seem to be possible. 

Because of PNS reheating by the energy released in its internal reconfiguration (contraction and/or release of latent heat) this phenomenon was proposed as a cause of a late neutrino burst after a short initial phase of PNS cooling and the 7\,s Kam-II gap~\cite{Takatsuka1987, Takahara+1988}. The total energy as well as the particle energies of the emitted neutrinos were estimated to be compatible with those of the last three Kam-II events. Detailed simulations with modern SN codes and physics for sufficiently long evolution periods, using progenitor and phase-transition models adjusted to the questions of SN~1987A, are, however, missing. Therefore a nuclear phase-transition scenario for the Kam-II and BUST gaps and subsequent events remains only speculative for the time being.  

\onecolumngrid

\section{Discussion and Outlook}

\label{sec:Discussion}
\twocolumngrid

We have taken a fresh look at the old question of interpreting the unique SN~1987A neutrino observations in terms of modern numerical models. From the outset it is clear that the sparse data do not provide very detailed information in time and energy space, for which reason one can limit the discussion to broad and overall signal features. For example, multi-dimensional, large-scale hydrodynamic phenomena such as the SASI and LESA modes can cause fast, high-amplitude time variations and can make the signal depend on the observer direction, all of which are unlikely to be diagnosed from the sparse SN~1987A data. 

Since such effects necessarily require 3D simulations, but leaving them out is no major shortcoming for our project goals, we have decided against considering a necessarily limited set of 2D or 3D simulations that cover only constrained post-explosion evolution periods. Instead, we opted for a more systematic approach by constructing a larger suite of spherically symmetric models in the parameter space of four different cases of the nuclear EoS and five choices of the final baryonic NS mass in the range of 1.36--1.93\,M$_\odot$. The NS masses are fixed by using different progenitors and by adjusting the end time $t_{\rm acc}$ of the post-bounce accretion phase, when the explosion is manually triggered. Our 1D models include lepton-number and energy transport by PNS convection via a mixing-length approximation, and they are computed until the end of the deleptonization and neutrino cooling of the PNS. 

Naturally, the detailed time evolution of the neutrino emission properties during the PNS accretion phase before and after the onset of the SN explosion is different in 3D compared to 1D models, because shock expansion or contraction phases influence the accretion emission by the PNS. In 3D compared to 1D, the accretion flow to the PNS and the corresponding accretion luminosity can be reduced by shock expansion before the explosion begins. But PNS accretion and associated neutrino emission continue for a few 100\,ms still after the shock expansion has begun. This is in contrast to the 1D situation, because by construction a 1D treatment excludes long-time accretion downflows and simultaneous neutrino-heated outflows of matter as well as the subsequent fallback of transiently ejected gas seen in multi-D simulations. Therefore we primarily focus on the signal connected to the long-time neutrino cooling of the PNS. However, despite the differences in the time structure around the onset of the explosion, one can expect that, overall, the integrated neutrino signal from post-bounce accretion, lasting typically $\sim$0.3--1\,s in 3D, is sufficiently well represented by 1D results that lead to the same final NS mass. For this reason we consider our 1D models also qualified for a comparison with the sparse SN~1987A neutrino data even in the first second of the measured signal.

Neutrino transport is implemented using the latest microphysics for six species and including muons in the EoS, but ignoring flavor conversion. Actually, the difference between the predicted $\bar\nu_\mu$ and $\bar\nu_\tau$ flux spectra is small so that in the analysis we usually consider either $\bar\nu_e$ or $\bar\nu_x$, an arithmetic average of $\bar\nu_\mu$ and $\bar\nu_\tau$. For detection, we only use inverse beta decay, ignoring the small contribution from elastic scattering on electrons. Our results depend only weakly on the events being caused by the primary $\bar\nu_e$ or $\bar\nu_x$ flux (after assumed complete swap) because the sparse data are not informative about the small differences. The true role of flavor evolution in SN physics is not fully understood, especially the impact of fast-flavor conversion deep inside the SN core.
While these effects could be quite relevant for the neutrino-driven explosion physics or neutrino-induced nucleosynthesis, their impact is relatively minor for the SN~1987A data interpretation.

Concerning the measurements, we use the data from all four relevant detectors IMB, Kam-II, BUST, and LSD, for the latter the non-observation during the detection period of the others. For the first time we have systematically included the reported uncertainties of the IMB trigger efficiencies, causing a rate uncertainty roughly on a $\pm50\%$ level. We have documented a large amount of scattered information about the data in Appendix~A. The largest detector (IMB) and the smallest (LSD) are essentially background free, Kam-II has a noticable low-energy background that, however, is energetically well separated from the expected signal, whereas the BUST background spectrum peaks in the same energy region as the signal. While we systematically include all detectors and take account of their backgrounds, the main information derives from IMB and Kam-II. Leaving out the small detectors BUST and LSD would not dramatically change our findings.

Within the perimeters of our model space, a number of interesting conclusions follow. Ignoring the time structure of the observations, the time-integrated event numbers and average neutrino energies broadly agree with expectations as illustrated, in particular, by Fig.~\ref{fig:overlap_simulations} for a common analysis of all experiments, although the intermediate-mass models are somewhat favored. A common analysis, however, somewhat hides the opposing trends particularly implied by the IMB data alone and the Kam-II data alone. The former prefers our higher-mass models, primarily caused by IMB's large detected energies. Such models, on the other hand, strongly overpredict the total Kam-II event rate. IMB itself, with its uncertain trigger rate, is much more adaptable to the overall flux. In the common analysis, Kam-II wins so that overall the 1.44\,M$_\odot$ models are favored as a compromise.

Concerning the time structure, within our model class, we do not find that the event number during the first second is strongly overpredicted, somewhat contradicting recent findings based on a heterogeneous selection of models taken from the literature~\cite{Li:2023ulf}. It is however true that complete flavor swap $\bar\nu_e\leftrightarrow\bar\nu_x$ during the first second makes the bulk of our models agree better with the cumulative distributions shown in the bottom rows of Fig.~\ref{fig:cumulative_distributions}, especially for Kam-II. This could be the same effect observed in Ref.~\cite{Li:2023ulf}. However, SN~1987A corresponds at most to one of our models, not an average, and also for the $\bar\nu_e$ emission we find models that match both the detected signals by Kam-II and IMB, for example our 1.62-DD2 case. Moreover, some of our low-mass models even undershoot the number of detected neutrino events in the first second for the original $\bar\nu_e$ spectra, and more matches could be expected if we had cooling simulations of PNSs with 1.50--1.55\,M$_\odot$ in our model set. 

It is true, however, that the average event energies in Kam-II during the first second are considerably smaller than predicted by any model~\cite{Li:2023ulf}. This effect is driven by the low energies of two of the five Kam-II events that were recorded within the first 0.5\,s. We look at this question in more detail in Appendix~\ref{app:first_second} and, using the approach of Ref.~\cite{Li:2023ulf} (with some improvements), we also find low p-values, but never as small as theirs. For many of our models, the effect is on the order of $2\sigma$, for the worst cases not even $3\sigma$. Most important in our opinion, however, is the fact that this effect is restricted to the first second and disappears for a broader time window. We interpret the low energies of events No.~3 and 4 in Kam-II as a local signal fluctuation or upward background fluctuation. Such local fluctuations should not be over-interpreted, keeping in mind the look-elsewhere effect. An anomaly in a chosen subset of data can look locally much more significant than it is globally. Therefore, based on the first second of data, there is no compelling indication that SN~1987A is outside of our model space.

This picture radically changes when we consider the full time structure. Our short cooling periods do not easily account for the late events in Kam-II and BUST, whereas there is no problem with the 5.6\,s burst duration at IMB. On the other hand, the triplet of late Kam-II events, beginning at 9.2~s after their first, and the doublet of BUST events after 7.7~s would each require an extreme signal fluctuation. In contrast to IMB, these detectors had non-negligible background rates, but the late events would also require an unlikely background fluctuation. As has been commented many times, besides the Kam-II and BUST signals lasting for a very long time, both had large gaps between their initial main burst and the late ones, giving the impression of two bursts.

The SN~1987A data are sparse and different experiments show various tensions with each other and unlikely features, such as the measured burst in LSD around five hours before the other experiments, the angle distribution in IMB, the large number of events in BUST, and the long time gaps in both the Kam-II and BUST data. Of course, any sparse data could show many anomalies, depending on what one would perceive as an ``anomaly'' without an objective definition. In this sense, this effect perhaps should not be over-interpreted.

However, if one attributes the late events to SN~1987A and not to background, one may first think that our mixing-length treatment of PNS convection is mainly responsible, but unphysically switching it off does not strongly improve the situation because the relatively small neutrino opacities, caused by standard nucleon-nucleon correlations, still engender short cooling times. Many SN models in the literature do not include PNS convection, nucleon correlation, or neither. While these are standard effects, it would be useful to dig deeper, and especially a more systematic evaluation of PNS convection in 3D studies is strongly mandated.

Searching for explanations beyond PNS cooling, but within standard physics, fallback of matter after the end of PNS neutrino cooling is one possibility. While we are confident that our 1D simulations account for the main PNS cooling emission, late and potentially episodic accretion flows or fallback after a more quiescent phase are distinct options outside of our original model space. A systematic quantitative 3D exploration is a compelling task for the future.

More exotic explanations could include late phase transitions in the nuclear medium. The release of gravitational energy by PNS contraction and/or of latent heat could power a late neutrino burst, but reliable quantitative interpretations of the SN~1987A data are missing and should be pursued.

Considering yet more exotic speculations, we mention secret neutrino-neutrino interactions \cite{Berryman:2022hds} that could cause neutrinos to behave as a relativistic fluid instead of an ideal fermion gas \cite{Dicus:1988jh}. Such discussions have been recently revived \cite{Chang:2022aas} and show that, despite the potentially dramatic modification of the neutrino behavior, the observable burst properties remain quite similar to the standard case and are regulated by PNS cooling \cite{Fiorillo:2023ytr, Fiorillo:2023cas}.
However, the secret interactions would couple all species to become a single fluid and the spectral mean opacity would increase, so it is conceivable that the cooling time could become somewhat larger. This may not be a huge effect and its impact also depends on the role of convection, but still this scenario deserves some attention if one were to entertain such ``crazy'' ideas.

Our analysis of the SN~1987A neutrino data is not conclusive concerning the best choice of nuclear EoS among the tested models. The cumulative energy distributions (Fig.~\ref{fig:cumulative_spectrum}) and cumulative event distributions (Fig.~\ref{fig:cumulative_distributions}) exhibit considerable degeneracy between NS mass and EoS, in particular for the predicted IMB signal. The test statistics for the individual experiments (Fig.~\ref{fig:likelihood_exps}) reveal mild preferences for different EoS models: SFHo and SFHx in the case of IMB, but DD2 and LS220 for Kam-II. However, the same analysis for all experiments combined (Fig.~\ref{fig:likelihood_plots}) suggests a slightly better acceptance of DD2 for the higher values of the tested NS masses, but it has no clear discriminatory power of the EoS in the cases of the most favored lower-mass NSs.

A future high-statistics galactic SN observation may go a long way to clarifying the many open questions, especially with regard to the late signal~\cite{Li:2020ujl}. Indeed, the interesting outlook for probing the {\em late} neutrino signal with the current and future generations of large detectors has been somewhat under-appreciated in the literature. While possible fallback effects would naturally vary from case to case, the PNS cooling signal should easily show up, but on the other hand, fallback could potentially cover more subtle effects of late phase transitions.

The SN~1987A signal duration has often been used to constrain the coupling strength of axions and other novel particles that extract energy from the deep interior of the PNS, energy that is subsequently missing to drive neutrino emission. Such constraints, based on the signal duration, are much more sensitive than using the overall energy measured in neutrinos. In an early paper co-authored by one of us \cite{Raffelt:1987yt} shortly after SN~1987A, actually the IMB signal duration was used as a yardstick for the cooling time, not the late Kam-II events. Still, over the years, the exact relation between such cooling arguments, realistic SN models, and the actual SN~1987A data was somewhat lost, especially in the later context of PNS convection and/or the correlation-reduced opacities, effects that are not included in all models that were used as reference cases. 

As a consequence of our present findings, we plan to return to a similar analysis as performed in this paper to study the sensitivity of the actually observed signal to non-neutrino energy losses, based on a set of self-consistent models that include the new effects. Of course, the high-statistics observation of the next nearby SN would go a long way to clarifying such questions on the data side of these arguments. From the particle-physics perspective, this is perhaps the most interesting question to be settled by the next SN neutrino observation.

\section*{Data Availability}

The model signals of this paper are available in our Garching
Core-Collapse Supernova Archive upon request (\href{https://wwwmpa.mpa-garching.mpg.de/ccsnarchive/}{https://wwwmpa.mpa-garching.mpg.de/ccsnarchive/}).

\section*{Acknowledgments}

We thank the authors of Ref.~\cite{Li:2023ulf} for comments that have led to a detailed
comparison in our Appendix D.
In Munich/Garching, this work was supported by the German Research Foundation (DFG) through the Collaborative Research Centre ``Neutrinos and Dark Matter in Astro- and Particle Physics (NDM),'' Grant SFB-1258\,--\,283604770, and under Germany’s Excellence Strategy through the Cluster of Excellence ORIGINS EXC-2094-390783311. DFGF is supported by the Villum Fonden under Project No.\ 29388 and the European Union's Horizon 2020 Research and Innovation Program under the Marie Sk{\l}odowska-Curie Grant Agreement No.\ 847523 ``INTERACTIONS.''  EV acknowledges support by the European Research Council (ERC) under the European Union’s Horizon Europe Research and Innovation Program (Grant No.\ 101040019). The calculations were performed at the Max Planck Computing and Data Facility (MPCDF). This article  is based upon work from COST Action COSMIC WISPers CA21106, supported by COST (European Cooperation in Science and Technology)

\onecolumngrid


\appendix
\section{SN~1987A Neutrino Observations}

\twocolumngrid

\begin{table*}[ht]
 \caption{Detector Parameters.\label{tab:Detectors}}
 \vskip4pt
    \begin{tabular*}{\textwidth}{@{\extracolsep{\fill}}lllll}
    \hline\hline
     & IMB \cite{Irvine-Michigan-Brookhaven:1983iap, Bionta:1987qt, IMB:1988suc, Matthews:1987} 
     & Kam-II \cite{Arisaka:1985lki, Hirata:Moriond, Kamiokande-II:1987idp, Hirata:1988ad, Nakahata:1988} 
     & BUST \cite{Pomansky:1978xc, Alekseev:1979pn, Chudakov:1979vh, Alekseev:1987ej, Alekseev:1988gp, Alekseev:1993dy, Kuzminov:2017, Novoseltsev:2019gdt, Novoseltseva:2022qic} 
     & LSD \cite{Badino:1984ww, Aglietta:1987it, Aglietta:1987we, Dadykin:1987ek, Saavedra:1987tz}\\
     \hline
     Target Material& Water  & Water  & Scintillator   & Scintillator    \\
                    & H$_2$O & H$_2$O & H$_{2n+2}$C$_n$ ($n\simeq 9$)$^a$  & H$_{2n+2}$C$_n$ ($n\simeq 10$)$^a$ \\
     Active target mass    & 6800~t & 2140~t & 200~t (280~t w.e.)$^b$& 90~t (126~t w.e.)$^b$\\
     \quad Number of protons    & $4.55\times10^{32}$& $1.43\times10^{32}$ & $1.87\times10^{31}$ & $8.43\times10^{30}$\\
     \quad Number of electrons  & $2.27\times10^{33}$& $7.15\times10^{32}$ & $5.26\times10^{31}$ & $2.37\times10^{31}$\\
     \quad Number of $^{16}$O or $^{12}$C   & $2.27\times10^{32}$& $7.15\times10^{31}$ & $8.47\times10^{30}$ & $3.81\times10^{30}$ \\
     Energy resolution $\sigma_E$ [Eq.~\eqref{eq:resolution}] &$\sqrt{1.27\,{\rm MeV}\, E_e}$&$\sqrt{0.60\,{\rm MeV}\, E_e}$&$\sqrt{0.72\,{\rm MeV}\, E_e}$&$\sqrt{0.15\,{\rm MeV}\, E_e}$ \\
     Geographic Location &$41.7^\circ$~N  &\hp$36.4^\circ$~N&$43.3^\circ$~N&$ 45.9^\circ$~N\\
              & $81.3^\circ$~W & $137.3^\circ$~E&$42.7^\circ$~E&\hp$6.9^\circ$~E\\
     Depth water equivalent & 1570~m & 2700~m & 850~m & 5200~m \\
     \quad Muon rate              & 2.7~Hz & 0.37~Hz & 15~Hz & $1.0\times10^{-3}$~Hz \\
     \quad Low-$E$ background rate& 0      & 0.187~Hz & 0.034~Hz & 0.012~Hz \\
     SN~1987A &       \\
     \quad Local time  & 2:35 am & 4:35 pm & 10:35 am & 8:35 am\\
     \quad Elevation below horizon   & $42.4^\circ$   & $19.7^\circ$   & $55.3^\circ$   & $65.5^\circ$\\
     \quad Azimuth relative to South & $28.0^\circ$~W & $14.0^\circ$~E & $26.3^\circ$~E &$37.5^\circ$~E \\
     \quad Neutrino path in Earth& 8592~km &  4295~km &  10476~km &  11595~km\\
     \quad Maximum depth in Earth& 1666~km &\hp373~km &\hp2744~km &\hp3729~km\\
     \hline
    \end{tabular*}
    \vskip2pt
    \vbox{\raggedright$^a$Apparently the scintillator was the same, but $n=9$, 9.6, or 10 are found in different papers. We use $n=9.5$ for both cases, the exact value making a $\pm0.5\%$ level difference.\\
    $^b$Equivalent mass for a water detector with the same number of protons.}
\end{table*}

\label{sec:SN-Observations}

In this appendix we collect the experimental data. Some of the information is scattered between papers, proceedings, and private communications cited in secondary papers, and there are various minor discrepancies.

\subsection{Location and Timing}

Supernova 1987A was the explosion of the blue supergiant Sanduleak $-69\,202$ in the Large Magellanic Cloud in the Southern Sky. The Equatorial Coordinates (epoch J2000.0) are:\footnote{ \href{https://simbad.cds.unistra.fr/simbad/sim-fbasic}{https://simbad.cds.unistra.fr/simbad/sim-fbasic} using the query SN~1987A} Right Ascension $05^\circ 35'28.020''$ and Declination $-69^\circ 16' 11.07''$. 

An often-cited distance determination of $51.4\pm1.2$~kpc relies on the light curve measurement and concomitant geometric size determination of the SN ring \cite{Panagia:1998,Panagia:2003}, which implies a distance to the LMC barycenter of $51.7\pm1.3$~kpc. In other words, SN~1987A is closer to us than the LMC barycenter by some 0.3~kpc, a value which itself is uncertain by $\pm0.3$~kpc. On the other hand, a recent geometric distance determination to the LMC is $49.59 \pm 0.09_{\rm stat} \pm 0.54_{\rm syst}$ kpc \cite{2019Natur.567..200P}. Overall, we adopt
\begin{equation}
    d_{\rm SN}=50~{\rm kpc}
\end{equation}
as a simple value that is consistent with these findings. Small distance uncertainties have no practical bearing on our study.

The first evidence for optical brightening was found on 23 February 1987 at 10:39 UT (Universal Time) on plates taken by McNaught \cite{IAUC4316,West:1992}. Based on the first neutrino in the IMB detector (see below), the explosion occurred around 3~hours earlier at UT 7:35:41.374${}\pm50$~ms. From this information follow the Horizontal Coordinates of SN~1987A at the four detector locations provided in Table~\ref{tab:Detectors}, i.e., expressed as the elevation below the horizon and the azimuth deviation from the southern direction.\footnote{\href{http://xjubier.free.fr/en/site_pages/astronomy/coordinatesConverter.html}{http://xjubier.free.fr/en/site\_pages/astronomy\\/coordinatesConverter.html}} At the time of explosion, the SN was at the zenith at geographic location $69^\circ 16'$~S and  $177^\circ 20'$~E in the Southern Ocean, north of McMurdo Station about one third the distance to New Zealand.

\subsection{IMB Detector}

\label{sec:IMB}

\subsubsection{General Description}

At the time of the explosion, there were four running experiments that were big enough that they could have detected the neutrino flux. The largest one was the Irvine-Michigan-Brookhaven (IMB) water Cherenkov detector, an experiment built to look for proton decay \cite{Irvine-Michigan-Brookhaven:1983iap}, that was located in the Morton-Thiokol salt mine (Fairport, Ohio, USA), with the geographic coordinates $41.7^\circ$~N and $81.3^\circ$~W at a depth of 1570~m w.e.\ (water equivalent) \cite{Bionta:1987qt}.

IMB was equipped with 2048 8-inch photomultiplier tubes (PMTs) such that 6,800 tons of water (of a total of 8,000 tons) were within the PMT planes, taken as the fiducial volume for the SN~1987A search \cite{IMB:1988suc}. A failure of a high-voltage power supply shortly before SN~1987A left a contiguous quarter of the PMTs off line with a geometric effect on the trigger efficiency that was later calibrated. The detector was triggered when at least 20~PMTs fired in 50~ns, corresponding to an energy threshold of 15--25~MeV for showering particles \cite{IMB:1988suc}.

The absolute time of an event was recorded with an uncertainty $\pm 50$~ms thanks to the WWVB clock, a time signal radio station operated by the National Institute of Standards and Technology~\cite{WWVB}. The first event occurred at UT 7:35:41.374 on 23 February 1987, corresponding to 2:35~am local time on an early Monday morning.

At its relatively shallow depth, the flux of atmospheric muons caused a trigger rate of 2.7~Hz. Muons were recognized by tracks entering the detector from the outside and of course coming mostly from above. The detector is dead for 35~ms after each trigger. The SN~1987A signal consisted of 8~events and in addition 15 muons were recorded over a period of 5.6~s \cite{Bionta:1987qt}. On average, one would expect 2.7~Hz $\times$ 5.6~s = 15.1 muon triggers, in agreement with the measured number. 

Atmospheric neutrinos are recognized as contained events and occurred at a rate of around 2/day in the energy range 20--2000~MeV~\cite{Bionta:1987qt}, i.e., they are irrelevant as a background.

\subsubsection{SN 1987A Signal}

The SN~1987A burst was found by looking in the recorded data for low-energy event clusters, where ``low energy'' was defined as fewer than 100~PMTs firing, corresponding roughly to a 75~MeV energy cut. A period of 6.4~h, beginning at UT 5:00:00, was subdivided in 2304 nonoverlapping 10~s intervals, showing events commensurate with a Poisson distribution of 0.077 events/s. There was no 10~s cluster with more than 5 events, and only one with 9 events. One of these was recognized as a muon, leaving a cluster of 8~events attributed to SN neutrinos listed in Table~\ref{tab:IMB-data} and shown in Fig.~\ref{fig:SNData}.

The cited background of 0.077~Hz refers to the raw rate relevant to this search for an event cluster. However, subsequently each event can be examined for its detailed properties and usually can be attributed to its origin such as corner-clipping through-going muons. Indeed, one of the originally nine events was thus recognized as a muon, not a SN neutrino. The probability that any of the remaining eight events was caused by some unidentifiable background is negligible.\footnote{Cited after the Addendum of Ref.~\cite{Kolb:1987dda} who attribute this information to a private communication from J.~van der Velde of the IMB collaboration.} In this sense, the IMB SN neutrino events are actually background free.

The SN~1987A events should be due mostly to inverse beta-decay with a practically isotropic distribution of final-state $e^+$. However, IMB found a conspicuous directional correlation in the opposite direction of SN~1987A, i.e., the events look ``forward peaked.'' This effect is not explained by the detector's geometrical bias due to the 25\% PMT failure, which however caused a bias in the azimuth distribution around the LMC direction \cite{IMB:1988suc}. We return to this question in Appendix~\ref{sec:AngularDistribution}.

\subsubsection{Trigger Efficiency}

\begin{figure}
\vskip4pt
    \centering
   \includegraphics[width=0.85\columnwidth]{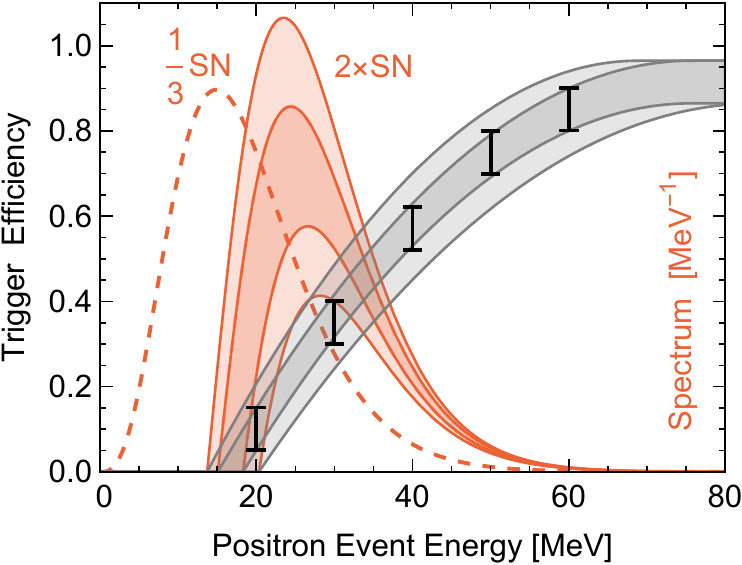}
    \caption{Trigger efficiency in IMB. Data points from Ref.~\cite{IMB:1988suc}, fit function according to Eq.~\eqref{eq:IMBTrigger} $\pm0.05$ systematic uncertainty (dark gray band). For the broader gray band, the energy calibration uncertainty of $\pm10\%$ is included. We also show the positron event spectrum from a fiducial SN with an assumed Maxwell-Boltzmann spectrum with $T=4$~MeV (red bands for the different trigger efficiencies) and for a hypothetical 100\% trigger efficiency (dashed). The positron spectra caused by the SN were multiplied by the indicated factors to fit the curves on a common scale. The hypothetical total event number under the dashed curve is 53.3, which is reduced to 6.4 after including the average trigger efficiency. These illustrative energy spectra do not include ``smearing'' by finite energy resolution.
    }\label{fig:IMBTrigger}
\end{figure}

The trigger efficiency as a function of $e^\pm$ energy $E_e$ was recalibrated in view of the partial detector outage \cite{IMB:1988suc}. In their Fig.~1, it is shown for $E_e=20, \ldots, 60$~MeV in steps of 10~MeV as reproduced in our Fig.~\ref{fig:IMBTrigger} with a systematic uncertainty of $\pm0.05$, implying a significant uncertainty at low energies. Burrows \cite{Burrows:1988ba} has provided an analytic fit for the average trigger efficiency that fits the data well\footnote{At the time of his writing, the second IMB paper \cite{IMB:1988suc} was not yet available, so the updated information must have come from~\cite{Matthews:1987} that is cited in his paper.} 
\begin{equation}\label{eq:IMBTrigger}
    \eta_{\rm IMB}=\!\begin{cases}0.3975\,x-0.02625\,x^2-0.59
    &\!\hbox{for $1.9<x<7.6$}\\
    0.915 &\!\hbox{for $7.6<x$}
    \end{cases}
\end{equation}
where $x=E_e/10$~MeV. The asymptotic value of 0.915 at large energies cannot be found in the IMB papers. However, the range above 60~MeV is of no significance for our study, so we do not worry about where this value came from.

Of greater significance is the systematic range $\pm0.05$ that makes a big difference at small energies. It appears that this issue has not been addressed in earlier SN~1987A analyses, also not by Loredo and Lamb \cite{Loredo:2001rx} who have performed the most detailed previous study. Moreover, there is an overall $\pm10\%$ systematic uncertainty of the overall energy scale \cite{IMB:1988suc} which also seems to have been ignored in previous works.

In Fig.~\ref{fig:IMBTrigger} we show the trigger efficiency Eq.~\eqref{eq:IMBTrigger} $\pm0.05$ as a dark gray band. We interpret the $\pm10\%$ uncertain energy calibration such that the trigger efficiency should be shifted relative to the true source spectrum, so the extreme trigger efficiencies would be $\eta_{\rm IMB}(0.9 x)$ or $\eta_{\rm IMB}(1.1 x)$ instead of $\eta(x)$. These two effects lead to nearly identical modified trigger efficiencies, i.e., applying the $\pm10\%$ energy uncertainty alone leads to range nearly identical to the dark gray band. Applying the energy modification to the $\pm0.05$ shifted cases leads to the wider gray band as an extreme range.

\subsubsection{Fiducial Supernova}

\label{sec:Fiducial-Supernova}

To gain some intuition for the impact of the trigger uncertainty we define a ``fiducial supernova'' that emits a total energy of $3\times10^{53}$~erg equipartitioned among six neutrino species and is at a distance of 50~kpc. The spectrum is taken to be of Maxwell-Boltzmann form with temperature $T$
\begin{equation}
    f_T(E_\nu)=\frac{E_\nu^2}{2T^3}\,e^{-E_\nu/T},
\end{equation}
which is normalized and has $\langle E_\nu\rangle=3T$. Then the $\bar\nu_e$ fluence at the detector is
\begin{equation}
    F_{\bar\nu_e}=\frac{3.49\times10^{10}~{\rm cm}^{-2}}{T/{\rm MeV}}.
\end{equation}
The main detection channel in all four experiments is inverse beta-decay (IBD). For a given experiment, the produced positron spectrum is therefore
\begin{equation}
\frac{dN_e}{dE_e}=n_p\,F_{\bar\nu_e}\,\sigma_{\bar\nu_ep}(E_e+Q)\,\eta(E_e)\,f_T(E_e+Q),
\end{equation}
where $Q=1.293$~MeV is the proton-neutron mass difference and $n_p$ is the number of proton targets in the detector. 

Assuming $T=4$~MeV, we show in Fig.~\ref{fig:IMBTrigger} the expected event spectrum in IMB if the trigger efficiency were 100\% (red dashed line), providing 53.3 expected positrons in the detector volume. Including the average trigger efficiency, this is reduced to 6.4 events. In Fig.~\ref{fig:IMBTrigger} we also show the event spectra for the $\pm0.05$ uncertainty (darker red band), corresponding to 5.1--8.0 events. Applying in addition the $\pm10\%$ energy uncertainty leads to the wider red band and an event range of 3.6--10.1.

We conclude that the uncertain IMB trigger efficiency has a strong impact on the expected number of events. On the other hand, the impact on the expected average event energy is much smaller as we glean from Fig.~\ref{fig:IMBTrigger}: The normalization of the red curves changes more strongly than their shape. For the intermediate trigger efficiency, the average event energy would be 30.1~MeV. Including $\pm0.05$ leads to 29.0--31.1~MeV, where the effect goes in the direction that a global increase by $+0.05$ leads to {\em more\/} events with somewhat {\em lower\/} average energies. Including the energy scale uncertainty provides the range 28.0--33.0~MeV.

To better understand the impact of the trigger uncertainty on our parameter estimates and model comparison, we derive the confidence regions in emitted SN~19987A energy and average energy $\bar{\epsilon}_{\bar{\nu}_e}$, assuming a Maxwell-Boltzmann spectrum. The details of how we construct such plots is explained in Appendix~\ref{app:methods_comparison}. We see that the confidence regions are strongly extended in the vertical direction (total energy or total event number), whereas the implied $\bar{\epsilon}_{\bar{\nu}_e}$ range is not strongly changed for plausible total energies around $3\times10^{53}$~erg.

\begin{figure}
\vskip-4pt
    \centering
   \includegraphics[width=0.85\columnwidth]{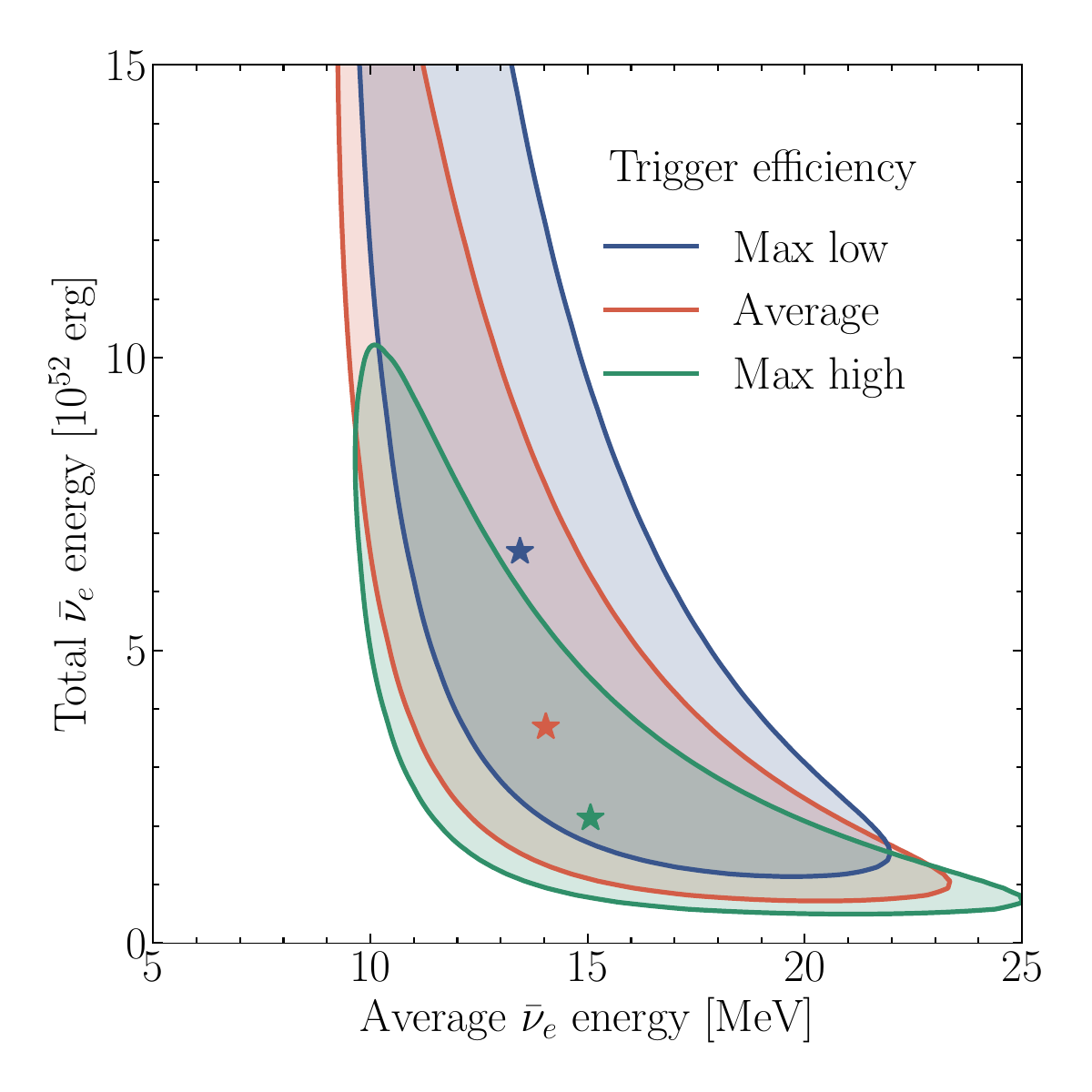}
    \caption{Confidence regions (95\%) and best-fit points for SN~1987A energy emission and average neutrino energy implied by the IMB data. The contours are for the maximally high or low and the average trigger efficiency.}\label{fig:IMBTiggerBanana.}
    \vskip-8pt
\end{figure}

\subsection{Kamiokande II}

\label{sec:Kam}

\subsubsection{General Description and Trigger Efficiency}

The second largest detector was the Kamiokande II water Cherenkov detector in the Mozumi Mine, Kamioka Section of Hida, Gifu Prefecture, Japan, with geographic coordinates $36.4^\circ$~N and $137.3^\circ$~E \cite{Nakahata:1988}. For the SN~1987A search, the fiducial mass was 2,140 metric tons, where the entire volume up to the plane of the PMTs was taken \cite{Hirata:Moriond, Kamiokande-II:1987idp, Hirata:1988ad}. This detector was built in 1983 to search for nucleon decay (Kamiokande = Kamioka Nucleon Decay Experiment) \cite{Arisaka:1985lki} and later upgraded to Kamiokande II (Kam-II) to search for solar $\nu_e$ in the 10~MeV range. The photo cathode coverage was increased and radioactive backgrounds decreased to lower the threshold and solar data were taken since the end of 1986. Despite its smaller mass, the low threshold made Kam-II competitive for the SN~1987A discovery. 

\begin{figure}[ht]
    \centering
   \includegraphics[width=0.85\columnwidth]{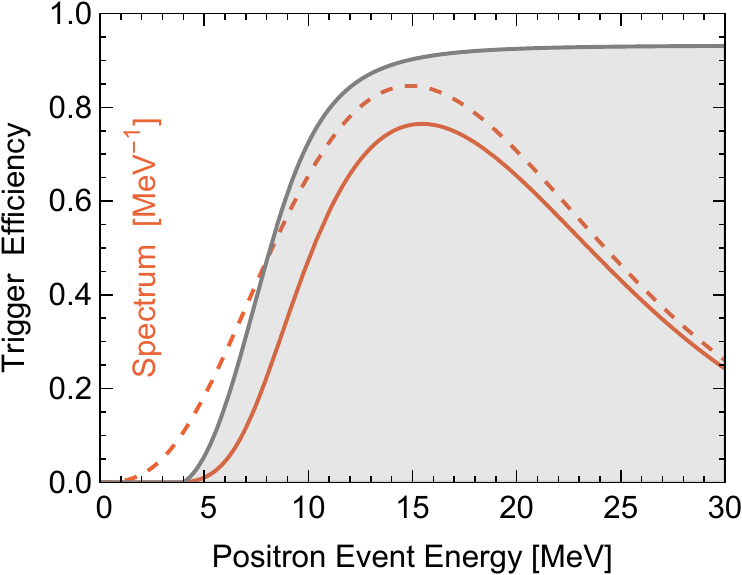}
    \caption{Trigger efficiency in Kam-II following our fit function Eq.~\eqref{eq:KamTrigger}. We also show the event spectrum from a fiducial SN with an assumed Maxwell-Boltzmann spectrum with $T=4$~MeV (red solid line) and for hypothetically 100\% trigger efficiency (dashed), leading to 14.0 (16.7) events. These illustrative energy spectra do not include ``smearing'' by finite energy resolution.}
    \label{fig:KamTrigger}
\end{figure}

In analogy to IMB, Burrows \cite{Burrows:1988ba} has also given an analytic approximation for the Kam-II trigger efficiency, which however we do not find satisfactory. Digitizing Fig.~3 of Ref.~\cite{Hirata:1988ad}, we find that a surprisingly simple fit function is
\begin{equation}\label{eq:KamTrigger}
    \eta_{\rm Kam}=0.932\Bigg/\sqrt{1+\left(\frac{34}{12-7x+x^2}\right)^2}
    \quad\hbox{for $x>4$}
\end{equation}
and zero for $x<4$, where $x=E_e/{\rm MeV}$. We have also compared with the trigger efficiency given in Ref.~\cite{Nakahata:1988}, which is very similar. We show our $\eta_{\rm Kam}$ in Fig.~\ref{fig:KamTrigger} together with the signal spectrum from a fiducial SN (as defined earlier) with an assumed Maxwell-Boltzmann $\bar\nu_e$ spectrum with $T=4$~MeV.

At a greater depth of 2700~m w.e., the atmospheric muon trigger rate was only 0.37~Hz and indeed, 8~muons were found in the 45~s time interval surrounding the SN~burst, four of them before the burst and another four roughly after event No.~12. Notice that in 45~s one would have expected $0.37\times45=16.7$ muons, so there was a significant downward fluctuation. Atmospheric neutrinos, in the form of fully contained events, show up once every few days. Low-energy radioactive backgrounds triggered with around 0.23~Hz. The trigger dead time is less than 50~ns after an event.

To find the SN~1987A burst, the Kam-II Collaboration pioneered the method also applied by IBM. The data recorded on a magnetic tape were searched for low-energy event clusters, where the definition was less than 170 PMTs firing ($E_e\alt 50$~MeV). We show the burst in Fig.~\ref{fig:SNData} as a function of time after the first SN event. 

In Table~\ref{tab:KamII-data} we list these events according to  Ref.~\cite{Hirata:1988ad} and in addition four later events that almost certainly should be attributed to background. These were shown in the Moriond (March 1987) proceedings \cite{Hirata:Moriond} and later by Krivoruchenko in a discussion of the scattering angle distribution \cite{Krivoruchenko:1988zg}, the information attributed to a private communication by Y.~Totsuka. Later they were shown again in the context of a detailed analysis of the SN~1987A neutrinos \cite{Loredo:2001rx}, attributed to a private communication by Y.~Totsuka, A.~Mann, and S.-B.~Kim, but the exact times and energies differ from the earlier listing. We show them based on the earlier paper \cite{Hirata:Moriond}, whereas the angles, not shown there, are taken from \cite{Krivoruchenko:1988zg}.

The absolute timing is poorly known, probably to within $\pm 15$~s based on comparing the computer clock with a wrist watch, but a conservative uncertainty of $\pm 1$~min was officially stated. A power outage in the mine on February 26 prevented a recalibration of the computer clock~\cite{Kajita:2012zz}. The signal arrived at 4:35 pm on Monday, 23 February 1987, but this was a substitute holiday. According to working-day schedule, the magnetic tape would have been exchanged at 4:30 pm and the signal might have been missed.

There is a conspicuous gap of 7.3~s between events 9 and 10. Recently, one member of the Kamiokande collaboration has speculated that the gap could have been caused by a fault of the magnetic tape drive~\cite{Oyama:2021oqp}. He noted that during that gap, there are also no other events (low-energy background or atmospheric muons) and that the probability for such a long gap was very small. However, according to a private communication by M.~Nakahata, this explanation is not viable because the event numbers were continuous across the gap. The event number was generated by the front-end electronics and the trigger system. When the number of hit PMTs within 100~ns was more than a given threshold value, a trigger was generated and the electronics system read out timing and charge information of each individual PMT. The event number was incremented by one whenever a trigger happened. If events had been lost by a tape-write error, there would have had to be an event-number gap as well.

\subsubsection{Low-Energy Background}

\begin{figure}[ht]
    \centering
   \includegraphics[width=0.85\columnwidth]{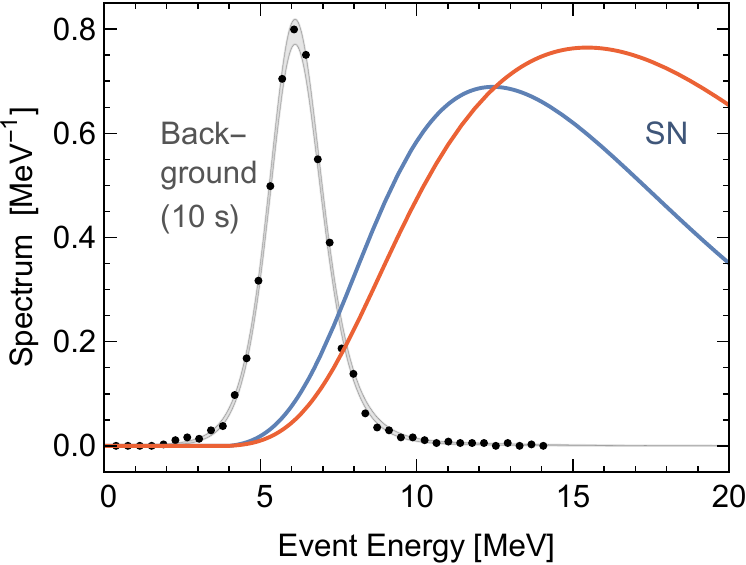}
    \caption{Background spectrum in Kam-II according to Ref.~\cite{Loredo:2001rx} (black dots) for a time interval of 10~s, yielding a total of 1.87 expected events. We also show our fit function Eq.~\eqref{eq:BKAM} together with its error Eq.~\eqref{eq:BKAM-err} as a one-sigma band, i.e., statistically around 68\% of the data points for the ten-hour measurement period should be found within the gray band.  (In Ref.~\cite{Loredo:2001rx} the corresponding error is shown for each data point.)  Also shown is a ``fiducial'' SN signal with an assumed Maxwell-Boltzmann spectrum with $T=3$~MeV (blue) and  4~MeV (red) that would yield a total of 9.0 (14.0) events. 
    These illustrative energy spectra do not include ``smearing'' by finite energy resolution.}
    \label{fig:KamBackground}
\end{figure}

Some of the low-energy events around the SN~1987A burst can be detector background. We 
treat this question in the maximum-likelihood analysis along the lines of Loredo and Lamb \cite{Loredo:2001rx}. They have published the Kam-II background spectrum in their Fig.~2a, citing a private communication, but without providing further details. The digitized data points from their Fig.~2a are shown in our Fig.~\ref{fig:KamBackground}, yielding a total rate of 0.187~Hz, in agreement with what they state.

In Kam-II, each event consists of a certain number $N_{\rm hit}$ of photomultipliers hit during a narrow time window corresponding to the time of flight of photons from the reconstructed vertex to the PMTs \cite{Nakahata:1988}. In Fig.~2a of Ref.~\cite{Loredo:2001rx}, the energy of each data point must correspond to a certain $N_{\rm hit}$. By inspection of this plot it is clear that, e.g., $N_{\rm hit}=16$ for the maximum at an energy near 6~MeV or $N_{\rm hit}=37$ for the point at 14.1~MeV. Indeed, the data points are uniformly spaced on the horizontal axis and correspond to the relation $\epsilon/{\rm MeV}=0.38 \Nhit$. This agrees within rounding errors with what is stated in the Kam-II paper \cite{Hirata:1988ad}, although for larger energies, the relation is not strictly linear and in detail, the reconstructed energy depends on vertex location. 

The shown data probably correspond to the total event numbers $n$ in each such channel and the shown uncertainty must indicate their $\pm\sqrt{n}$ range. Taking $N_{\rm hit}=16$ (energy near 6~MeV) as an example, the uncertainty shown in Fig.~2a of Ref.~\cite{Loredo:2001rx} is a fractional error $\pm0.0305$. If this is interpreted as $\pm1/\sqrt{n}$, this measurement was based on $n=0.0305^{-2}=1075$ events. The rate is $0.080~{\rm s}^{-1}~{\rm MeV}^{-1}$ and the bin width is 0.38~MeV, so the rate in this channel is $0.030~{\rm s}^{-1}$, implying that it took around 35400~s to accrue these events, which is nearly exactly ten hours. Similar numbers follow from the other data points. We conclude that the background shown in Fig.~2a of Ref.~\cite{Loredo:2001rx} corresponds to the background partly shown in Fig.~4 of the Kam-II paper \cite{Hirata:1988ad} for $\pm5$~hours around the SN burst.

Next we compare with information provided directly in the Kam-II paper \cite{Hirata:1988ad}. From their Fig.~10(a) we learn that around the time of the SN, the rate for events with $\Nhit\geq20$ was 0.0219~Hz. (Notice that in this figure they give a rate per 10~s.) From our digitized data we find a slightly smaller value 0.0203~Hz. From their Fig.~12 we learn that the rate for $\Nhit\leq20$ was $5.37/30\,{\rm s}=0.179$~Hz, to be compared with our 0.174~Hz. Within the uncertainties caused by data extracted from a plot, these values are all consistent. 

The tails of the shown background spectrum depends on few events per $N_{\rm hit}$ channel and suffers from digitization uncertainties, so the tails are not reliable. Therefore, instead of an interpolating function between these fluctuating data points as used in Ref.~\cite{Loredo:2001rx} one may equally use a convenient fit function such as the one shown in Fig.~\ref{fig:KamBackground}. It is
\begin{equation}\label{eq:BKAM}
    B_{\rm Kam}=\left(\frac{0.84\,x}{[4+(x-6)^2]^3}+\frac{0.001}{5+(x-12)^2} \right)~{\rm MeV}^{-1}~{\rm s}^{-1},
\end{equation}
where $x$ is the energy in MeV, and integrates to a total rate of 0.187~Hz. Notice that the background has fat tails and is not well approximated by a Gaussian.

If this were the exact background spectrum, we can imagine to measure it in bins of width 0.38~MeV for ten hours, leading to data that should fluctuate in similar ways to those shown in Fig.~\ref{fig:KamBackground}. Then the $\sqrt{n}$ Gaussian fluctuation from bin to bin should be given by
\begin{equation}\label{eq:BKAM-err}
    {\rm Err}(B_{\rm Kam})
    =\sqrt{\frac{B_{\rm Kam}}{0.38~{\rm MeV}\,10~{\rm h}}}.
\end{equation}
In other words, around 68\% of all data points should lie between the curves
$B_{\rm Kam}\pm {\rm Err}(B_{\rm Kam})$ shown in Fig.~\ref{fig:KamBackground} as a gray band, which is very approximately the case.

We also show the signal of a fiducial SN defined earlier in Sec.~\ref{sec:Fiducial-Supernova}. It assumes a Maxwell-Boltzmann spectrum with $T=3$~MeV (blue curve) or 4~MeV (red) that would yield 9.0 (14.0) events in Kam-II, compared with a background in 10~s of 1.87 events. We glean from this figure that background and expected signal are well separated. It is clear, for example, that event No.~6 and the ones beyond No.~12, that have energies around 6~MeV, are very likely background.

The background with $N_{\rm hit}\leq20$ ($\epsilon<7.5$~MeV) originates primarily from $^{214}$Bi ($\beta$-decay endpoint 3.26 MeV), which itself derives from $^{222}$Rn decays. Their overall rate in the Kam-II volume is huge, perhaps some $10^4$~Hz \cite{Nakahata:1988}, but the trigger efficiency is low, leading to the shown background spectrum that peaks around 6~MeV reconstructed energy, which however is caused by the $\beta$ spectrum with an endpoint of 3.26~MeV true energy.

For larger energies ($N_{\rm hit}>23$), the background is caused by radioactive decays outside the detector, including $\gamma$ rays from the surrounding rock. These backgrounds are concentrated near the detector surface. Therefore, the probability that a given higher-energy event might be background depends strongly on the vertex location and cannot be captured by a volume-averaged background rate in a quantitatively meaningful way. For example, the late event No.~11 with $N_{\rm hit}=37$ is far away from the walls and as such does not look like background.

From Fig.~10(b) of Ref.~\cite{Hirata:1988ad} we learn that the rate of events with $\Nhit\geq30$ was $1.18\times10^{-3}$~Hz. (They only give two digits, but to reproduce ``the expected $8\times10^{-11}$ entries in 2.7 days'', one can reconstruct the third digit.) We have adjusted our fit-function Eq.~\eqref{eq:BKAM} to reproduce this higher-energy background.

Returning to the total background rate of 0.187~Hz, it is smaller than 0.23~Hz stated elsewhere in the Kam-II paper, which we believe should be interpreted as the raw trigger rate. A trigger is formed if 20 PMTs fire within a 100~ns time window. After the vertex has been reconstructed, only those PMTs are counted into $N_{\rm hit}$ that could have contributed to the signal by virtue of the time of flight from the vertex to the PMT, so $N_{\rm hit}$ will often be smaller and can lead to the many observed events with $N_{\rm hit}<20$. The PMTs have a large dark current \cite{Kume:1983hs}
and low-energy backgrounds can trigger PMTs that are unrelated to the Cherenkov ring of a given event. Apparently not every raw trigger leads to a clearly reconstructed event, probably explaining why the background rate of 0.187~Hz in the $N_{\rm hit}$ distribution is smaller than the raw trigger rate of 0.23~Hz.

We also stress that the raw trigger threshold of 20 PMTs in 100~ns is not the same as an energy threshold on observed SN neutrinos, i.e., it is not a threshold for the reconstructed $N_{\rm hit}$, in contrast to what appears to have been assumed in Ref.~\cite{Li:2023ulf}, e.g.\ in their Eq.~(D2).

\subsection{BUST (Baksan)}

\label{sec:Baksan}

\subsubsection{General Description and Trigger Efficiency}

The third experiment to observe SN~1987A neutrinos was the Baksan Scintillator Underground Telescope (BUST), operated by the Institute of Nuclear Research (Moscow)
\cite{Pomansky:1978xc, Alekseev:1979pn, Chudakov:1979vh, Alekseev:1987ej, Alekseev:1988gp, Alekseev:1993dy, Kuzminov:2017, Novoseltsev:2019gdt, Novoseltseva:2022qic}. This instrument is located in the Baksan underground laboratory under Mount Andyrchi in the North Caucasus. The tunnel entrance is near the village ``Neutrino'' in the Baksan valley and geographically located at $43.3^\circ$~N and $42.7^\circ$~E. BUST is located in a chamber 550~m from the entrance and thus at a depth of only 850~m~w.e., whereas other chambers are much deeper in the mountain.

BUST consists of 3150 separate elements\footnote{This number is given as 3132 in \cite{Alekseev:1979pn} and as 3156 in \cite{Alekseev:1987ej}, whereas in later papers one reads 3150.} of dimension $70\times70\times30$~cm, filled with scintillator based on ``white spirit'' (C$_n$H$_{2n+2}$ with $n\simeq9$) and is viewed by one FEU-49 photomultiplier with a photocathode diameter 15~cm. The elements are arranged in eight planes, four horizontal and four vertical. The atmospheric muon rate is around 15~Hz \cite{Chudakov:1979vh}. The search for low-energy neutrinos uses the inner part of the detector (1200 segments) with a mass of 130~t and looks for events that trigger one and only one individual segment with $E_e\alt 50$~MeV. For the period around SN~1987A, the rate for such singles events was 0.0127~Hz \cite{Alekseev:1987ej, Alekseev:1988gp}. BUST has surveyed the galaxy for SN neutrino bursts since mid-1980 and is still operating today \cite{Alekseev:1993dy, Novoseltsev:2019gdt, Novoseltseva:2022qic}, but no signal was observed other than SN~1987A. To search for a suspected signal, the sensitive mass can be increased to 200~t, which increases the singles rate to 0.033~Hz. 

\begin{figure}[ht]
    \centering
   \includegraphics[width=0.85\columnwidth]{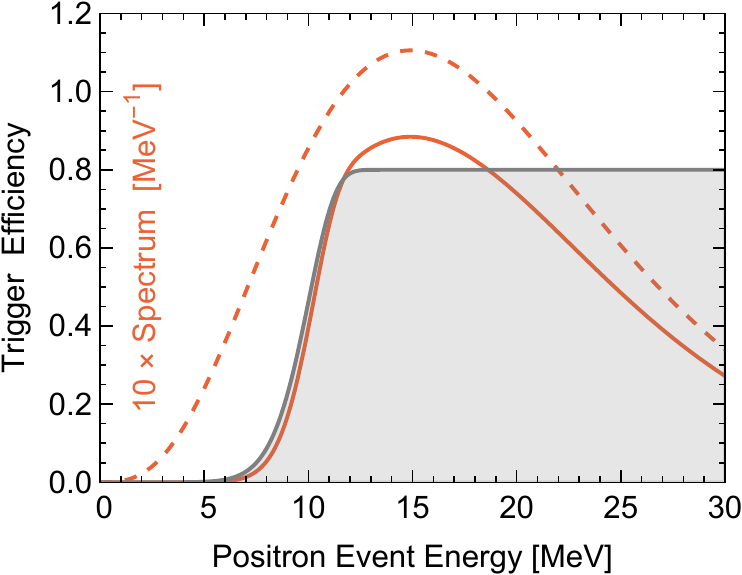}
    \caption{Trigger efficiency in BUST (Baksan) following the fit function Eq.~\eqref{eq:BakTrigger}. We also show the event spectrum from a fiducial SN with an assumed Maxwell-Boltzmann spectrum with $T=4$~MeV (red solid line) and for hypothetically 100\% trigger efficiency (dashed), leading to 1.52 (2.19) events.
    These illustrative energy spectra do not include ``smearing'' by finite energy resolution.}
    \label{fig:BakTrigger}
\end{figure}

The average trigger efficiency was provided as a fit function ($x=E_e/{\rm MeV}$) in Ref.~\cite{Alekseev:1988gp}
\begin{equation}\label{eq:BakTrigger}
    \eta_{\rm Bak}=0.8\,\left[1-e^{-(x/10.17)^9}\right],
\end{equation}
shown in Fig.~\ref{fig:BakTrigger}. In analogy to Kam-II, we also show the signal spectrum of a fiducial SN that would lead to 1.5 events, or 2.2 events if the trigger efficiency were 100\%.

\subsubsection{SN~1987A Signal}

Around the time of SN~1987A, a bunch of 6~events was found with the properties shown in Table~\ref{tab:Baksan-data}, apparently around 30~s later than the IMB signal. However, while the clock synchronization with UT is usually $\pm2$~s, the clock was observed to have shifted forward by 54~s between February 17 and March 11 for unknown reasons. So the observed signal is probably contemporaneous with IMB and Kam~II. 

In the first BUST paper reporting the SN~1987A burst \cite{Alekseev:1987ej}, the event No.~0 in Table~\ref{tab:Baksan-data} was attributed to background on the basis of the time structure, i.e., it was assumed that the events No.~1--3 should have been near the SN bounce time, but such an identification on the basis of the assumed source properties is not justified as a prior assumption. In later publications, this event was never mentioned again. 

Loredo and Lamb \cite{Loredo:2001rx} have performed a detailed analysis of the SN~1987A signal, stressing that data should not be censored, but backgrounds should be identified based on the relevant likelihood, and for that reason have included events Nos.~13--16 in Kam-II as well as detailed background models for both Kam-II and BUST. On the other hand, they do not mention event No.~0 in BUST and thus must have censored it after all.

Unfortunately, it is not documented in which part of the detector the signals occurred. Using 200~t detector mass, the additional 70~t of outer segments contribute a background of 0.071~Hz and so the possible attribution to background of a given event strongly depends on location within the entire BUST. 

Censoring event No.~0, a 5-event burst within 9~s would randomly occur around 0.7/day. Therefore, on the basis of this signal alone, one could not claim the observation of core-collapse neutrinos. On the other hand, the observation around that time do constrain the properties of the source that caused the Kam-II and IMB events. One cannot ignore the BUST observations on the basis that they alone would not have been a convincing detection.

\subsubsection{Low-Energy Background}

\begin{figure}[ht]
    \centering
   \includegraphics[width=0.85\columnwidth]{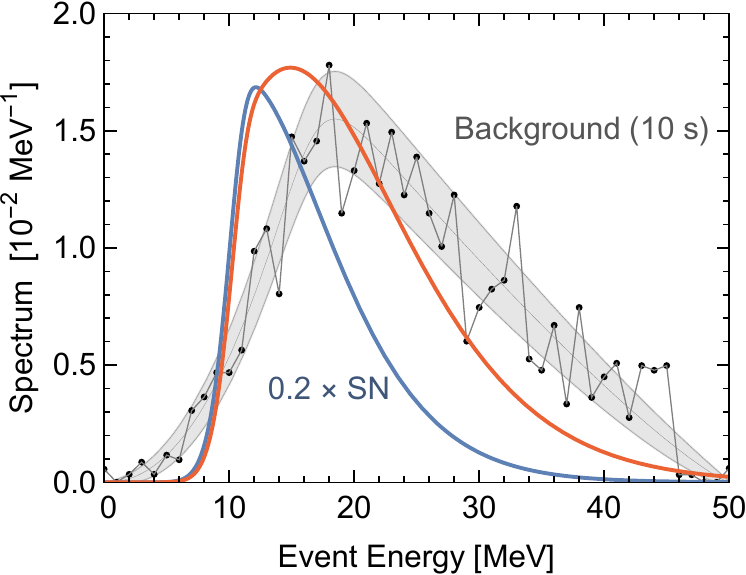}
    \caption{Background spectrum in BUST according to \cite{Loredo:2001rx} (black dots) for a time interval of 10~s, yielding a total 0.34 expected events. We also show our fit function Eq.~\eqref{eq:BBak} with its error given in Eq.~\eqref{eq:BBak-err} as a one-sigma gray band, i.e., the expected scatter of the data points for the ten-hour measurement period. (In Ref.~\cite{Loredo:2001rx} the corresponding error is shown for each data point.)  Also shown is a ``fiducial'' SN signal with an assumed Maxwell-Boltzmann spectrum with $T=3$~MeV (blue) and  4~MeV (red) that would yield a total of 0.93 (1.52) events. The illustrative predicted SN energy spectra do not include ``smearing'' by finite energy resolution. The background events refer directly to the reconstructed energy and no such smearing would be applied.}
    \label{fig:BakBKD}
\end{figure}

In analogy to Kam-II, Loredo and Lamb \cite{Loredo:2001rx} have provided the BUST background spectrum in their Fig.~2b, once more attributed to a private communication. We show these data points, referring to bins of width 1~MeV, in our Fig.~\ref{fig:BakBKD} as black dots. They also show errors for each point, which we find to be exactly reproduced by $\sqrt{{\rm Rate}}/0.1935$ in the units of that figure, so these errors once more are simply $\sqrt{n}$ fluctuations of the number of events in each such bin. From these errors we conclude that unexpectedly this background once more corresponds to a measurement period of 10~hours as in \hbox{Kam-II}. These data add to a background rate of 0.0345~Hz, in agreement with what Loredo and Lamb state, whereas the BUST paper states 0.033~Hz.

There is no objective way of smoothing these noisy data. Loredo and Lamb \cite{Loredo:2001rx} have used an interpolation based on a three-point running average, but have found that the exact representation of the background makes no practical difference. In this spirit we choose a somewhat arbitrary fit function
\begin{equation}\label{eq:BBak}
    B_{\rm Bak}= 10^{-3}~{\rm MeV}^{-1}~{\rm s}^{-1}\,
    \left[\left(\frac{13.3}{x}\right)^{11}
    +\left(\frac{20.9}{50-x}\right)^{6}\right]^{-1/5}
\end{equation}
for $0<x<50$ and zero otherwise, where $x=E_e/{\rm MeV}$. This function indeed integrates to 0.0345~Hz. In analogy to Kam-II, we can imagine to measure it in bins of width 1~MeV for ten hours, leading to data that should fluctuate in similar ways to those shown in Fig.~\ref{fig:BakBKD}. Then the $\sqrt{n}$ Gaussian fluctuation from bin to bin should be given by
\begin{equation}\label{eq:BBak-err}
    {\rm Err}(B_{\rm Bak})
    =\sqrt{\frac{B_{\rm Bak}}{1~{\rm MeV}\,10~{\rm h}}}.
\end{equation}
In Fig.~\ref{fig:BakBKD} we show $B_{\rm Bak}\pm{\rm Err}(B_{\rm Bak})$ as a gray band. Around 68\% of all data points should be found within this band, which is approximately the case.

For comparison, we also show the expected positron spectrum for a fiducial SN with $T=3$ and 4~MeV, respectively, that would yield a total of 0.93 (1.52) events. In contrast to Kam-II, signal and background cannot be separated by an energy cut.

\subsection{LSD (Mont Blanc)}

\label{sec:LSD}

A fourth instrument was the Liquid Scintillation Detector (LSD), located in the gallery of the Mont Blanc tunnel, between Italy and France~\cite{Badino:1984ww, Aglietta:1987it} at the approximate geographic location given in Table~\ref{tab:Detectors}. It has vertical depth of 5200~m~w.e., the atmospheric muon rate in the entire detector is only around 3.5/hour.

LSD was specifically built to search for a galactic SN burst with a typical assumed distance of 10~kpc. It worked similar to the BUST detector and indeed was a collaboration between researchers from the University of Torino (Italy) and the Institute of Nuclear Research in Moscow. LSD used 72 liquid scintillator modules of dimension $100\times 150\times 100 \,\rm cm^3$, arranged in three horizontal layers for a total mass of 90~tons. Each module was equipped with three PMTs of the same type as in BUST. The scintillator was prepared at INR with the same properties as the one used in BUST. 

The low-energy radioactive background from the rock is shielded with more than 200~tons of iron slabs, reducing the trigger rate to 0.012~Hz. In an IBD reaction $\bar\nu_e+p\to n+e^+$, on average after 170~$\mu$s the neutron is captured as $n+p\to d+\gamma$(2.2 MeV). Thanks to the iron shielding, this 2.2~MeV signature can also be detected. On average, this signature is seen in 40\% of all cases in the same module where the IBD has taken place.

\begin{figure}[ht]
    \centering
   \includegraphics[width=0.85\columnwidth]{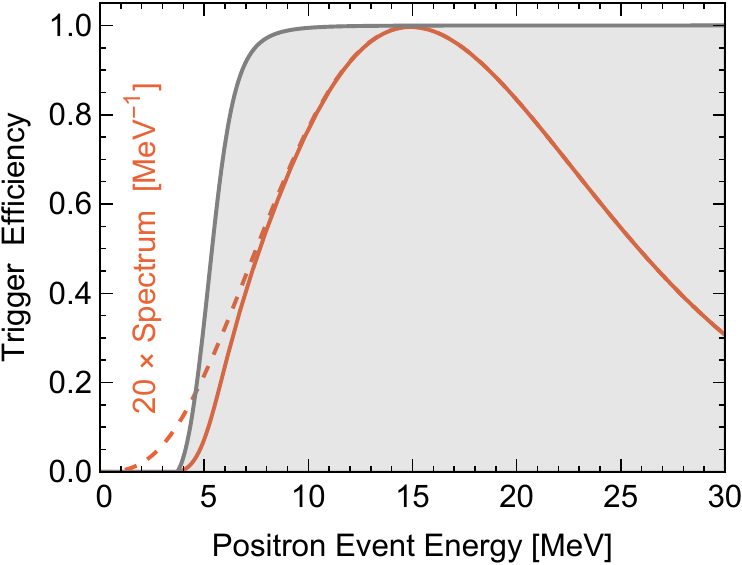}
    \caption{Trigger efficiency in LSD (Mont Blanc) following the fit function Eq.~\eqref{eq:LSDTrigger}. We also show the event spectrum from a fiducial SN with an assumed Maxwell-Boltzmann spectrum with $T=4$~MeV (red solid line) and for hypothetically 100\% trigger efficiency (dashed), leading to 0.962 (0.987) events. These illustrative energy spectra do not include ``smearing'' by finite energy resolution.}
    \label{fig:LSDTrigger}
\end{figure}

The combined signal of the three photomultipliers is recorded if they are in 3-fold coincidence within 150~ns. A 1~MeV energy loss yields on average 15 photoelectrons in one PMT. From Fig.~2 of Ref.~\cite{Saavedra:1987tz} we infer the approximate trigger efficiency ($x=E_e/{\rm MeV}$)
\begin{equation}\label{eq:LSDTrigger}
    \eta_{\rm LSD}= \left[1+\left(1.9684-1.124\,x+0.16\,x^2\right)^{-2.3}\right]^{-1/2.3},
\end{equation}
for $x>3.7$ and zero otherwise. We show this function together with the spectrum of a fiducial SN in Fig.~\ref{fig:LSDTrigger}. The expected number of events is about 2/3 that of BUST.

The LSD collaboration was the first to declare the possible discovery of SN neutrinos due to the detection of 5~events in an interval of 7~seconds, beginning at UT 2:52:36.79 \cite{Aglietta:1987it, Aglietta:1987we, Dadykin:1987ek}, almost five hours earlier than the other detectors which observed nothing special at the LSD time. The five events were in different segments, three of them interior. In one case, an accompanying 1.2~MeV pulse was observed 278~$\mu$s later, possibly caused by the subsequent neutron capture.

No similar high-multiplicity event was found during the entire LSD operation beginning around 1984 and ended with the devastating fire in the Mont Blanc Tunnel on March 24, 1999, although after 1988, the background had been further reduced.

The absolute event timing at LSD was accurate to $\pm2$~ms. At the time of the IMB burst, LSD did not observe any events, but it found one event at UT 7:36:00.5 (10 MeV) and one at 7:36:18.9 (9~MeV), i.e., 19 and 38~s after the first IMB event. Based on the LSD efficiency, the nonobservation of events at the IMB time provides some constraints on the neutrino signal properties.

The LSD collaboration has also studied time correlations with the background events in other detectors at the LSD time as well as two small gravitational wave detectors that were operating at that time, see e.g.\ \cite{Aglietta:1989tw, Chudakov:1988gu}, without leading to a tangible physical interpretation. No satisfactory explanation of the LSD burst has been proposed because even a double-bang SN scenario, apart from its astrophysical problems, leaves open why the other detectors saw nothing at the LSD time. Therefore, the community has settled for the LSD event as not related to SN~1987A.

Later, a ``double bang'' scenario based on a rotating collapsar model was proposed that suggests a large $\nu_e$ flux from deleptonization as a first burst, but few antineutrinos \cite{Imshennik:2004iya}. The authors pointed out that a large $\nu_e$ burst could have caused charged-current events in the iron shielding and subsequent production of multi-MeV gamma-rays that could have entered the scintillator and caused the observed events. The CC cross section on iron is very large relative to that on oxygen or carbon. Such a burst would have been missed by the other~detectors.

\subsection{Energy Resolution}
\label{sec:EnergyResolution}

The neutrino energy is estimated from the number of PMTs hit or the number of photoelectrons triggered in the PMTs. The fluctuation of this number is roughly Gaussian. Therefore, the energy resolution $\sigma_E$ must be proportional to $\sqrt{E_e}$. While the true energy resolution depends on the exact pattern of the Cherenkov ring and detector location of the event, we use an average resolution that we write in the form
\begin{equation}\label{eq:resolution}
    \sigma_E=\sqrt{E_\sigma E_e}.
\end{equation}
From the stated energy uncertainties of the events listed for the different detectors, we extract the values shown in Table~\ref{tab:Detectors}. 

In the spirit of using an average trigger efficiency for the entire detector, it is consistent to also use an average energy resolution in the maximum likelihood analysis instead of the resolution stated for each specific event.

For IMB, $E_\sigma\simeq1.3$~MeV for most events and the average is $1.36$. However, No.~5 is an outlier with $E_\sigma=2.25$ and taking it out reduces the average to 1.23. No.~6 is also a less extreme outlier with $E_\sigma=1.0$. Taking this one out as well increases the average to $1.27$ which is the number we will use. Of course, few-percent variations of $E_\sigma$ make no tangible difference.

For Kam-II, $E_\sigma\simeq0.63$~MeV using all listed 16 events. However, No.~7 is an outlier with $E_\sigma=1.81$ and taking it out reduces the average to 0.55. Taking also out Nos.~13--16, that were not formally published, increases the average again to 0.61. On the other hand, using only the 11~events usually attributed to the SN, including the outlier, one finds 0.74, very close to what was used in Ref.~\cite{Jegerlehner:1996kx}. We adopt $\sigma_E=0.60$ as a compromise between these results.

For BUST, we use all six listed events, providing $E_\sigma=0.72$~MeV, similar to Kam-II.

For LSD, the event energies for the five events of the ``early burst'' are very similar, on average 6.7~MeV. An average precision of energy determination of 15\% was stated, implying $E_\sigma=0.15^2\times6.7\,{\rm MeV}=0.15$~MeV, much better than in BUST. It means that they typically picked up about five times as many scintillation photos for a given event.

In our likelihood analysis, we construct the expected event spectrum for a given SN model and detector in that we begin with the $\bar\nu_e$ spectrum, apply the IBD cross section to obtain the positron spectrum, apply the detector efficiency curve, and finally smear the positron spectrum with a Gaussian that depends on positron energy $E_e$ as described here. In other words, we interpret the trigger efficiencies to be given as functions of the true $E_e$ that is subsequently reconstructed with imperfect precision.

We also mention that the Gaussian energy smearing of the event spectrum leaves the total event number unchanged, but also the average expected event energy. The smearing modifies the variance and higher moments of the event spectrum, which however are here not interesting given the sparse data. The energy smearing is a small effect in our overall analysis.

\subsection{Angular Distribution}
\label{sec:AngularDistribution}

\begin{figure}
    \centering
    \includegraphics[width=0.85\columnwidth]{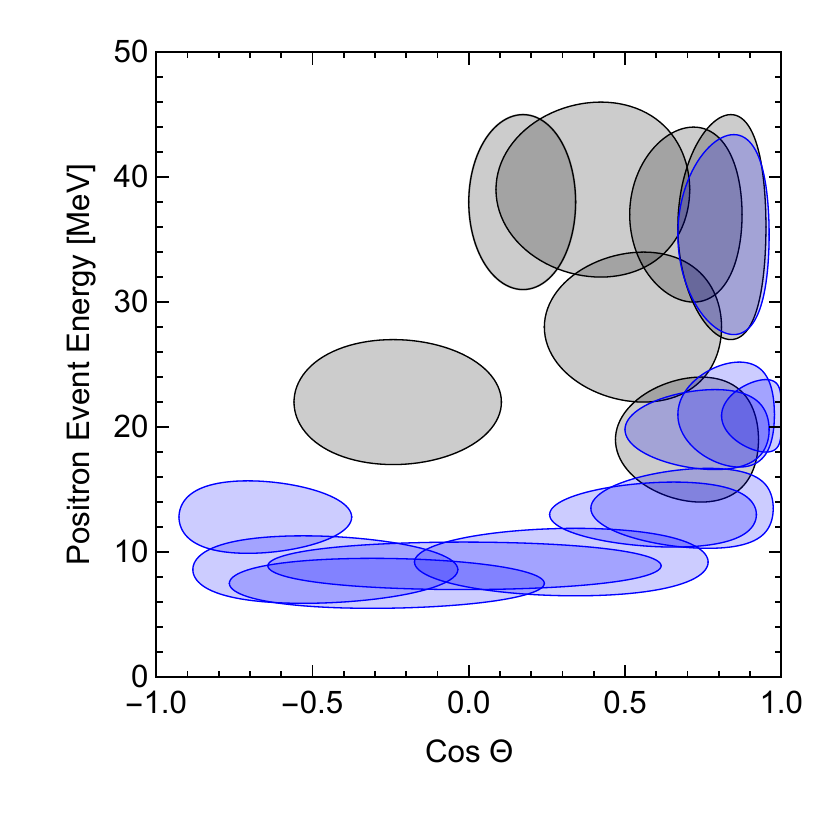}
    \caption{Distribution of scattering angles and energies of the IMB (black) and Kam-II (blue) events with approximate $1\sigma$ error regions, excluding the Kam-II event No.~6 which is likely background. Notice that the directional errors are for the scattering angle $\theta$, whereas here we show $\cos\theta$, leading to $1\sigma$ error ellipses that are strongly distorted near the forward or backward directions.
    \label{fig:angle-energy}}
\end{figure}

The main detection process is IBD that should provide a nearly isotropic event distribution, although at higher energies it is slightly forward biased (Section~\ref{sec:IBD}). In Fig.~\ref{fig:angle-energy} we show the distribution of the scattering angles and event energies for IMB and Kam-II, the only detectors that provide directional information. It has been noted many times that the distribution looks forward peaked, especially the higher-energy events \cite{LoSecco:1988hb, VanDerVelde:1989xb}. An early detailed analysis was performed by Krivoruchenko~\cite{Krivoruchenko:1988zg} who found a very small probability that the signal is a random manifestation of an essentially isotropic distribution. In principle, electron-scattering events are strongly forward peaked, so one can speculate, see e.g. Ref.~\cite{Costantini:2004ry}, that one or a few of the events could come from this process. However, the detected electron has less energy than the primary neutrino, so such events would be expected at lower, not at higher, energies. Moreover, the electron-scattering distribution should be more strongly forward peaked. In general, due to the low cross sections of neutrino-electron scattering at the relevant energy scale, the expected number of events from this process is always much smaller than $1$ for reasonable energies injected from the supernova.

Konishi et al.\ \cite{Konishi:1993np} have argued that the probability for a directional clustering is much larger if one considers spherical tests, i.e., considers angular clustering in all possible directions, not just a forward-backward bias relative to the source. On the other hand, if the events were clustered in some other direction would be less significant for the interpretation as some other process or some other particle. It would signify perhaps a detector bias as indeed IMB had because of the failed part of the detector which indeed introduced an azimuthal bias, but no strong forward-backward preference.

One idea held that the signal was not caused by neutrinos but instead some new $X^0$ bosons that scatter coherently on oxygen and thus generate the observed angular characteristic \cite{VanDerVelde:1989xb}. However, the required cross section is excluded by stellar energy-loss bounds from the reverse process \cite{Raffelt:1988gv}. 

No viable explanation other than a rare statistical fluctuation is available.

\subsection{Neutrino Path in the Earth}

The distance that neutrinos travel through the Earth matter affects the flavor regeneration effect. If the SN occurred at an elevation $\delta$ below the horizon at a given detector location, the distance traveled through Earth matter and the greatest depth from the surface are
\begin{subequations}
\begin{eqnarray}
    L_\nu&=&R_\oplus\,2\sin\delta,
    \\
    D_\nu&=&R_\oplus\,(1-\cos\delta),
\end{eqnarray}
\end{subequations}
where $R_\oplus=6371$~km is the average Earth radius. In this way we find these parameters given in Table~\ref{tab:Detectors}.

Slightly different distances are found in the literature. Lunardini and Smirnov \cite{Lunardini:2000sw} state 8535~km for IMB. The IMB Collaboration stated $\delta=42^\circ$, compared to our $42.4^\circ$, a rounding difference that would mostly explain the small difference to our 8592~km. The Kam-II Collaboration states $\delta=19.7^\circ$, identical to our value. Lunardini and Smirnov state a distance of 4363~km, compared to our 4295~km, so they would require $\delta\simeq20.1^\circ$, but still a rounding error to $20^\circ$ would go in the right direction. Of course, these are all small differences of no practical significance.

\onecolumngrid
{\ }

\section{Detection cross sections}
\twocolumngrid

\begin{figure}
    \centering
    \includegraphics[width=0.48\textwidth]{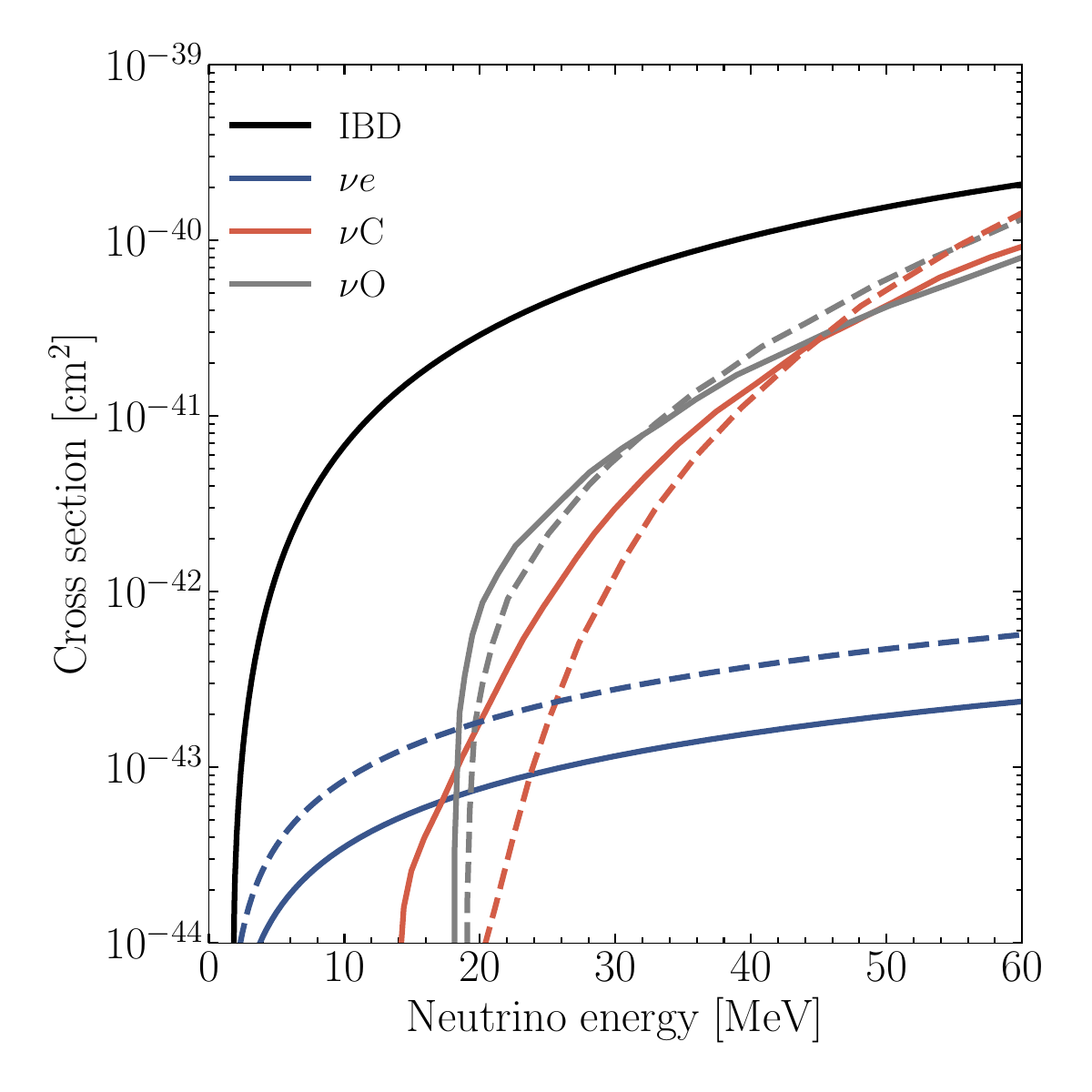}
    \caption{Charged-current neutrino cross sections per target particle. Solid lines for $\bar\nu_e$, dashed lines for $\nu_e$.
    }
    \label{fig:crosssections}
\end{figure}

\subsection{Inverse Beta Decay (IBD)}

\label{sec:IBD}

The primary channel for neutrino detection from SN 1987A was inverse beta-decay (IBD) $\bar{\nu}_e + p \to e^+ + n$ on the hydrogen nuclei of water or oil. Neglecting the recoil of the proton, the final positron has an energy $E_e=E_\nu-Q$, with $Q=1.293$~MeV, and it emits Cherenkov or scintillation light visible in the detector. At typical SN energies, $\bar{\nu}_\mu$ and $\bar{\nu}_\tau$ are kinematically unable to interact via charged current (CC). 

An analytic approximation for the IBD cross section, precise on the few per mille level for $E_\nu\alt300$~MeV, was given in Eq.~(25) of Ref.~\cite{Strumia:2003zx}
\begin{equation}\label{eq:cross_section}
    \sigma_{\bar\nu_ep}=10^{-43}~\mathrm{cm}^2~(E_\nu-Q) \sqrt{(E_\nu-Q)^2-m_e^2}\,E_\nu^\beta
\end{equation}
where
\begin{equation}
\beta=-0.07056+0.02018\ln(E_\nu)-0.001953\ln^3(E_\nu)
\label{eq:ebeta}
\end{equation}
and energies are in MeV. The cross section is shown in Fig.~\ref{fig:crosssections}. 

Near threshold, the usual lowest-order result is $0.952\times10^{-43}~\mathrm{cm}^2~(E_\nu-Q)\sqrt{(E_\nu-Q)^2-m_e^2}$ that was used in some SN~1987A analysis papers. The modification arises from recoil effects and in particular from weak magnetism \cite{Vogel:1983hi, Fayans:1985uej}. At the energies of the IMB events of around 40~MeV, the naive result overestimates the true cross section by some~35\%.

Recoil and weak magnetism also modify the angular distribution of the positron relative to the incoming $\bar\nu_e$. Using only the largest corrections, Vogel and Beacom find for the average scattering angle
\cite{Vogel:1999zy}
\begin{eqnarray}
    \langle\cos\theta\rangle&=&\frac{f^2-g^2}{3(f^2+3g^2)}\,v_e+
    \left[1+\frac{4(f+f_2)g}{3(f^2+3g^2)}\right]\frac{E_\nu}{m_N}
    \nonumber\\[2ex]
    &=&-0.034\,v_e+2.4\,\frac{E_\nu}{m_N},
\end{eqnarray}
where $f=1$, $g=1.26$, and $f_2=\mu_p-\mu_n=3.706$ are the different form factors and $v_e$ is the positron velocity. From their numerical result, one finds approximately
\begin{equation}
    \langle\cos\theta\rangle=-0.035+0.0024\,E_\nu/{\rm MeV}.
\end{equation}
At low energies, the cross section slightly favors backward directions, at roughly 15~MeV it is isotropic, and at the IMB energies of roughly 40~MeV, the forward bias is $\langle\cos\theta\rangle\simeq0.06$. 

If the angular distribution is roughly $(1+a\, \cos\theta)/2$, then $a=3\,\langle\cos\theta\rangle$. Therefore, the angular bias is not very small at the IMB energies.

\subsection{Oxygen and Carbon}

Above about 70~MeV, neutrino interactions in a water Cherenkov detector are dominated by CC reactions on oxygen of the form $\nu_e+{}^{16}{\rm O}\to e^- + {\rm X}$, where X is an excited nuclear state dominated by ${}^{16}{\rm F}^*$ \cite{Haxton:1987kc,Kolbe:2002gk,Scholberg:2012id} and a similar reaction for antineutrinos, where the dominant final state is ${}^{16}{\rm N}^*$. The final state $e^\pm$ retains memory of the initial neutrino energy.  While we do not account for this reaction in our analysis, we show the corresponding cross sections in Fig.~\ref{fig:crosssections}, extracted from Ref.~\cite{Scholberg:2012id}.
We do not show the corresponding neutral current cross sections, which are even smaller in the range of interest.

For the BUST and LSD scintillator detectors, the additional processes $\nu_e + {}^{12} {\rm C} \to {}^{12}{\rm N}^{(*)} + e^-$ and $\bar{\nu}_e + {}^{12} {\rm C} \to {}^{12}{\rm B}^{(*)} + e^+$ occur with an approximate threshold of
30~MeV. We also do not account for this reaction, which would be relevant at higher energies than the range we are interested in, and we limit ourselves to showing the corresponding cross sections in Fig.~\ref{fig:crosssections}.

\subsection{Neutrino-Electron Scattering}

Elastic scattering on electrons is another detection channel both in water Cherenkov and scintillator detectors. The cross section is well known \cite{Okun:1982ap} and shown for $\nu_e$ and $\bar\nu_e$ in Fig.~\ref{fig:crosssections}. For $\nu$ and $\bar\nu$ of the other flavors, it is a factor 6--7 smaller than for $\nu_e$. While there are ten electrons per water molecule, but only two protons, the detection rate through electron scattering is much smaller than IBD. Notice also that in IBD, the positron takes essentially the full $\bar\nu_e$ energy, whereas the electron energy in elastic scattering is significantly degraded on average.

The higher-energy detected SN~1987A neutrinos are somewhat forward peaked
(Appendix~\ref{sec:AngularDistribution}), suggesting at first a connection to elastic electron scattering. However, they are not forward peaked enough, so independently of the small cross section, this channel is not a good explanation for the angle distribution. In our analysis, finally we will only use IBD.


\onecolumngrid

\section{Likelihood Analysis}
\label{app:methods_comparison}
\twocolumngrid

\subsection{Likelihood function}

\label{eq:Likelihood}

Our goal is to compare the SN~1987A data with theoretical models and select among alternatives by means of a maximum likelihood analysis. Our models can be generic representations that depend on a number of parameters (such as the total emitted energy and the neutrino spectral shape) or can be the output of numerical simulations, which also depend on parameters such as the EoS, the progenitor mass, the presence or absence of PNS convection, as well as neutrino mixing parameters.

The only relevant detection reaction is inverse beta-decay (IBD). Therefore, the only relevant output of the models for a given choice of SN and neutrino parameters is the spectral $\bar\nu_e$ flux $\Phi_{\bar\nu_e}(E_\nu,t)$, in units of ${\rm cm}^{-2}\,{\rm s}^{-1}\,{\rm MeV}^{-1}$, at a given detector location. The spectral rate of visible positrons caused by the SN in each detector therefore is
(units ${\rm s}^{-1}\,{\rm MeV}^{-1}$)
\begin{equation}
    R_e^\mathrm{SN}(E_e,t)=
    \Phi_{\bar{\nu}_e}(E_\nu,t)\sigma_{\bar{\nu}_ep}(E_\nu)\Big|_{E_\nu=E_e+Q} n_p\eta(E_e),
\end{equation}
where $\sigma_{\bar{\nu}_ep}$ is the IBD cross section from Eq.~\eqref{eq:cross_section}, $\eta(E_e)$ the trigger efficiency, $n_p$ the number of target protons given in Table~\ref{tab:Detectors} for each experiment, and $Q=1.293$~MeV the neutron-proton mass difference.

The positron spectrum is not directly observable. Each positron excites a certain number of photomultipliers $N_{\rm hit}$ which translates nearly linearly to the positron energy with Poisson fluctuations. However, for the SN~1987A data, each event was individually studied, and the energy determined not by a global scaling to $N_{\rm hit}$, but determined on the basis of its vertex location and direction.  For example, event No.~7 in Kam-II is an outlier with a larger stated energy than behooves its $N_{\rm hit}$ and a larger stated energy uncertainty than corresponds to the Poisson fluctuation of $N_{\rm hit}$. In the small scintillator detectors BUST and LSD, the energy follows from the photoelectrons created in one (BUST) or three (LSD) PMTs observing a given cell of scintillator. 

Our main approximation is to neglect the dependence on the vertex position in the detector and positron direction. In other words, we do not use the energy uncertainty stated for each event, but rather a global Gaussian energy resolution according to Eq.~\eqref{eq:resolution}. Therefore, we smear the spectrum according to
\begin{equation}
    R_\mathrm{det}^\mathrm{SN}(E_{\rm det},t)=
    \int dE_e\, R_e^\mathrm{SN}(E_e,t)
    \frac{\exp\left[-\frac{(E_e-E_\mathrm{det})^2}{2\sigma_E^2}\right]}{\sqrt{2\pi}\sigma_E},
\end{equation}
where $\sigma_E=\sqrt{E_\sigma E_e}$ according to Eq.~\eqref{eq:resolution} with a parameter $E_\sigma$ for each detector as discussed in Sec.~\ref{sec:EnergyResolution} and listed in Table~\ref{tab:Detectors}. 

During the SN detection, some events may have been caused by background that we assume is constant during the short detection period. In IMB, there was no background, whereas for Kam-II and BUST we use the approximate spectra provided in Eqs.~\eqref{eq:BKAM} and \eqref{eq:BBak}. Therefore, overall
\begin{equation}
    R_\mathrm{det}(E_{\rm det},t)=R_\mathrm{det}^{\rm SN}(E_{\rm det},t)+B(E_{\rm det})
\end{equation}
is the expected event rate.

In comparing the observed data in a given detector with the model flux, we define a likelihood
\begin{equation}\label{eq:full_likelihood}
    \mathcal{L}=\exp\left[-\int R_\mathrm{det}\,dtdE_\mathrm{det}\right]\prod_i 
    R_\mathrm{det}(E_\mathrm{det}^i,t_i+\delta t).
\end{equation}
Here the index $i$ refers to the observed events, and we account for a time delay $\delta t$ which measures the temporal offset between the first measured event at each experiment and the SN bounce time, namely the zero time of our models. Due to the clock uncertainties, $\delta t$ is an independent parameter for each experiment and will be used as a nuisance parameter in our studies. 

The term in the exponent is the expected number of events, which for the SN source is an integral over the SN duration, in practice from the bounce time to the end of the simulation. The background contribution to the event number is $\mathcal{B}\,\tau$, where $\mathcal{B}=\int dE_{\rm det} B(E_{\rm det})$ is the energy-integrated background rate (units ${\rm s}^{-1}$). For $\tau$ we somewhat arbitrarily use the time between the first and last recorded event in a given detector. However, $e^{-\mathcal{B}\tau}$ is a factor in the likelihood that does not depend on SN model parameters and as such drops out from model comparison.  

The clock of LSD relative to IMB was fixed, but LSD had no events, obviating a $\delta t$ for this small detector. The absence of events also provides constraining information. In this case the likelihood is
\begin{equation}
    {\cal L}= \exp\left[-\int R_\mathrm{det}\,dtdE_\mathrm{det}\right],
\end{equation}
namely the probability of observing no event during the supernova explosion.

An additional complication is the possible dead-time effect after each recorded event, which is negligible except for IMB, where the detector cannot record fresh events for 35~ms after each trigger. Over the 5.6~s between first and last SN event, there were 15 recorded muons, but their exact timing is not documented. Otherwise one could set the trigger rate to zero for 35~ms after each muon. Moreover, after each of the 8~SN events, the detector is also dead, and this would cause a significant temporal anti-correlation if the event rate were large. Within the recorded burst duration of 5.6~s, there were 23 triggers, causing an overall dead time of $23\times35~{\rm ms}=0.805~{\rm s}$ or 14\% of 5.6~s. Therefore, as a pragmatic recipe we include an overall reduction factor 0.86 on the event rate, or equivalently, multiply the trigger efficiency with $0.86$. (The IMB paper \cite{IMB:1988suc} mentions an overall 13\% dead time effect, presumably based on a rounded duration of 6~s.)

Yet another complication arises for IMB because the trigger efficiency is small in the SN energy range so that the trigger uncertainty has a strong impact on the expected event rate. The trigger efficiency has an uncertainty of $\pm0.05$ and a relative uncertainty of $10\%$ is reported for the energy scale for which the efficiency is given. Therefore, we write the trigger efficiency in terms of two nuisance parameters $-1\leq\xi\leq+1$ and $-1\leq\zeta\leq+1$ in the form
\begin{equation}\label{eq:IMB-nuisance}
    \eta_{\rm IMB}^{\rm eff}(E_e)=0.05\,\xi+\eta_{\rm IMB}[(1+0.1\,\zeta)E_e],
\end{equation}
where $\eta_{\rm IMB}$ is the function given in Eq.~\eqref{eq:IMBTrigger} with a small modification. The behavior at small energies is continued to negative values and $\eta_{\rm IMB}^{\rm eff}(E_e)$ is taken to be zero whenever it would become negative. In this way one obtains the gray band in Fig.~\ref{fig:IMBTrigger} that continue to zero trigger efficiency without a cutoff. To speed up the computation, for the $\zeta$ parameter we only compute  the spectrum for $\zeta=-1,0,1$ and use a quadratic interpolation for intermediate values.

\subsection{Energy and Time-Integrated Analysis}

We also perform a separate time-integrated, energy-dependent analysis. To this end we define the detected event spectrum 
\begin{equation}
    N_{\rm det}'(E_{\rm det})=\int R_{\rm det}(E_{\rm det},t)\,dt.
\end{equation}
However, it is not necessarily obvious of how to treat the detector background, i.e, over which duration the background event spectrum should be taken, a question mainly relevant for Kam-II and BUST. Depending on context, we will make different assumptions about this question. The likelihood is
\begin{equation}\label{eq:energy_likelihood}
    \mathcal{L}_E=\exp\left[-\int  N_{\rm det}'dE_\mathrm{det}\right]
    \prod_i N_\mathrm{det}'(E_\mathrm{det}^i),
\end{equation}
where again in the exponential is the total event number.

For the energy-integrated, time-dependent analysis, we instead define the energy-integrated event rate
\begin{equation}
    \dot N_{\rm det}(t)=\int R_{\rm det}(E_{\rm det},t)\,dE_{\rm det}
\end{equation}
and
\begin{equation}\label{eq:time_likelihood}
    \mathcal{L}_t=\exp\left[-\int  \dot N_{\rm det} dt\right]
    \prod_i \dot N_\mathrm{det}(t_i+\delta t).
\end{equation}
is the likelihood for a given experiment.

\subsection{Test Statistic}

In comparing among different models, we use as a measure of comparison the test statistic (TS)
\begin{equation}\label{eq:TS_definition}
    \Lambda=-2\log(\mathcal{L}/\hat{\mathcal{L}}),
\end{equation}
where $\hat{\mathcal{L}}$ denotes the maximum likelihood among the chosen class of models. In the limit of a large number of data, the asymptotic distribution of this TS in the null hypothesis is known by Wilk's theorem to be a chi-squared distribution, with a number of degrees of freedom equal to the number of parameters of the model space. We use this result both for our phenomenological fits with a pinched energy distribution, and for the comparison with the SN models. The asymptotic distribution corresponding to a large number of data is of course not exactly realized in the measurements of SN~1987A; we will nonetheless use this approach, with the caveat that our confidence levels may be slightly overestimated because of the limited statistics.

\subsection{Goodness of Fit}

We also want to provide a measurement of the goodness of fit that a certain model provides to the data themselves. For this purpose, we define as a TS the logarithm of the likelihood to be tested itself
\begin{equation}\label{eq:abs_TS_definition}
    \chi=-2\log\mathcal{L}=2\int f(x)dx -2\sum_i \log\left[f(x_i)\right],
\end{equation}
where for simplicity of notation we denote by $f(x)$ the distribution of events according to a generic variable, which could be either energy or time or both. Notice that $f(x)$ is normalized to the total number of expected events.

Under the null hypothesis that the events are distributed according to the distribution to be tested, the expected value of the TS is
\begin{equation}
    \langle\chi\rangle=2\int f(x)dx-2\int f(x) \log f(x)\, dx.
\end{equation}
The spread of the TS around this value $\sigma_\chi^2=\langle \chi^2\rangle-\langle\chi\rangle^2$ can be evaluated to be
\begin{equation}
    \sigma^2_\chi= 4\int f(x)\log^2 f(x)\,dx.
\end{equation}
For values of $\chi$ not too far from the expected $\langle \chi\rangle$, the distribution of $\chi$ can be approximated as a normal distribution. As an indicative measure of the goodness of fit, we will use this approximation; notice that this does not rely on the assumption of large statistics, but only on the assumption that the observed $\chi$ does not lie too far from the expected $\langle\chi\rangle$.

\onecolumngrid

\vskip36pt

\section{First second of emission}
\twocolumngrid
\label{app:first_second}

Recently, a large heterogeneous suite of SN models with a variety of physical assumptions was compared with the neutrino measurements of SN 1987A \cite{Li:2023ulf}. The analysis was limited to the ``first second'' of emission with the philosophy that the models tend to exhibit basic agreement in this short period that largely excludes PNS cooling, so this would be a useful test of theory vs.\ observation, and moreover, many multi-D models in the literature have not been evolved far beyond this period. Of course, we have taken the opposite approach and primarily focused on PNS cooling. Their main finding is a strong tension between data and models, whereas we have stated that we do not see a significant discrepancy between our models and the first-second data. What is the origin of the difference in conclusion? Our following discussion refers to version~1 of the arXiv posting of Ref.~\cite{Li:2023ulf}.

As a first remark, they have ignored the BUST data, whereas the IMB data during this period is too sparse to play any role. So effectively, only the first five events in Kam-II are used, i.e., the period of 0--0.507\,s after the first event; No.~6 was interpreted as background and left out, and No.7 (which they call No.~6) at 1.541\,s is already beyond their test window, in which No.~1 was somewhat shifted in time relative to the SN bounce time. Ref.~\cite{Li:2023ulf} identifies two main tensions with these five Kam-II events in that the models overpredict the number of events and their average energy. 

Visual inspection of our Fig.~\ref{fig:differential_distributions} reveals that the Kam-II data indeed do not look like a ``typical'' representation of the model, and interpreting No.~6 as background, in particular leaves Nos.~3 and 4 looking as ``outliers'' relative to the signal prediction, but also relative to the other data. Of course, one or both could be background, although this would have to be a significant upward fluctuation. The question of these low-energy events was discussed a long time ago in a dedicated paper~\cite{Costantini:2006xd}. Ref.~\cite{Li:2023ulf} does not include background in their analysis, but assumes a sharp cut between signal and background at $E_e=7.5$~MeV, leaving No.~3 exactly on threshold. We mention that these two events are located directly at the detector's cylindrical surface near the floor, making a background interpretation somewhat more plausible.

It is clear that this feature of the Kam-II data, which is the main source of discrepancy with models, cannot be explained by plausible SN models as it would require a sudden dip of the $\bar\nu_e$ signal or its energies. Our interpretation is that of a local signal or background fluctuation of the sparse data. While it may look locally significant, such fluctuations somewhere in the data can well be expected in the spirit of the ``look-elsewhere effect.'' The overall SN~1987A data are full of anomalies, like the notable deviation from the isotropy of events in both detectors we have discussed in Sec.~\ref{sec:AngularDistribution}. However, it is hard to make them statistically objective in the absence of a definition of what is an anomaly.

A goodness-of-fit test of data vs.\ models should include simultaneously the time and energy distributions, as a two-dimensional Kolmogorov-Smirnov test does. Instead, in Ref.~\cite{Li:2023ulf} the number of events and their energies were separately tested, which means, among other issues, that the event energies of a SN model for the first 1.5\,s were compared with data within an 0.5\,s period. The absence of events within 1\,s of the test period is in itself a large fluctuation.

The number of events predicted in the first second is found by Ref.~\cite{Li:2023ulf} to be larger than the observed one for all of their models. However, even within these models, there are some that are in a less than $2\sigma$ tension with the data, so this finding is not statistically significant. Within our set of models, not only do we not obtain a significant tension, but we even find models that \textit{underpredict} the number of events in the first second, as shown in Fig.~\ref{fig:cumulative_distributions}, depending mostly on the final NS mass. 

A more significant tension is found in Ref.~\cite{Li:2023ulf} with the neutrino energy spectrum at Kam-II, driven by the low energies of events Nos.~3 and 4. A simple test to quantify this impression would be the average event energies. However, Ref.~\cite{Li:2023ulf} opted for a one-dimensional Kolmogorov-Smirnov test for the spectral shape of the observed neutrinos compared with the theoretically expected one. For each model,
the predicted spectrum was determined from the entire simulation period; this varies for different models between 0.38--1.37\,s. For all of the models, the p-values from these KS tests (determined by a Monte Carlo) are lower than 5\%, which is interpreted as a statistically significant tension between all of the SN models analyzed with the data.

The time and energy structure of the early Kam-II data implies that this discrepancy must be present in any SN model that is even only vaguely similar to those used in Ref.~\cite{Li:2023ulf}, which certainly applies to our models. We illustrate this point by following the approach of Ref.~\cite{Li:2023ulf} for the event spectrum, although with some differences in detail. We include the detector background and do not impose an unphysical threshold at 7.5~MeV on the signal prediction, but instead follow our usual treatment for predicting the detection signal from a given model. We then perform Kolmogorov-Smirnov tests on the Kam-II data, comparing them with all of our models, taken only up to a time $t_\mathrm{cutoff}$, and we let the latter vary. We neglect any possible offset between the first Kam-II event and the bounce time of the models. (Instead, in Ref.~\cite{Li:2023ulf} the offset was chosen to minimize the discrepancy.)

\begin{figure}
    \vskip6pt
    \centering
    \includegraphics[width=0.90\columnwidth]{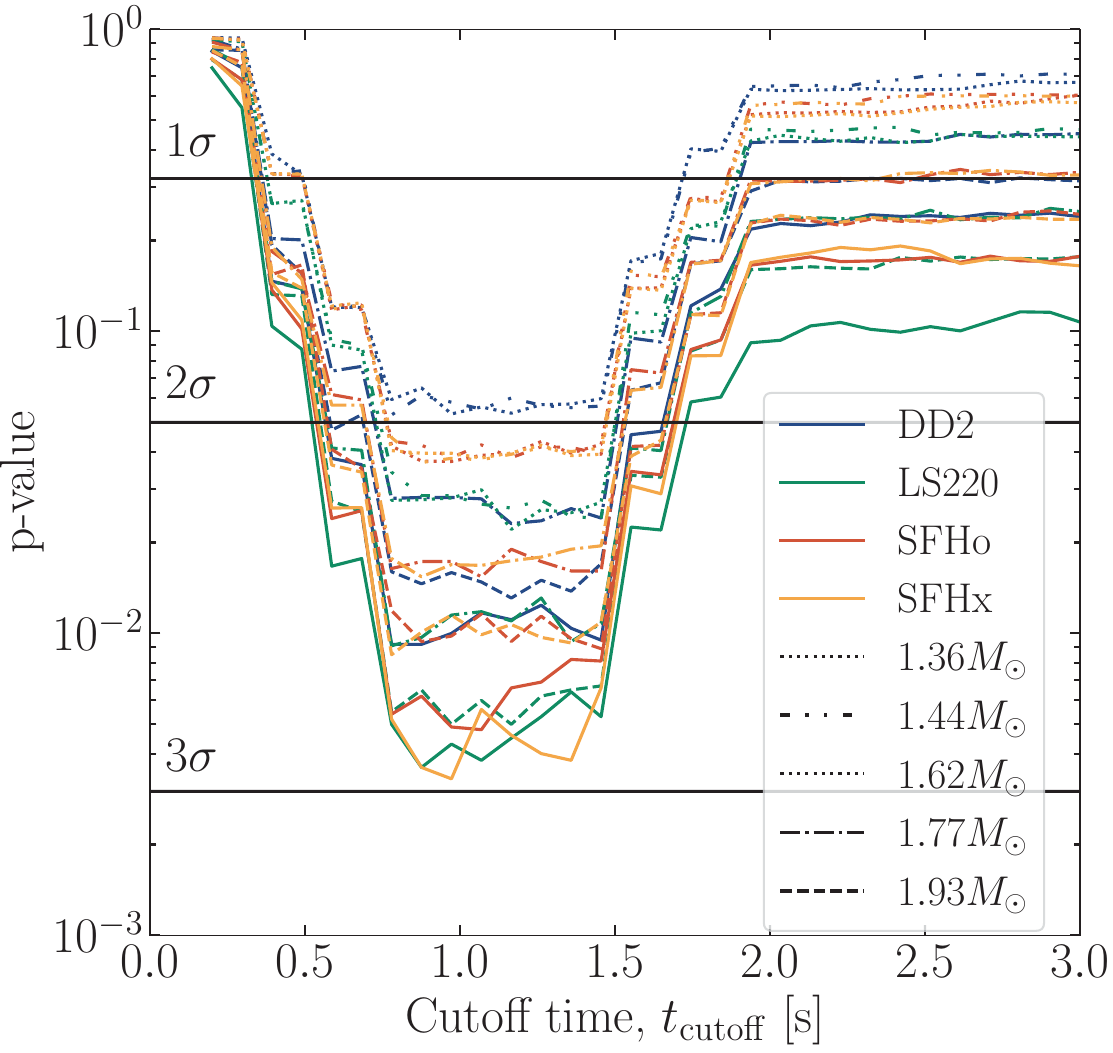}
    \caption{Results of the Kolmogorov-Smirnov test for the spectral shape of Kam-II neutrinos with our suite of models, up to a maximum time $t_\mathrm{cutoff}$. We assume no offset between the first event and $t=0$ of our models.
    }
    \label{fig:ks_pvalues}
\end{figure}

\vbox{
In practice, for each model we compare the predicted cumulative event distribution as a function of positron energy with the observed cumulative event distribution, using as a test statistic (TS) the maximum distance between the two distributions. For each model, we then perform a Monte Carlo sampling from the theoretical distribution of 10000 samples with the same size as the observed one, and extract the expected distribution of the TS; we then define as p-value the probability that a data sample extracted from the expected distribution would sport a TS larger than the observed value. Figure~\ref{fig:ks_pvalues} shows the corresponding p-values as a function of $t_\mathrm{cutoff}$ for our entire suite of models.
}

Cutting the signal at $t_\mathrm{cutoff}$ around $1$~s indeed leads to a tension with the data, with the p-values for most models being around or lower than the $2\sigma$ threshold of $5\%$. However, crucially, if one extends the cutoff time $t_\mathrm{cutoff}$, the p-value rapidly returns to reasonable values which are not in tension with the data. We recall that there are no Kam-II events in the period $0.5<t<1.5$\,s, explaining the 1\,s width of the depression in these curves. While none of our models, with our procedure, crosses the $3\sigma$ threshold, the general feature observed in Ref.~\cite{Li:2023ulf} is unavoidably present as it is caused by a real feature of the data. 

However, given that the global p-value even at $t_\mathrm{cutoff}\simeq 2$\,s is already large, we interpret this effect as a local fluctuation of the data, and the relatively small p-values as a local significance that should not be over-interpreted if one pays attention to the look-elsewhere phenomenon.

\onecolumngrid
\section{Gamma Distribution}
\twocolumngrid
\label{sec:Gamma}

Neutrinos emerging from a CCSN typically follow a quasi-thermal spectrum that can be approximately represented by a so-called Gamma distribution, colloquially sometimes referred to as $\alpha$-fit
\cite{Keil+2003,Tamborra:2012ac}
\begin{equation}
    f_\nu(\epsilon_\nu)=
    \frac{(1+\alpha)^{1+\alpha}}{\Gamma_{1+\alpha}} 
    \frac{\epsilon_\nu^\alpha}{\bar\epsilon^{\,1+\alpha}} e^{-(1+\alpha)\epsilon_\nu/\bar\epsilon},
\end{equation}
where $\Gamma_x$ is the Gamma function at argument $x$, not to be confused with our Gamma distribution. This function is normalized and has the first two moments
\begin{eqnarray}
    \langle\epsilon\rangle&=&\bar\epsilon,
    \\
    \langle\epsilon^2\rangle&=&\frac{2+\alpha}{1+\alpha}\,
    \bar\epsilon^2.
\end{eqnarray}
Conversely, for a given empirical or numerical result for $f_\nu(\epsilon_\nu)$, the parameter $\bar\epsilon$ is given as the average energy $\langle\epsilon_\nu\rangle$ and the ``pinching parameter'' as
\begin{equation}
    \alpha=\frac{-2\langle\epsilon\rangle+\langle\epsilon^2\rangle}
    {\langle\epsilon\rangle-\langle\epsilon^2\rangle}.
\end{equation}
Besides the overall normalization, the approximate representation is chosen to provide the exact first and second energy moments.

A function of this form is a surprisingly good fit for the instantaneous emerging SN neutrino fluxes \cite{Tamborra:2012ac}. A Maxwell-Boltzmann spectrum corresponds to $\alpha=2$, whereas $\alpha>2$ to ``pinching,'' where the variance is smaller. The instantaneous spectra are more or less strongly pinched, whereas the time-integrated ones come fairly close to the Maxwell-Boltzmann case.

The neutrino fluxes coming from our simulations are represented by their instantaneous luminosity (energy flux) as well as the instantaneous $\bar\epsilon$ and $\alpha$, separately for all six species.

The time-integrated spectra are obtained by expressing the instantaneous ones through the tabulated parameters $(L,\bar\epsilon,\alpha)$, integrating over time, and finding the corresponding parameters for the integrated spectrum.

\onecolumngrid


\section{Flavor Conversion}
\label{sec:FC}
\twocolumngrid

\subsection{Flavor-Dependent Spectral Hierarchy?}

Flavor conversion of SN~1987A neutrinos is a topic with a chequered history. Numerical SN modeling was still in its infancy at that time. The now-standard neutrino mechanism for driving the explosion had been discovered and explained by Bethe and Wilson only a few years earlier \cite{Bethe+1985}. The standard paradigm at that time held that the heavy-lepton neutrino species, collectively called $\nu_x$, had much larger average energies, perhaps twice higher, than the electron-flavored ones.\footnote{Several examples for such models and their flavor-dependent spectral properties can be found in Table~I of Ref.~\cite{Smirnov:1993ku}.} An example for the time-dependent neutrino signal representative of that generation of numerical models has circulated even in the recent literature under the name of ``Livermore model'' \cite{Totani:1997vj}. Moreover, the only solar neutrino data came from the Homestake experiment and the solar neutrino problem had several flavor-conversion solutions based on the MSW effect, which itself had just been discovered around the same time \cite{Mikheyev:1985zog, Mikheev:1986wj}. The low energies of the SN~1987A events clearly excluded the Large Mixing Angle (LMA) solution \cite{Smirnov:1993ku}, which later turned out to be the correct one \cite{Workman:2022ynf}.
The hopes for neutrinos from the next galactic SN as a laboratory for flavor conversion largely rested on the expectation of a large flavor dependence of the flux spectra.

The standard physical explanation for the expected large $\nu_x$ energies was that heavy-flavor neutrinos decouple deeper inside, where the ambient temperature is larger and thus the $\nu_x$ should emerge with larger spectral energies. However, $\nu_x$ continue to scatter on nucleons without significant energy exchange even after pair processes, then taken to be $e^-e^+$ annihilation, had frozen out. The quadratic energy dependence of the scattering cross section lets lower-energy $\nu_x$ pass the ``scattering atmosphere'' more easily, tilting the spectrum to lower energies relative to that at the ``energy sphere.'' Overall, simple transport models suggest that there could not be a large hierarchy of average energies \cite{Raffelt:2001kv}. The appearance of such large hierarchies, e.g.\ in the Livermore model, probably has to be attributed to a simplified neutrino transport scheme and incomplete sets of neutrino interaction processes. For example, heavy-flavor pairs are mostly produced by nucleon bremsstrahlung, not by $e^-e^+$ annihilation. 

Be that as it may, present-day SN simulations produce flavor-dependent fluxes that are broadly consistent between all groups and with our results. On the other hand, flavor-conversion is not included in numerical simulations. One justification is that large matter effects in the dense SN medium ``demix'' neutrinos: The propagation and interaction eigenstates very nearly coincide.

\subsection{Fast Flavor Conversion (FFC)}

This logic was challenged by the insight that neutrino-neutrino refraction can cause collective flavor conversion potentially even in high-density regions. In particular, the phenomenon of fast flavor conversion (FFC) features a classical instability of the mean field of neutrino flavor driven by crossed angular distributions of the flavor lepton number fluxes. FFC has been a topic of vivid discussion in the recent literature, ranging from conceptual issues (mean-field treatment vs.\ many-body quantum entanglement) to phenomenological studies concerning the neutrino angle distribution and concerning the impact on SN explosion physics and nucleosynthesis in SNe and NS mergers. Recent parametric studies, assuming a maximum effect at densities below some chosen threshold value, reveal that the modified neutrino spectral fluxes can both help and hinder the explosion~\cite{Ehring+2023a, Ehring+2023}. References to the current state of the FFC discussion are provided in these papers.

However, currently there exists no practical and well-justified recipe for reliably implementing FFCs in SN simulations. In 1D models, the conditions for FFC probably would not arise \cite{Tamborra:2017ubu} and so in this sense our models are self-consistent. However, if FFC is a real physical effect, it would have to be included in some effective way in analogy to convection that also would not arise in spherical symmetry, but is included according to a mixing-length prescription. Therefore, the flavor dependence of the output of current-generation simulations is not well justified.

\subsection{Spectral Swaps by Collective Effects}

If the FFC phenomenon did not operate after all and neutrinos would emerge with their ``naive'' spectra from the decoupling regions, the ``neutrino spheres'', flavor conversion would still occur on the way out. Neutrino-neutrino refraction would still be important and can drive what nowadays would be called ``slow'' flavor conversion.\footnote{For a review of SN flavor oscillations effects circa 2015, before the era of FFC, see Ref.~\cite{Mirizzi+2016}.} Here the mean-field instability would be related to the species-dependent spectra (not the angular distribution), depends on neutrino masses, and operates on much larger scales, leading to flavor swaps of some part of the spectrum. On the other hand, there are many unresolved complications related to instabilities on smaller scales or the ``halo effect'', the refractive effect of back-scattered neutrinos that have a large refractive effect based on their broad angular distribution relative to those emitted from the small PNS surface. The examination of these phenomena led to the study of FFC without reaching a final conclusion about the impact of slow conversions. In general, both slow and fast conversions would be expected to occur and the two phenomena do not necessarily factorize. Based on the current state of the art, the outcome of any kind of collective neutrino flavor conversion cannot be asserted with confidence.

\subsection{MSW Conversion}

Whatever neutrino fluxes emerge from the neutrino spheres, and their collective modification, still need to stream away through the density gradient provided by the progenitor star. Here they would likely undergo adiabatic flavor evolution, which in general is a three-flavor effect. In general, of course, collective and MSW evolution may not be spatially separated, depending on the progenitor's matter profile.

Considering only the matter effect, if the propagation begins at sufficiently large density, the neutrino flavor eigenstates are essentially propagation eigenstates and emerge as such from the surface of the star, then corresponding to mass eigenstates. These would propagate as such to the detector and then need to be projected back to the detected $\bar\nu_e$ state. A neutrino state beginning as $\bar\nu_e$ at high density would emerge as $\bar\nu_1$ (the mass eigenstate No.~1) if the neutrino mass ordering is normal, and as $\bar\nu_3$ if it is inverted~\cite{Dighe:1999bi,Vitagliano:2019yzm}.\footnote{This is seen most easily in the level diagram Fig.~5 of Ref.~\cite{Dighe:1999bi}. Notice, however, that the asymptotic behavior at large density is not correctly shown, cf.\ Fig.~35 of Ref.~\cite{Vitagliano:2019yzm}.}

Assuming that the heavy-lepton flavor spectra are similar and can be expressed as a common $\bar\nu_x$ spectrum, the flux detected by IBD is
\begin{equation}
    F_{\bar\nu_e}^{\rm det}=\bar p F_{\bar\nu_e}^{0}+(1-\bar p)\, F_{\bar\nu_x}^{0},
\end{equation}
where the fluxes ``0'' are the ones produced at high density in the SN, before adiabatic conversion, and $\bar p$ is the $\bar\nu_e$ survival probability. Depending on the mass ordering, one finds~\cite{Dighe:1999bi}
\begin{equation}
    \bar p=\begin{cases}
        \cos^2\theta_\odot\simeq 0.68 &\hbox{Normal},\\
        |U_{e3}|^2~~\simeq         0.02 &\hbox{Inverted}.\\
    \end{cases}
\end{equation}
In this scenario, one would never expect to observe the original $\bar\nu_e$ flux. In the main text, we will also refer to $\bar p$ as the ``swap parameter.''

\subsection{Earth Effect}

Whatever happens in terms of flavor evolution within the SN and during propagation through the progenitor star, the distance to the SN is so large that the different mass eigenstates would have decohered even if they do not emerge as mass eigenstates. In this sense, what arrives at Earth is an incoherent superposition of fluxes of mass eigenstates. These evolve through the Earth matter to the detectors over different distances 
(Table~\ref{tab:Detectors}) so that the different detectors could not have measured the same fluxes. In the above MSW scenario, such Earth matter effects would primarily show up for the case of normal mass ordering. This effect can cause different spectra measured at different detectors and might have explained the spectral tension between the Kam-II and IMB measurements~\cite{Lunardini:2000sw}. 

However, the solar mass difference is today known to be larger than thought at that time, reducing the impact of the Earth matter effect. (The propagation through the Earth is more similar to vacuum.) Moreover, the expected flavor-dependent spectral differences are much smaller. Therefore, while the Earth effect could still be of interest in the context of a high-statistics future SN neutrino observation (cf.\ Fig.~39 of Ref.~\cite{Mirizzi+2016}), its impact on the SN~1987A signal interpretation is minimal for any realistic scenario.

\subsection{Summary}

The flavor-dependent output of current-generation numerical SN models is not well justified due to a lack of theoretical understanding and practical algorithm for implementing collective neutrino flavor evolution. It is also not obvious if an average flux between the $\bar\nu_e$ and $\bar\nu_x$ fluxes of our models is a better approximation or even if such a procedure can bracket the possible extremes. On the other hand, the flavor-dependent impact on the SN~1987A signal interpretation is fairly small. Therefore, in the absence of practical alternatives, we often show results as if the signal were caused by $\bar\nu_e$ alone (no flavor conversion) or by $\bar\nu_x$ alone (full flavor swap). Our conclusions do not strongly depend on this question.

\onecolumngrid


\begin{table}[ht]
\section{Supernova Models. Additional Tables}
  \caption{Time-integrated neutrino emission properties for all of our CCSN models and all neutrino species. For each $\bar\nu$ flavor also characteristic properties of the predicted signals in the three detectors are listed, assuming that all measured events are caused by IBD reactions of this flavor alone, provided complete flavor conversion took place.  The detailed numerical neutrino outputs are available at the Garching Core-Collapse Supernova Archive \cite{JankaWeb}.
 \label{tab:Neutrino-flux-properties}}
 \vskip4pt
    \begin{tabular*}{\textwidth}{@{\extracolsep{\fill}}llllllllllllllllll}
    \hline\hline
    &&&&&&\multicolumn{4}{l}{Kam-II}&\multicolumn{4}{l}{IMB}&\multicolumn{4}{l}{BUST}\\
Model      & &$E_{\rm tot}^{\rm end}$& $\langle \epsilon_\nu\rangle$&$\alpha$&$\tau_E$&$N$&$\langle\epsilon_e\rangle$&$\tau$&$\delta t$&$N$&$\langle\epsilon_e\rangle$&$\tau$&$\delta t$&$N$&$\langle\epsilon_e\rangle$&$\tau$&$\delta t$\\
           & & [B]         & [MeV]                        &        &[s]&&[MeV]&[s]&[s]&&[MeV]&[s]&[s]&&[MeV]&[s]&[s]\\
\hline
1.36-DD2&$\bar{\nu}_e$&28.79&12.47&2.95&5.05&7.85&18.64&4.68&0.06&2.42&28.01&4.39&0.08&0.85&19.09&4.64&0.08\\
&$\bar{\nu}_\mu$&30.86&12.47&2.33&5.27&8.77&19.85&4.81&0.03&3.36&29.54&4.29&0.04&0.96&20.28&4.77&0.04\\
&$\bar{\nu}_\tau$&30.16&12.44&2.33&5.28&8.55&19.82&4.83&0.03&3.27&29.54&4.32&0.04&0.93&20.26&4.79&0.04\\
&$\nu_e$&32.43&9.70&3.09&4.86&-&-&-&-&-&-&-&-&-&-&-&\\
&$\nu_\mu$&29.09&11.97&2.38&5.28&-&-&-&-&-&-&-&-&-&-&-&\\
&$\nu_\tau$&29.01&11.97&2.38&5.29&-&-&-&-&-&-&-&-&-&-&-&\\
1.36-LS220&$\bar{\nu}_e$&28.98&12.38&2.45&6.35&8.09&19.43&4.47&0.06&2.89&29.04&3.23&0.08&0.88&19.87&4.29&0.08\\
&$\bar{\nu}_\mu$&32.56&12.72&1.97&6.44&9.78&21.12&4.43&0.03&4.52&30.99&2.99&0.04&1.07&21.53&4.24&0.04\\
&$\bar{\nu}_\tau$&31.69&12.65&1.95&6.53&9.46&21.08&4.50&0.03&4.36&30.98&3.02&0.04&1.04&21.49&4.31&0.04\\
&$\nu_e$&33.16&9.85&2.81&6.07&-&-&-&-&-&-&-&-&-&-&-&\\
&$\nu_\mu$&30.43&12.14&2.02&6.61&-&-&-&-&-&-&-&-&-&-&-&\\
&$\nu_\tau$&30.40&12.13&2.02&6.59&-&-&-&-&-&-&-&-&-&-&-&\\
1.36-SFHo&$\bar{\nu}_e$&31.00&12.57&2.83&5.97&8.62&18.99&5.40&0.06&2.83&28.42&4.87&0.09&0.94&19.42&5.35&0.08\\
&$\bar{\nu}_\mu$&33.81&12.68&2.26&6.32&9.89&20.32&5.58&0.03&4.08&30.08&4.72&0.08&1.08&20.74&5.51&0.04\\
&$\bar{\nu}_\tau$&33.01&12.65&2.25&6.31&9.62&20.29&5.62&0.03&3.95&30.07&4.77&0.08&1.05&20.71&5.54&0.04\\
&$\nu_e$&34.85&9.92&3.03&5.74&-&-&-&-&-&-&-&-&-&-&-&\\
&$\nu_\mu$&31.86&12.16&2.30&6.33&-&-&-&-&-&-&-&-&-&-&-&\\
&$\nu_\tau$&31.72&12.16&2.31&6.33&-&-&-&-&-&-&-&-&-&-&-&\\
1.36-SFHx&$\bar{\nu}_e$&31.09&12.58&2.83&6.05&8.65&19.00&5.47&0.06&2.85&28.43&4.91&0.09&0.94&19.43&5.41&0.08\\
&$\bar{\nu}_\mu$&33.99&12.68&2.26&6.43&9.93&20.33&5.67&0.03&4.10&30.10&4.75&0.08&1.09&20.75&5.59&0.04\\
&$\bar{\nu}_\tau$&33.19&12.65&2.25&6.42&9.67&20.30&5.70&0.03&3.97&30.09&4.80&0.08&1.06&20.72&5.63&0.04\\
&$\nu_e$&34.88&9.92&3.03&5.80&-&-&-&-&-&-&-&-&-&-&-&\\
&$\nu_\mu$&32.04&12.16&2.31&6.43&-&-&-&-&-&-&-&-&-&-&-&\\
&$\nu_\tau$&31.89&12.16&2.31&6.44&-&-&-&-&-&-&-&-&-&-&-&\\
\hline
 \end{tabular*}
 \vskip-3pt
 \vskip4pt
\hbox{\hbox to3em{\hfil$E_{\rm tot}^{\rm end}$~~}\hbox to7.5cm{Total energy emitted in neutrinos up to $t_{\rm end}$\hfil}
\hbox to3em{\hfil$\langle \epsilon_\nu\rangle$~~}\hbox to8.0cm{Average neutrino energy of the time-integrated spectrum\hfil}}
\vskip1pt
\hbox{\hbox to3em{\hfil$\alpha$~~}\hbox to7.5cm{Pinch parameter of the time-integrated spectrum\hfil}
\hbox to3em{\hfil$\tau_E$~~}\hbox to8.0cm{Period over which 95\% of $E_\mathrm{tot}^{\rm end}$ are emitted\hfil}}
\vskip1pt
\hbox{\hbox to3em{\hfil$N$~~}\hbox to7.5cm{Number of detected events\hfil}
\hbox to3em{\hfil$\langle\epsilon_e\rangle$~~}\hbox to8cm{Average energy of the detected positrons\hfil}}
\vskip1pt
\hbox{\hbox to3em{\hfil$\tau$~~}\hbox to7.5cm{Period over which 95\% of the total events are detected\hfilneg}
\hbox to3em{\hfil$\delta t$~~}\hbox to8cm{Best-fit offset time between bounce and first detected event\kern-3em\hfil}
}
\end{table}

\clearpage

\begin{table}[ht]
\vskip-5pt
TABLE~\ref{tab:Neutrino-flux-properties} (continued). Time-integrated neutrino flux properties of our supernova models.
\vskip4pt
    \begin{tabular*}{\textwidth}{@{\extracolsep{\fill}}llllllllllllllllll}
    \hline\hline
    &&&&&&\multicolumn{4}{l}{Kam-II}&\multicolumn{4}{l}{IMB}&\multicolumn{4}{l}{BUST}\\
Model      & &$E_{\rm tot}^{\rm end}$& $\langle \epsilon_\nu\rangle$&$\alpha$&$\tau_E$&$N$&$\langle\epsilon_e\rangle$&$\tau$&$\delta t$&$N$&$\langle\epsilon_e\rangle$&$\tau$&$\delta t$&$N$&$\langle\epsilon_e\rangle$&$\tau$&$\delta t$\\
           & & [B]         & [MeV]                        &        &[s]&&[MeV]&[s]&[s]&&[MeV]&[s]&[s]&&[MeV]&[s]&[s]\\
\hline
1.44-DD2&$\bar{\nu}_e$&32.26&12.48&2.90&5.35&8.84&18.75&4.96&0.05&2.78&28.16&4.66&0.09&0.96&19.19&4.92&0.06\\
&$\bar{\nu}_\mu$&35.53&12.59&2.32&5.56&10.24&20.03&5.07&0.02&4.04&29.72&4.53&0.05&1.12&20.46&5.03&0.03\\
&$\bar{\nu}_\tau$&34.71&12.56&2.31&5.57&9.98&20.01&5.10&0.02&3.92&29.71&4.56&0.05&1.09&20.44&5.05&0.03\\
&$\nu_e$&35.98&9.70&3.07&5.17&-&-&-&-&-&-&-&-&-&-&-&\\
&$\nu_\mu$&33.49&12.08&2.36&5.57&-&-&-&-&-&-&-&-&-&-&-&\\
&$\nu_\tau$&33.37&12.08&2.36&5.58&-&-&-&-&-&-&-&-&-&-&-&\\
1.44-LS220&$\bar{\nu}_e$&32.83&12.42&2.40&6.79&9.25&19.59&4.74&0.06&3.39&29.24&3.46&0.10&1.01&20.02&4.55&0.08\\
&$\bar{\nu}_\mu$&37.91&12.87&1.94&6.83&11.59&21.37&4.62&0.03&5.54&31.21&3.18&0.08&1.27&21.76&4.43&0.04\\
&$\bar{\nu}_\tau$&36.86&12.80&1.93&6.93&11.20&21.33&4.69&0.03&5.33&31.20&3.21&0.08&1.23&21.73&4.50&0.04\\
&$\nu_e$&37.07&9.87&2.78&6.50&-&-&-&-&-&-&-&-&-&-&-&\\
&$\nu_\mu$&35.37&12.27&2.00&7.04&-&-&-&-&-&-&-&-&-&-&-&\\
&$\nu_\tau$&35.31&12.27&2.00&7.00&-&-&-&-&-&-&-&-&-&-&-&\\
1.44-SFHo&$\bar{\nu}_e$&34.88&12.61&2.78&6.34&9.76&19.12&5.74&0.05&3.29&28.59&5.19&0.11&1.07&19.55&5.68&0.06\\
&$\bar{\nu}_\mu$&39.08&12.82&2.24&6.67&11.60&20.54&5.89&0.02&4.94&30.29&4.98&0.06&1.27&20.95&5.81&0.03\\
&$\bar{\nu}_\tau$&38.13&12.79&2.24&6.67&11.29&20.51&5.92&0.02&4.78&30.27&5.04&0.06&1.24&20.92&5.85&0.03\\
&$\nu_e$&38.83&9.94&3.00&6.12&-&-&-&-&-&-&-&-&-&-&-&\\
&$\nu_\mu$&36.81&12.28&2.29&6.68&-&-&-&-&-&-&-&-&-&-&-&\\
&$\nu_\tau$&36.62&12.29&2.29&6.69&-&-&-&-&-&-&-&-&-&-&-&\\
1.44-SFHx&$\bar{\nu}_e$&34.93&12.61&2.78&6.44&9.77&19.12&5.82&0.05&3.29&28.59&5.24&0.11&1.07&19.54&5.76&0.06\\
&$\bar{\nu}_\mu$&39.25&12.82&2.24&6.80&11.64&20.54&5.98&0.03&4.96&30.30&5.02&0.09&1.28&20.95&5.90&0.03\\
&$\bar{\nu}_\tau$&38.30&12.78&2.24&6.79&11.33&20.51&6.02&0.03&4.80&30.28&5.08&0.09&1.24&20.92&5.94&0.03\\
&$\nu_e$&38.81&9.94&3.00&6.19&-&-&-&-&-&-&-&-&-&-&-&\\
&$\nu_\mu$&36.99&12.28&2.29&6.80&-&-&-&-&-&-&-&-&-&-&-&\\
&$\nu_\tau$&36.78&12.28&2.29&6.81&-&-&-&-&-&-&-&-&-&-&-&\\
\hline
1.62-DD2&$\bar{\nu}_e$&43.36&13.01&2.86&5.79&12.59&19.49&5.35&0.08&4.50&28.83&4.97&0.07&1.38&19.88&5.31&0.05\\
&$\bar{\nu}_\mu$&44.61&12.90&2.28&6.10&13.30&20.55&5.57&0.03&5.66&30.24&4.96&0.04&1.46&20.95&5.52&0.03\\
&$\bar{\nu}_\tau$&43.54&12.86&2.27&6.11&12.94&20.52&5.59&0.03&5.48&30.23&5.00&0.04&1.42&20.92&5.54&0.03\\
&$\nu_e$&47.18&10.24&2.80&5.59&-&-&-&-&-&-&-&-&-&-&-&\\
&$\nu_\mu$&42.00&12.36&2.32&6.11&-&-&-&-&-&-&-&-&-&-&-&\\
&$\nu_\tau$&41.83&12.36&2.32&6.13&-&-&-&-&-&-&-&-&-&-&-&\\
1.62-LS220&$\bar{\nu}_e$&43.99&13.13&2.36&7.00&13.35&20.64&4.85&0.05&5.74&30.22&3.57&0.15&1.47&21.02&4.67&0.05\\
&$\bar{\nu}_\mu$&47.10&13.32&1.92&7.26&15.03&22.08&4.94&0.03&7.84&31.91&3.47&0.06&1.65&22.45&4.75&0.04\\
&$\bar{\nu}_\tau$&45.67&13.23&1.90&7.40&14.48&22.03&5.02&0.03&7.51&31.89&3.50&0.06&1.59&22.40&4.82&0.04\\
&$\nu_e$&48.43&10.56&2.47&6.65&-&-&-&-&-&-&-&-&-&-&-&\\
&$\nu_\mu$&43.73&12.68&1.97&7.45&-&-&-&-&-&-&-&-&-&-&-&\\
&$\nu_\tau$&43.70&12.66&1.97&7.47&-&-&-&-&-&-&-&-&-&-&-&\\
1.62-SFHo&$\bar{\nu}_e$&47.20&13.18&2.71&6.91&14.06&19.99&6.21&0.05&5.46&29.40&5.53&0.14&1.54&20.36&6.15&0.05\\
&$\bar{\nu}_\mu$&49.54&13.18&2.18&7.38&15.28&21.17&6.51&0.03&7.09&30.92&5.50&0.05&1.68&21.55&6.44&0.04\\
&$\bar{\nu}_\tau$&48.30&13.14&2.18&7.38&14.84&21.13&6.56&0.03&6.85&30.90&5.56&0.05&1.63&21.51&6.48&0.04\\
&$\nu_e$&51.30&10.52&2.71&6.68&-&-&-&-&-&-&-&-&-&-&-&\\
&$\nu_\mu$&46.63&12.61&2.23&7.39&-&-&-&-&-&-&-&-&-&-&-&\\
&$\nu_\tau$&46.33&12.61&2.23&7.40&-&-&-&-&-&-&-&-&-&-&-&\\
1.62-SFHx&$\bar{\nu}_e$&47.43&13.19&2.71&7.04&14.15&20.01&6.31&0.05&5.52&29.44&5.58&0.13&1.55&20.39&6.24&0.05\\
&$\bar{\nu}_\mu$&49.98&13.18&2.17&7.54&15.43&21.20&6.62&0.03&7.19&30.96&5.54&0.04&1.70&21.58&6.53&0.04\\
&$\bar{\nu}_\tau$&48.71&13.14&2.17&7.54&14.98&21.15&6.67&0.03&6.94&30.94&5.61&0.04&1.65&21.53&6.58&0.04\\
&$\nu_e$&51.44&10.52&2.70&6.77&-&-&-&-&-&-&-&-&-&-&-&\\
&$\nu_\mu$&47.05&12.61&2.22&7.54&-&-&-&-&-&-&-&-&-&-&-&\\
&$\nu_\tau$&46.72&12.61&2.22&7.56&-&-&-&-&-&-&-&-&-&-&-&\\
\hline
\end{tabular*}
 \end{table}

\clearpage

\begin{table}[ht]
\vskip-5pt
TABLE~\ref{tab:Neutrino-flux-properties} (continued). Time-integrated neutrino flux properties of our supernova models.
\vskip4pt
    \begin{tabular*}{\textwidth}{@{\extracolsep{\fill}}llllllllllllllllll}
    \hline\hline
    &&&&&&\multicolumn{4}{l}{Kam-II}&\multicolumn{4}{l}{IMB}&\multicolumn{4}{l}{BUST}\\
Model      & &$E_{\rm tot}^{\rm end}$& $\langle \epsilon_\nu\rangle$&$\alpha$&$\tau_E$&$N$&$\langle\epsilon_e\rangle$&$\tau$&$\delta t$&$N$&$\langle\epsilon_e\rangle$&$\tau$&$\delta t$&$N$&$\langle\epsilon_e\rangle$&$\tau$&$\delta t$\\
           & & [B]         & [MeV]                        &        &[s]&&[MeV]&[s]&[s]&&[MeV]&[s]&[s]&&[MeV]&[s]&[s]\\
\hline

1.77-DD2&$\bar{\nu}_e$&52.72&13.30&2.80&6.20&15.81&20.00&5.72&0.06&6.14&29.39&5.27&0.15&1.74&20.37&5.68&0.08\\
&$\bar{\nu}_\mu$&53.28&13.11&2.25&6.57&16.26&20.91&6.00&0.03&7.28&30.61&5.34&0.10&1.79&21.29&5.94&0.04\\
&$\bar{\nu}_\tau$&51.96&13.07&2.24&6.59&15.80&20.88&6.03&0.03&7.04&30.59&5.38&0.10&1.73&21.26&5.97&0.04\\
&$\nu_e$&56.45&10.54&2.64&5.99&-&-&-&-&-&-&-&-&-&-&-&\\
&$\nu_\mu$&50.13&12.56&2.29&6.57&-&-&-&-&-&-&-&-&-&-&-&\\
&$\nu_\tau$&49.88&12.55&2.29&6.61&-&-&-&-&-&-&-&-&-&-&-&\\
1.77-LS220&$\bar{\nu}_e$&53.88&13.46&2.25&7.58&16.99&21.36&5.08&0.06&8.07&31.00&3.75&0.15&1.87&21.72&4.89&0.09\\
&$\bar{\nu}_\mu$&56.60&13.53&1.84&7.98&18.52&22.59&5.24&0.03&10.25&32.41&3.75&0.10&2.04&22.94&5.05&0.05\\
&$\bar{\nu}_\tau$&54.78&13.43&1.82&8.17&17.80&22.53&5.32&0.03&9.79&32.39&3.78&0.10&1.96&22.89&5.12&0.05\\
&$\nu_e$&58.30&10.90&2.26&7.16&-&-&-&-&-&-&-&-&-&-&-&\\
&$\nu_\mu$&52.39&12.88&1.89&8.22&-&-&-&-&-&-&-&-&-&-&-&\\
&$\nu_\tau$&52.36&12.84&1.89&8.26&-&-&-&-&-&-&-&-&-&-&-&\\
1.77-SFHo&$\bar{\nu}_e$&57.63&13.53&2.65&7.42&17.81&20.59&6.67&0.06&7.58&30.08&5.87&0.15&1.96&20.95&6.60&0.09\\
&$\bar{\nu}_\mu$&59.51&13.45&2.16&7.96&18.86&21.62&7.04&0.03&9.27&31.37&5.94&0.10&2.08&21.98&6.95&0.04\\
&$\bar{\nu}_\tau$&57.98&13.40&2.15&7.97&18.30&21.57&7.09&0.03&8.94&31.35&6.01&0.10&2.01&21.93&7.01&0.04\\
&$\nu_e$&61.68&10.87&2.55&7.17&-&-&-&-&-&-&-&-&-&-&-&\\
&$\nu_\mu$&55.98&12.86&2.20&7.96&-&-&-&-&-&-&-&-&-&-&-&\\
&$\nu_\tau$&55.56&12.85&2.20&7.99&-&-&-&-&-&-&-&-&-&-&-&\\
1.77-SFHx&$\bar{\nu}_e$&57.98&13.53&2.63&7.60&17.93&20.62&6.80&0.06&7.66&30.11&5.93&0.15&1.97&20.98&6.72&0.08\\
&$\bar{\nu}_\mu$&60.17&13.45&2.14&8.17&19.08&21.66&7.18&0.03&9.43&31.43&5.99&0.10&2.10&22.02&7.08&0.04\\
&$\bar{\nu}_\tau$&58.59&13.40&2.13&8.18&18.50&21.60&7.23&0.03&9.08&31.40&6.07&0.10&2.04&21.97&7.15&0.04\\
&$\nu_e$&61.91&10.86&2.53&7.32&-&-&-&-&-&-&-&-&-&-&-&\\
&$\nu_\mu$&56.60&12.86&2.19&8.17&-&-&-&-&-&-&-&-&-&-&-&\\
&$\nu_\tau$&56.14&12.84&2.18&8.20&-&-&-&-&-&-&-&-&-&-&-&\\
\hline
1.93-DD2&$\bar{\nu}_e$&63.66&13.56&2.74&6.66&19.62&20.45&6.13&0.05&8.16&29.85&5.61&0.10&2.16&20.80&6.08&0.08\\
&$\bar{\nu}_\mu$&63.63&13.33&2.20&7.07&19.88&21.31&6.45&0.03&9.39&31.02&5.73&0.10&2.19&21.68&6.40&0.04\\
&$\bar{\nu}_\tau$&62.00&13.28&2.20&7.09&19.29&21.27&6.48&0.03&9.07&31.00&5.78&0.10&2.12&21.64&6.43&0.04\\
&$\nu_e$&67.22&10.80&2.54&6.44&-&-&-&-&-&-&-&-&-&-&-&\\
&$\nu_\mu$&59.83&12.76&2.25&7.07&-&-&-&-&-&-&-&-&-&-&-&\\
&$\nu_\tau$&59.46&12.74&2.24&7.11&-&-&-&-&-&-&-&-&-&-&-&\\
1.93-LS220&$\bar{\nu}_e$&65.64&13.80&2.15&8.25&21.46&22.06&5.35&0.06&11.12&31.69&4.00&0.10&2.37&22.39&5.19&0.08\\
&$\bar{\nu}_\mu$&68.38&13.79&1.76&8.82&23.00&23.19&5.56&0.03&13.58&32.99&4.06&0.10&2.54&23.52&5.38&0.04\\
&$\bar{\nu}_\tau$&66.05&13.67&1.74&9.04&22.05&23.12&5.65&0.03&12.93&32.95&4.09&0.10&2.43&23.45&5.45&0.04\\
&$\nu_e$&69.93&11.24&2.11&7.75&-&-&-&-&-&-&-&-&-&-&-&\\
&$\nu_\mu$&63.12&13.11&1.81&9.03&-&-&-&-&-&-&-&-&-&-&-&\\
&$\nu_\tau$&63.04&13.06&1.80&9.15&-&-&-&-&-&-&-&-&-&-&-&\\
1.93-SFHo&$\bar{\nu}_e$&69.96&13.83&2.57&8.05&22.31&21.15&7.20&0.05&10.26&30.66&6.25&0.10&2.46&21.48&7.13&0.08\\
&$\bar{\nu}_\mu$&71.72&13.70&2.09&8.66&23.34&22.11&7.63&0.03&12.19&31.88&6.40&0.10&2.57&22.46&7.54&0.04\\
&$\bar{\nu}_\tau$&69.80&13.65&2.09&8.67&22.61&22.05&7.68&0.03&11.73&31.85&6.48&0.10&2.49&22.40&7.60&0.04\\
&$\nu_e$&73.83&11.16&2.42&7.79&-&-&-&-&-&-&-&-&-&-&-&\\
&$\nu_\mu$&67.42&13.09&2.14&8.65&-&-&-&-&-&-&-&-&-&-&-&\\
&$\nu_\tau$&66.81&13.07&2.14&8.70&-&-&-&-&-&-&-&-&-&-&-&\\
1.93-SFHx&$\bar{\nu}_e$&70.70&13.84&2.55&8.32&22.59&21.20&7.41&0.05&10.46&30.72&6.37&0.10&2.49&21.54&7.33&0.08\\
&$\bar{\nu}_\mu$&72.88&13.70&2.08&8.96&23.74&22.17&7.85&0.03&12.48&31.96&6.52&0.10&2.62&22.51&7.76&0.04\\
&$\bar{\nu}_\tau$&70.88&13.64&2.07&8.97&22.98&22.10&7.92&0.03&11.99&31.92&6.62&0.10&2.53&22.45&7.83&0.04\\
&$\nu_e$&74.37&11.15&2.39&8.03&-&-&-&-&-&-&-&-&-&-&-&\\
&$\nu_\mu$&68.51&13.09&2.12&8.95&-&-&-&-&-&-&-&-&-&-&-&\\
&$\nu_\tau$&67.84&13.06&2.12&9.00&-&-&-&-&-&-&-&-&-&-&-&\\
\hline
 \end{tabular*}
 \textbf{\vskip-5pt}
\end{table}

\clearpage

\begin{table}[ht]
\vskip-5pt
TABLE~\ref{tab:Neutrino-flux-properties} (continued). Time-integrated neutrino flux properties of our supernova models.
\vskip4pt
    \begin{tabular*}{\textwidth}{@{\extracolsep{\fill}}llllllllllllllllll}
    \hline\hline
    &&&&&&\multicolumn{4}{l}{Kam-II}&\multicolumn{4}{l}{IMB}&\multicolumn{4}{l}{BUST}\\
Model      & &$E_{\rm tot}^{\rm end}$& $\langle \epsilon_\nu\rangle$&$\alpha$&$\tau_E$&$N$&$\langle\epsilon_e\rangle$&$\tau$&$\delta t$&$N$&$\langle\epsilon_e\rangle$&$\tau$&$\delta t$&$N$&$\langle\epsilon_e\rangle$&$\tau$&$\delta t$\\
           & & [B]         & [MeV]                        &        &[s]&&[MeV]&[s]&[s]&&[MeV]&[s]&[s]&&[MeV]&[s]&[s]\\
\hline
1.62-DD2-c&$\bar{\nu}_e$&42.30&12.51&2.62&8.12&11.84&19.31&6.84&0.08&4.13&28.90&5.59&0.14&1.29&19.74&6.70&0.08\\
&$\bar{\nu}_\mu$&43.59&12.07&2.15&8.93&12.04&19.77&7.57&0.03&4.58&29.71&5.91&0.03&1.31&20.24&7.41&0.04\\
&$\bar{\nu}_\tau$&42.83&12.01&2.15&9.06&11.76&19.71&7.70&0.03&4.44&29.69&6.00&0.03&1.28&20.19&7.54&0.04\\
&$\nu_e$&45.89&9.90&2.59&8.09&-&-&-&-&-&-&-&-&-&-&-&\\
&$\nu_\mu$&41.04&11.56&2.19&9.09&-&-&-&-&-&-&-&-&-&-&-&\\
&$\nu_\tau$&41.26&11.58&2.19&9.09&-&-&-&-&-&-&-&-&-&-&-&\\
1.61-LS220-c&$\bar{\nu}_e$&43.43&12.27&2.03&12.07&12.30&20.39&7.83&0.05&5.14&30.44&4.77&0.15&1.35&20.82&7.51&0.08\\
&$\bar{\nu}_\mu$&45.11&11.99&1.74&12.77&12.72&20.92&8.41&0.03&5.76&31.19&5.01&0.11&1.39&21.38&8.09&0.04\\
&$\bar{\nu}_\tau$&44.02&11.89&1.73&12.98&12.29&20.83&8.56&0.03&5.51&31.14&5.12&0.11&1.34&21.30&8.23&0.04\\
&$\nu_e$&47.76&10.03&2.21&11.99&-&-&-&-&-&-&-&-&-&-&-&\\
&$\nu_\mu$&42.00&11.43&1.79&13.08&-&-&-&-&-&-&-&-&-&-&-&\\
&$\nu_\tau$&42.35&11.44&1.80&13.07&-&-&-&-&-&-&-&-&-&-&-&\\
1.62-SFHo-c&$\bar{\nu}_e$&45.66&12.43&2.40&11.39&12.79&19.69&8.92&0.08&4.78&29.49&6.67&0.14&1.40&20.12&8.73&0.08\\
&$\bar{\nu}_\mu$&47.99&11.98&2.04&12.62&13.14&20.02&10.01&0.03&5.22&30.12&7.30&0.03&1.43&20.49&9.80&0.04\\
&$\bar{\nu}_\tau$&47.15&11.91&2.03&12.75&12.82&19.95&10.18&0.03&5.03&30.08&7.45&0.03&1.40&20.42&9.98&0.04\\
&$\nu_e$&49.52&10.01&2.42&11.19&-&-&-&-&-&-&-&-&-&-&-&\\
&$\nu_\mu$&45.12&11.44&2.07&12.86&-&-&-&-&-&-&-&-&-&-&-&\\
&$\nu_\tau$&45.39&11.47&2.08&12.79&-&-&-&-&-&-&-&-&-&-&-&\\
1.62-SFHx-c&$\bar{\nu}_e$&45.36&12.40&2.39&11.88&12.64&19.70&9.13&0.08&4.73&29.55&6.74&0.14&1.38&20.14&8.94&0.08\\
&$\bar{\nu}_\mu$&47.84&11.90&2.04&13.18&12.94&19.95&10.29&0.03&5.08&30.09&7.47&0.03&1.41&20.43&10.08&0.04\\
&$\bar{\nu}_\tau$&47.06&11.83&2.03&13.31&12.63&19.88&10.44&0.03&4.91&30.04&7.62&0.03&1.37&20.36&10.24&0.04\\
&$\nu_e$&49.21&9.99&2.40&11.52&-&-&-&-&-&-&-&-&-&-&-&\\
&$\nu_\mu$&45.02&11.38&2.07&13.37&-&-&-&-&-&-&-&-&-&-&-&\\
&$\nu_\tau$&45.31&11.40&2.08&13.36&-&-&-&-&-&-&-&-&-&-&-&\\
\hline
1.62-DD2-m&$\bar{\nu}_e$&43.38&13.05&2.93&5.48&12.59&19.43&5.13&0.05&4.45&28.74&4.82&0.07&1.38&19.82&5.09&0.05\\
&$\bar{\nu}_\mu$&43.40&12.95&2.31&5.77&12.98&20.55&5.36&0.03&5.52&30.22&4.86&0.04&1.42&20.94&5.32&0.04\\
&$\bar{\nu}_\tau$&43.40&12.95&2.31&5.77&12.98&20.55&5.36&0.03&5.52&30.22&4.86&0.04&1.42&20.94&5.32&0.04\\
&$\nu_e$&46.88&10.24&2.87&5.32&-&-&-&-&-&-&-&-&-&-&-&\\
&$\nu_\mu$&41.68&12.44&2.36&5.79&-&-&-&-&-&-&-&-&-&-&-&\\
&$\nu_\tau$&41.68&12.44&2.36&5.79&-&-&-&-&-&-&-&-&-&-&-&\\
1.62-SFHo-m&$\bar{\nu}_e$&47.14&13.22&2.78&6.47&14.03&19.91&5.89&0.09&5.37&29.29&5.32&0.07&1.54&20.28&5.84&0.04\\
&$\bar{\nu}_\mu$&48.17&13.22&2.21&6.94&14.89&21.15&6.22&0.03&6.89&30.88&5.35&0.05&1.64&21.53&6.15&0.04\\
&$\bar{\nu}_\tau$&48.17&13.22&2.21&6.94&14.89&21.15&6.22&0.03&6.89&30.88&5.35&0.05&1.64&21.53&6.15&0.04\\
&$\nu_e$&50.90&10.51&2.77&6.28&-&-&-&-&-&-&-&-&-&-&-&\\
&$\nu_\mu$&46.20&12.68&2.27&6.96&-&-&-&-&-&-&-&-&-&-&-&\\
&$\nu_\tau$&46.20&12.68&2.27&6.96&-&-&-&-&-&-&-&-&-&-&-&\\
\hline
 \end{tabular*}
  \textbf{\vskip-5pt}
%
\vskip12pt
\caption{Temporal properties of the signal for our 1.62\,M$_\odot$ models, with and without convection. For each experiment, the signal is computed assuming $\bar{\nu}_e$ alone without accounting for any flavor conversion. For model 1.62-LS220, the corresponding model without convection is 1.61-LS220-c. 
\label{tab:with-without-convection}}
\vskip2pt
\begin{tabular*}{\textwidth}{@{\extracolsep{\fill}}lllllllll}
\hline\hline
    &\multicolumn{2}{l}{Cooling timescale $\tau_E$ [s]}&\multicolumn{2}{l}{Kam-II duration $\tau$ [s]}&\multicolumn{2}{l}{IMB duration $\tau$ [s]}&\multicolumn{2}{l}{BUST duration $\tau$ [s]}\\
Model      & w conv. & w/o conv. & w conv. & w/o conv.& w conv. & w/o conv.& w conv. & w/o conv.\\
1.62-DD2&5.99&8.78&5.35&6.84&4.97&5.59&5.31&6.70\\
1.62-LS220&7.22&12.67&4.85&7.83&3.57&4.77&4.67&7.51\\
1.62-SFHo&7.21&12.35&6.21&8.92&5.53&6.67&6.15&8.73\\
1.62-SFHx&7.36&12.87&6.31&9.13&5.58&6.74&6.24&8.94\\
\hline
\end{tabular*}
 \hbox to\textwidth{\hbox to3em{\hfil$\tau_E$~~}\hbox to15cm{Period over which 95\% of the total energy ($E_\mathrm{tot}^{\rm end}$) are emitted\hfil}\hfil}
 \hbox to\textwidth{\hbox to3em{\hfil$\tau$~~}\hbox to15cm{Period over which 95\% of the total events are detected\hfil}\hfil}
 \vskip-30pt
\end{table}

\clearpage

\twocolumngrid

\bibliographystyle{bibi}
\bibliography{References}

\end{document}